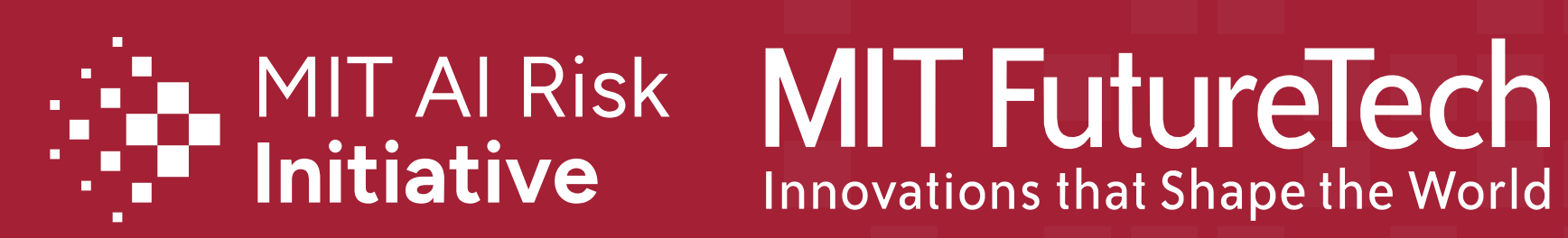

# Prioritization of Risks from Artificial Intelligence

## A Delphi Study of 272 International Experts

**June 2026**

# Prioritization of Risks from Artificial Intelligence: A Delphi Study of 272 International Experts

Alexander K Saeri, Jess Graham, Michael Noetel*, Peter Slattery, Dennis Ah-king, Edla Aittokallio, Ibitola Akindehin, Abbas Al Mahdi, Elie Alhajjar, Rafael Andersson Lipcsey, Gary Ang, Catherine M Azam, Amos Azaria, Rishal Balkissoon, Isabel Barberá, Claudio Bareato, Jonathan Barry, Michael Basehart, Andrew M Bean, Danny Belitz, Samantha Augusta Bennett, Kayla Blomquist, Damian Borstel, Ben Bucknall, Tomas Bueno Momcilovic, Aurelie Bugeau, Nicholas Caputo, Stephen Casper, Gulam Chagani, Ze Shen Chin, Jiyeon Cho, Jay Chooi, Joel N Christoph, Dmytro Chumachenko, Kieran Conboy, Elizabeth M Daly, Tom David, Paul de Font-Reaulx, Antonio De Santis, Fabrizio Degni, Christopher W DiCarlo, Yawen Duan, Janet Egan, Ian W Eisenberg, Sherif M Elsafty, Adam Ennamli, Mark Esposito, Nicola Fabiano, Gallo Fall, Neil R Fernandes, Pip Foweraker, Chiara Gallese, Sandra Galletti, Andrew Gamino-Cheong, Rokas Gipiškis, Gwyn Glasser, Delaram Golpayegani, Jeff Grayson, Hans Gundlach, Josiah Hagen, Alexander Hagenah, Amelia S Haines, The Anh Han, Yixiong Hao, Kasii Harris, Tianxing He, Koen Holtman, Giorgos Iacovides, Kenneth L Ingham, Krystal Jackson, Adam Jones, Himanshu Joshi, Brian Judge, Arturs Kanepajs, Shreya Kapoor, Win Myat Nwe Khine, Aidan Kierans, Aleksandra Korolova, Markus Krebsz, Nicholas Kruus, Joe Kwon, Valeria Lazzaroli, Ray X Lee, Evelina Leivada, Stephan Lewandowsky, Michael B Li, Xiaojian Li, Geunsik Lim, Henrique Lisakowski, Fabio Lonardoni, Todd C Lowe, Jackson G Lu, Alexander Lyzhov, Nada Madkour, Parv Mahajan, David Manheim, Kareem Mathias, Claudio Mayrink Verdun, Sean McGregor, Scott McLean, Matthew J McMahon, Minas Megalokonomos, Nicolas Moës, Fernando Mourao, Yaroslav Mukhin, Malcolm Murray, Simon Mylius, Neeraj Nagpal, Koichi Nakada, Anna Neumann, Jessica Newman, Kwan Yee Ng, Minh N Nguyen, Quynh Phuong Nguyen, Seán S Ó hÉigeartaigh, Daria Onitiu, Kelly Onu, Oscar Oviedo-Trespalacios, Ugur Ozer, Chanwoo Park, M. Alejandra Parra-Orlandoni, Patricia Paskov, Anna M Pastwa, Burak Piskin, Jacob Pratt, Claudiu A Predincea, Marjana Prifti Skenduli, Kenneth Priore, Mukunda Madhab Pujari, Zhenting Qi, Preethi Raghunathan, Robi Rahman, Deepika Raman, Max Reddel, Jyoti Ruparel, Emma B Ruttkamp-Bloem, Tiffany Saade, Greg Sadler, Said Saillant, Paul M Salmon, Ayrton San Joaquin, Lama Saouma, Maziya Sarangpurwala, Supheakmungkol Sarin, Daniel S Schiff, Anna D Schilling, Chris Schmitz, Reva Schwartz, Abeer Sharma, Tianhao Shen, Kehan Sheng, Maury D Shenk, Eli Sherman, Chandler Smith, Julie M Smith, Estevenson Solano, Oliver Sourbut, Madhulika Srikumar, Ryan Stendall, Jakob Stenseke, Michael Stern, Joshua Sternfeld, Nikko Stevens, Ilia Sucholutsky, Yuanyuan Sun, Mariami Tkeshelashvili, Cristian Trout, Brian Tse, Nikolaos Tsinganos, Michelle Vaccaro, Anthony R Valiaveedu, Ramakrishnan Veeramony, Jeremy Verdo, Pulkit Verma, Andrea Luigi Vitali, Jinge Wang, JR Washebek, Yonah Welker, George F Westerman, James Williams, Tristan Williams, Rongwu Xu, Mick Yang, Xuemeng Yang, Sander Zeijlemaker, Jingyu Zhang, Marta Ziosi, Neil Thompson

Affiliations and ORCiDs at the end of the document

* Corresponding Author: m.noetel@uq.edu.au; The University of Queensland, St Lucia, Queensland, 4072, Australia

# Abstract

Artificial intelligence poses many risks, ranging from familiar present-day harms to unprecedented and potentially catastrophic ones. Effective risk management requires prioritization: we must understand which risks are most severe, who is most vulnerable, and who is most responsible for addressing them. We report results from a three-round Delphi study conducted late 2025 with 272 international AI experts. Experts rated 24 AI risks on harm probability and severity, sector and actor vulnerability, actor responsibility, and overall concern. Experts estimated the five most severe harms in the next 5 years were likely to come from dangerous capabilities, competitive dynamics, weapons & cyberattacks (including CBRNE), power centralization, and false information. In a business-as-usual scenario, experts judged 18 of 24 risks as having a more than 10% probability of catastrophic outcomes (e.g., more than 1 million deaths or more than USD 100B in financial loss) in the next 5 years (2025-2030). In a scenario where pragmatic mitigations are implemented, experts still judged five risks as having a more than 10% probability of catastrophic outcomes: dangerous capabilities, weapons & cyberattacks, environmental harm, inequality & unemployment, and power centralization. All 24 risks were judged as being more than 5% likely to cause catastrophic outcomes. AI users and the general public were judged the most vulnerable to these risks, but experts assigned the highest responsibility for addressing them to general-purpose AI developers and governance actors (including governments, regulators, and standards bodies). Across most risks, experts identified information, finance, and national security as the most vulnerable sectors. These findings can guide AI risk prioritization and clarify expert expectations about who should bear responsibility for mitigation.

## Highlights

- Experts assessed at least 10% probability of catastrophic harm (e.g., more than 1 million human deaths or more than USD $100B loss) from 18 of 24 AI risk domains under business-as-usual trajectories over the next 5 years
- AI users and affected stakeholders were judged most vulnerable to AI risks; while general-purpose AI developers, governments, regulators, and standards bodies were judged most responsible for addressing AI risks
- Information, finance, and national security were judged as the most vulnerable sectors across risks
- Experts judged pragmatic mitigations would reduce the severity of AI harms, but the likelihood of catastrophic harm from dangerous capabilities, weapons & cyberattacks, environmental harm, inequality & unemployment, and power centralization remained above 10% in a scenario with pragmatic mitigations; all 24 risks had >5% likelihood of catastrophic harm

# Statements and Declarations

## Competing Interests

The core research team (Saeri, Graham, Noetel, Slattery, Thompson) declare no competing interests. Some contributing authors who participated as expert panelists may hold positions in organizations with commercial or policy interests related to AI development or governance. However, all expert contributions were collected anonymously, de-identified, and reported only in aggregate, limiting the potential for individual interests to influence results. No authors with competing interests were involved in the design or analysis of findings.

## Author Contribution Statement (CRediT)

**Alexander K. Saeri:** Conceptualization, Project administration, Investigation, Data curation, Writing – review & editing. **Jess Graham:** Software, Data curation, Investigation, Project administration, Formal analysis, Writing – review & editing. **Michael Noetel:** Methodology, Formal analysis, Writing – original draft, Visualization, Data curation, Validation, Project administration. **Peter Slattery:** Conceptualization, Investigation, Methodology, Writing – review & editing. **Neil Thompson:** Conceptualization, Supervision, Funding acquisition. **All other authors:** Investigation, Writing – review & editing.

## Data and Code Availability

All data is publicly available at https://osf.io/pj2qr

## Generative AI Use

Large language models were used to assist in the presentation and summarization of qualitative expert feedback, with code for data analysis, and with manuscript editing. All content was reviewed and verified by the authors, who take full responsibility for the published article.

## Funding

This work was supported by Commonwealth Bank of Australia, who reviewed the design, but did not influence the collection, analysis, interpretation or reporting of the data.

## Acknowledgements

Many of our authors participated as experts in the Delphi study. In addition, we wish to acknowledge the contributions of the following experts who also generously gave their time to participate: Aithan Shapira, Alex Boussetta, Alexander Meinke, Anka Reuel, Ansgar R Koene, April Chin, Austin Crumpton, Ben Wilkinson, Bill Black, Bram Rijsbosch, Bronte Pendergast, Cameron F Kerry, Catherine Barrett, Daniel J Ragsdale, Daniela Elia, Dimitris Alchatzidis, Elena Gurevich, Elyn Yun Ling Lee, Eyup Engin Kucuk, Fazl Barez, Fion Lee-Madan, Graham H Ryan, Hernan Huwyler, Jimena Sofia Viveros Alvarez, Jodie Levy, Lauriane Aufrant, Layne L Morrison, Lisa Soder, Lucas G Uberti-Bona Marin, Luis E Urtubey, Lynna Leong, Markus P Luchsinger, Matt MacDermott, Matthew Bedsole, Michael J Howell, Minseok Jung, Noam Kolt, Pablo Rice, Patrick Butlin, Paul Kedrosky, Paul Röttger, Possum Michael Hodgkin, Puneet Sondh, Rajiv Dattani, Raz Karmi, Robert Cunningham, Rozita Dara, Samuel T Segun, Simon Goldstein, Stephen Cave, Strahinja Janjusevic, Timo K Harakka, Uma Kalkar, Uzma Chaudhry, and all experts who chose to remain anonymous.

# Executive summary

There are many AI risks. Some are familiar: discrimination, loss of privacy, and fraud. Others are emerging: overreliance, dangerous capabilities being (mis)used in weapons or cyberattacks, and AI systems pursuing unintended goals. To effectively manage these risks, we must understand which are most important to address and who should be responsible for addressing them.

We asked 272 AI experts to prioritize these risks by judging their expected *severity and likelihood*, who is most *vulnerable*, and who should be *responsible*. The highest-severity risks are AI-enabled weapons and cyberattacks, dangerous AI capabilities, extreme power centralization, and false information. In a business-as-usual scenario, experts judged a more than 10% chance of catastrophic outcomes (i.e., 'more than 1 million human deaths or more than a USD 100B in financial loss or civilizational-scale intangible impacts') from 18 of our 24 AI risk domains over the next five years. They judge that five risks—dangerous capabilities, weapons and cyberattacks, power centralization, inequality & unemployment, and environmental harm—have a greater than 10% chance of catastrophic outcomes even if pragmatic mitigations are put in place. And even in the scenario where pragmatic mitigations are in place, experts judge more than 5% chance of catastrophic outcomes from all 24 risk domains surveyed.

Experts identified asymmetries in how vulnerability to AI risks and responsibility for addressing AI risks are distributed across the AI ecosystem. According to experts, general-purpose AI developers and governance actors such as governments, regulators, and standards bodies hold primary responsibility for addressing AI risks. In contrast, AI system users and affected stakeholders are most vulnerable to AI risks. This mismatch means that those who are most responsible for addressing AI risks are not those who are most vulnerable, leading to misaligned incentives in addressing the most important AI risks.

# Introduction

There are many AI risks. The International AI Safety Report has cataloged threats ranging from algorithmic bias to catastrophic loss-of-control [1,2]. Slattery and colleagues synthesized 74 existing frameworks and found most of the 1,725 documented risks could be categorized into 24 risk subdomains [3]. Some of these risks look grave but have been judged to be relatively unlikely. Forecasters in the Existential Risk Persuasion Tournament estimated roughly a 3% chance of human extinction from AI this century [4]. Other risks are already materializing: privacy violations, discriminatory outputs, and deepfake fraud already harm people daily [5].

To effectively manage these risks, we must understand which are most important to address and who should be responsible for addressing them. However, there have been limited attempts to systematically prioritize risks and provide clear guidance on which risks are relevant to whom and to what degree. The NIST AI Risk Management Framework notes that organizations must triage when conducting risk management because resources are finite and risks vary in severity [6]. The ISO 31000 standard for risk management advises prioritizing risks by likelihood-weighted impact, then allocating mitigation effort in proportion to expected harm [6,7].

Risk governance scholarship further argues that translating analysis into action requires identifying who has the capability, obligation, and causal influence to reduce a risk, and holding those actors accountable for implementing safeguards [8–11]. Even well-understood risks can persist when responsibility is diffuse, misaligned with capability, or disconnected from those who bear the harm.

Prior AI risk surveys have assessed the risk from AI as an industry [4,12,13], but few have systematically compared the different AI risks (e.g., misuse vs. malfunction). We aim to fill this gap by asking domain experts to estimate the probability of different harms for 24 different AI risks [3] over the next five years. This requires answering at least three questions: Which AI risks are most severe? Who is most vulnerable to them? Who is most responsible for addressing them? These three questions map onto the structure of standard risk governance: severity tells us which risks to prioritize, vulnerability tells us where harm will land, and responsibility tells us who should act. These estimates can help orient a whole-of-society response toward the risks most likely to cause the greatest harm, and help decision-makers allocate finite resources accordingly.

Prioritization should also extend to *who* is exposed. Vulnerability tells us who is exposed to risks, and how badly. Many claims about AI's effects operate at a whole-of-economy level, or focus on the capabilities of a particular system; there is less discussion of how different stakeholders and sectors are vulnerable, and to which risks. Severity also depends on context: the same technical failure can be negligible in one sector and catastrophic in another. Given the dozens of risks to consider [3], we aimed to help actors prioritize by estimating the size of the risks and judging who is exposed.

However, those who are most vulnerable to AI risks are not always those best positioned to mitigate them, which constitutes a form of moral hazard.[14,15] This is common in many areas of society, where the public are most vulnerable to safety failures, but engineers and governance actors usually bear responsibility for upholding standards.[11] But, to date, efforts to manage many AI risks have been led by voluntary self-governance from frontier AI developers [16,17] and by regulations targeting those developers [5]. As AI is adopted across more sectors of the economy and more incidents happen [5,18], it becomes more important to know who is responsible, in part to help society and regulators determine appropriate accountability frameworks (e.g., liability and insurance). We define responsibility along three dimensions:[8,9] capability (who *can* address it), obligation (who *should* address a risk), and causal influence (who, if they fail to act, allows the harm to materialize). An actor may have substantial capability to address risks from environmental harm but limited capacity to address AI misalignment. By mapping expert judgments about responsibility alongside severity and vulnerability, we can identify where accountability is concentrated, and where there may be gaps between who bears the risk and who is expected to manage it. Commentators have argued that the emerging regulatory landscape resembles a patchwork rather than a coordinated framework [19,20], and it remains unclear which risks governments should prioritize [21].

To answer these questions, we surveyed experts using the Delphi method. This method builds consensus through iterative anonymous consultation [22]. It is well suited to AI risk assessment, where empirical data on many risks is limited and the landscape is changing rapidly [23]. In each round, experts provide ratings and rationales, see peers' responses, and can update their ratings from previous rounds. This serves two purposes. First, it surfaces genuine disagreement between experts, allows experts to learn from each other's reasoning, and mitigates common biases like conformity and status effects [24]. Second, it reduces artificial disagreement: experts who initially differ because of misunderstood definitions or insufficient reasoning can revise over rounds, distinguishing stable consensus from premature convergence [25]. The qualitative rationales also help explain why experts hold these views. In this Delphi, we aimed to prioritize among the many AI risks, and clarify who is seen as most vulnerable and responsible.

# Results

## Risk Severity

### Business as Usual Scenario

The business-as-usual scenario assumes organizations & governments continue their existing practices but do not implement additional AI-specific risk mitigations. Under this scenario, experts assigned highest expected severity to dangerous capabilities, competitive dynamics, weapons & cyberattacks[a], power centralization, and false information (Table 1; Figure 1, left panel).

The severity scale ran from 1 (negligible) to 5 (catastrophic), with each level defined across multiple harm dimensions. Catastrophic harm meant, for example, more than 1 million human deaths (physical harm), more than USD $100B damage (financial loss), or civilization-scale intangible harms (e.g., democratic norms, privacy).[b] Experts gave 18 of 24 risks at least a 10% probability of catastrophic outcomes over the next 5 years (late 2025–late 2030; Figure 1, right panel). Definitions of the 24 AI risks we presented [3] are shown in supplementary materials. Experts allocated substantial probability to both the upper tail (catastrophic) and the middle range (substantial to severe)[c], suggesting uncertainty about the magnitude of harm rather than its occurrence.[d] Quotes justifying their judgments are available in supplementary materials.

Risk to AI welfare was measured but is omitted from Figure 1. None of the 10 areas of harm in the severity rubric directly included AI systems as an entity that could experience harm. Survey guidance instructed experts to "use the Human/Civil Rights column and adjust for AI systems

---

[a] Weapons & Cyberattacks and Dangerous capabilities answer different questions. The first covers humans deliberately using or misusing AI as a tool: building malware, designing bioweapons, deploying autonomous weapons. The second covers the inherent danger from AI capabilities themselves, such as deception, persuasion, cyber-offence, weapons design, and self-proliferation. Those capabilities can cause mass harm through misuse, through misalignment with human goals, or through other failures no one intended. If harm requires a human to wield AI as a weapon, it sits in Weapons & Cyberattacks. If it could occur through misuse, misalignment, or accident alike, it sits in Dangerous capabilities. The two overlap in practice. An AI helping synthesize a bioweapon plausibly belongs in both. Slattery et al. [3] coded each risk into the single most relevant domain rather than treating them as mutually exclusive.

[b] Severity was aggregated across multiple harm areas (e.g., physical harm, economic loss, social harm, institutional disruption). The full list of harm areas and the severity scale thresholds is described in the supplementary files. This included both quantifiable thresholds for tangible harms (e.g., 1 million deaths or USD $100B in financial loss) and qualitative thresholds for intangible harms ("Global democratic collapse or authoritarian lock-in"). However, experts may have focused on one of these harm domains at the expense of others (e.g., if financial harm was easy to quantify). For a given risk, they may have envisioned a different causal pathway to that threshold. Also, anchors at each level might not be equivalent (e.g., if 'authoritarian lock in' would be closer to '>100 million deaths', not '>1 million'). Harms to non-human entities such as AI systems and animals were not directly included. Nevertheless, a rubric was required to get severity estimates across diverse harm areas, so we drew on the best available rubric we could find (Center for Security and Emerging Technology AI Harm Framework[26] and Mylius [27]).

[c] As with all severity ratings, these were defined across many areas of harm (see Supplementary File 4). For example, 'Substantial' defined physical harm as 1-99 casualties and financial loss as USD $1M-$100M; 'Severe' was up to 1 million deaths and $100M-$10B in financial loss.

[d] These values are panel means of subjective probability distributions elicited against our severity rubric, five-year horizon, and scenario framing; they summarize expert belief rather than calibrated real-world frequencies. We use 10% as a descriptive reference level for comparing twenty-four distributions, not an objective threshold. Small changes in the rubric anchors could shift individual domains across the threshold.

where appropriate", but qualitative responses confirmed experts found this translation difficult. Several experts noted their ratings reflected uncertainty about the framing rather than a substantive judgment about the risk's importance (see supplementary materials).

**Table 1. | Five most severe AI risks.**

| Risk name | Short label † | Risk description | Mean severity / 5 | Likelihood of catastrophic harm [95% CI] ^ |
|---|---|---|---|---|
| AI possessing dangerous capabilities | *Dangerous capabilities* | AI systems that develop, access, or are provided with capabilities that increase their potential to cause mass harm through deception, weapons development and acquisition, persuasion and manipulation, political strategy, cyber-offense, AI development, situational awareness, and self-proliferation. These capabilities may cause mass harm due to malicious human actors, misaligned AI systems, or failure in the AI system. | 3.49 | 21.5% [16.9, 26.4] |
| Competitive dynamics | *Competitive dynamics* | Competition by AI developers or state-like actors in an AI "race" by rapidly developing, deploying, and applying AI systems to maximize strategic or economic advantage, increasing the risk they release unsafe and error-prone systems. | 3.49 | 16.6% [12.0, 21.6] |
| Cyberattacks, weapon development or use, and mass harm | *Weapons & Cyberattacks* | Using AI systems to develop cyber weapons (e.g., by coding cheaper, more effective malware), develop new or enhance existing weapons (e.g., Lethal Autonomous Weapons or chemical, biological, radiological, nuclear, and high-yield explosives), or use weapons to cause mass harm. | 3.49 | 21.0% [15.1, 27.5] |
| Power centralization and unfair distribution of benefits | *Power centralization* | AI-driven concentration of power and resources within certain entities or groups, especially those with access to or ownership of powerful AI systems, leading to inequitable distribution of benefits and increased societal inequality. | 3.47 | 18.0% [12.1, 24.8] |
| False or misleading information | *False information* | AI systems that inadvertently generate or spread incorrect or deceptive information, which can lead to inaccurate beliefs in users and undermine their autonomy. Humans that make decisions based on false beliefs can experience physical, emotional, or material harms | 3.44 | 12.8% [8.9, 18.1] |

*Notes.* † The short label is used in figures and text throughout this article; the full risk name and description were reproduced from Slattery et al. and presented to experts in the Delphi study.
^ Refers to mean likelihood of catastrophic harm over the period 2025-2030 under a Business-as-Usual scenario. For all bootstrapped 95% confidence intervals, see supplementary materials.

**Fig. 1 | Expert probability distributions for harm severity.** Side-by-side comparison of experts' average perceived severity for each AI risk under Business as Usual (left) and Pragmatic Mitigations (right) scenarios.[e] Ridgelines show mean probability distributions across experts. Lines show experts' mean catastrophic risk probability 2025–2030. See supplementary materials for ratings of AI welfare risks.

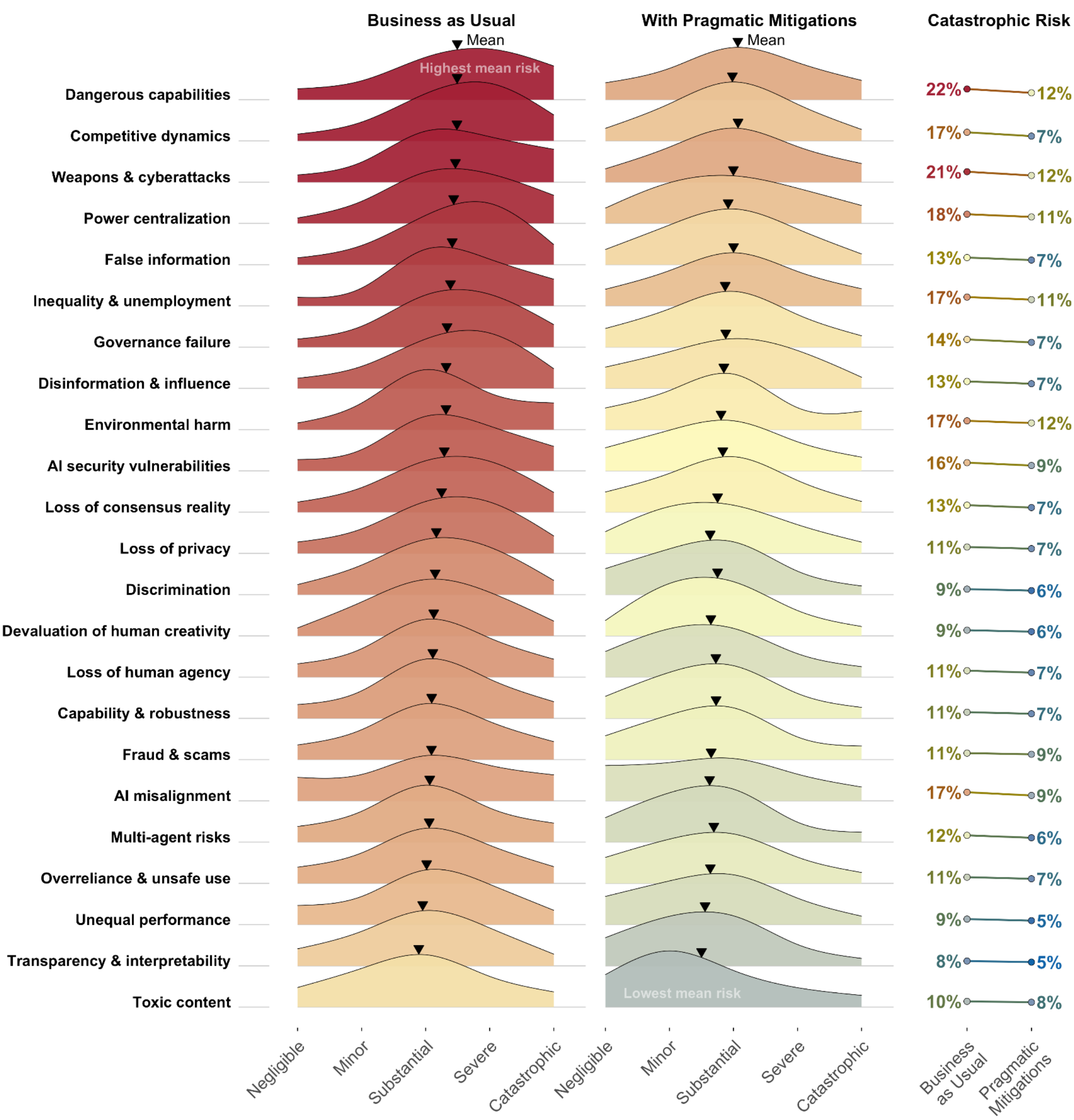

[e] Business as Usual "assumes organizations & governments continue their existing practices but do not implement additional AI-specific risk mitigations". In contrast, Pragmatic Mitigations "assumes organizations & governments make pragmatic and cost-effective efforts to address risks from AI." We kept this description deliberately brief because more specific framings (e.g., a percentage of GDP) risked anchoring respondents on numbers whose real-world meaning would be unclear even to domain experts. We acknowledge that this brevity means different experts may have imagined different policy packages, introducing noise into comparisons between the Business as Usual and Pragmatic Mitigations scenarios.

## Pragmatic Mitigations Scenario

The Pragmatic Mitigations scenario assumes organizations & governments make pragmatic and cost-effective efforts to address risks from AI. Accounting for pragmatic mitigation efforts shifted expected severity downward for all risks, but the magnitude of reduction varied between risks (Figure 1, right panel). Mean severity fell by 0.36–0.53 points on the 1–5 scale (median reduction = 0.44).[f] The largest shifts occurred for risks judged to have a relatively high chance of catastrophic outcomes under business as usual: dangerous capabilities (-10 percentage points [pp] down to 12%), weapons and cyberattacks (-9 pp down to 12%), competitive dynamics (-10 pp down to 7%), AI misalignment (-8 pp down to 9%). Three effects likely contribute to the larger absolute reductions in higher-baseline risks: experts may believe pragmatic mitigations are more tractable for these risks, mitigations might already be underway for those risks, or the reduction may partly reflect mathematical headroom (risks rated higher under Business as Usual have more room to decrease). Our design cannot distinguish these mechanisms, and the difference in scenarios could therefore be interpreted as signals of either perceived tractability or baseline severity.

However, all 24 risks retained at least a 5% catastrophic probability by 2030, even under a Pragmatic Mitigations scenario. Experts still judged many risks to be above 10% probability of catastrophic outcomes even assuming pragmatic mitigations: dangerous capabilities (12%), weapons and cyberattacks (12%), environmental harm (12%), inequality & unemployment (11%), and power centralization (11%).[g] For example, experts argued:

> *Dangerous capabilities emerge as an emergent property of scaling AI systems, which makes them difficult to predict and control. Even with safety measures in place, the potential for catastrophic misuse remains.*
>
> *Cyberattacks from both state and non state actors will be a permanent fixture of AI related risks.*
>
> *Heavy data infrastructure and large models will continue [to] consume energy, cause resource pressure.*
>
> *Inequality is deeply entrenched in economic and social systems. AI may exacerbate existing inequalities through automation-driven job losses in certain sectors while creating wealth for those who own and control AI systems.*
>
> *Power centralization is perhaps the most stubbornly persistent risk because the same entities developing AI are often best positioned to capture its benefits, creating self-reinforcing dynamics that are difficult to reverse through technical interventions alone.*

[f] Paired Wilcoxon signed-rank tests were significant for all 24 risks after Holm correction ($p_{adj} < 0.001$), with uniformly large effect sizes ($r = 0.73$–$0.84$)

[g] The fact that experts ascribe ≥10% catastrophic probability to 18 domains should not be interpreted as experts assessing a high probability that at least one catastrophic outcome will occur; this joint probability was not assessed, and catastrophic outcomes are likely correlated.

# Actor Vulnerability and Responsibility

## Vulnerability Patterns

Experts assessed how vulnerable each of seven AI ecosystem actors were to AI risks using a 5-point scale (1 = Not at all vulnerable to 5 = Extremely vulnerable). Vulnerability was defined as the exposure and sensitivity of an actor to a specific AI risk[h]. Figure 2 displays vulnerability assessments with bolded value-labels indicating consensus (90% within ± 1 of median).

The most vulnerable actors were judged to be AI Users and Affected Stakeholders (members of the public, consumers, employees subject to AI decisions, but who are not necessarily direct users; full definitions in supplementary materials). These two actor categories received median vulnerability ratings of 4–5 (Highly to Extremely vulnerable) across nearly all 24 risks. Experts reached consensus on this assessment for most actor-risk pairs (e.g., that AI Users were extremely vulnerable to power centralization). As one expert argued: “Affected stakeholders bear the impact at population scale.”

The least vulnerable actor was judged to be AI Infrastructure Providers (entities that provide compute, cloud infrastructure, and/or data to train and run AI). This actor received median vulnerability ratings of 2 (Minimally vulnerable) across most risks, with expert consensus. The pattern suggests that providers of compute, cloud, and data infrastructure are largely insulated from many of the risks of AI, with the exception of environmental harm, weapons & cyberattacks, AI security vulnerabilities, and dangerous capabilities.

AI Developers were judged to be moderately vulnerable to AI risks. Both General-Purpose Developers and Specialized AI Developers typically received median ratings of 3–4 (Moderately to Highly vulnerable), reflecting their exposure to liability, reputational harm, and regulatory action, but not the direct harms borne by end users and stakeholders.

Consensus was more common at the extremes. Eighty one percent of "extremely vulnerable" (median = 5) cells reached consensus versus only 14% at “moderately vulnerable”. Experts agree most on who is highly vulnerable and diverge on mid-level judgments.

Actor vulnerability to AI welfare risks was measured, but is omitted from Figure 2. AI systems were not represented as an entity that could be vulnerable, only human institutions or roles. Qualitative responses confirmed experts did not interpret the question consistently (see supplementary materials).

[h] We defined exposure as “the extent to which the actor’s people, operations, and assets interact with or are dependent on AI systems”, and sensitivity as “the extent to which an actor would be harmed if the hazard materialized. The harm may be direct or indirect”.

**Fig. 2 | Expert consensus on actor vulnerability and responsibility for AI risks.** Diverging bar showing expert-median responsibility (left) and vulnerability (right). See supplementary materials for consensus areas and ratings of AI welfare risks.

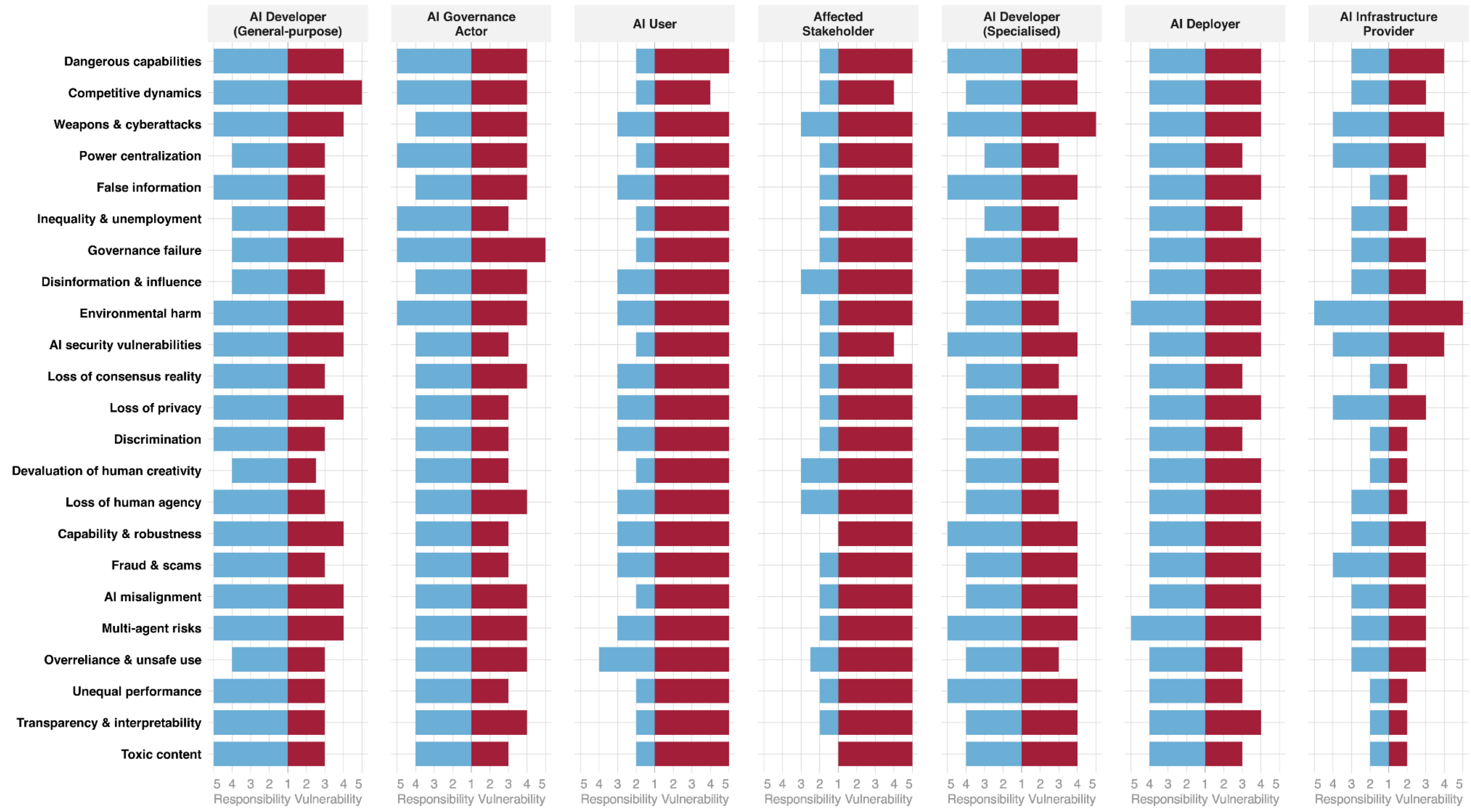

## Responsibility Patterns

Experts also assessed which actors bear responsibility for addressing each risk (1 = Not at all responsible to 5 = Primarily responsible). We defined responsibility as the obligation, capability, and causal influence of an actor for a specific AI risk[i]. Figure 2 also displays responsibility assessments for each actor.

Affected stakeholders and users were judged to have at most 'moderate' responsibility. Despite being most vulnerable, affected stakeholders received median responsibility ratings of 2–3 (Minimally to Moderately responsible) across most risks. Experts clearly distinguish between exposure to harm and obligation to mitigate. As one expert argued:

> *Affected stakeholders lack both the agency and systemic leverage to mitigate risk, and assigning responsibility to them risks reinforcing harm by misplacing accountability.*

In contrast, General-Purpose AI Developers bear highest responsibility. Across nearly all 24 risks, experts assigned median responsibility ratings of 4–5 (Highly to Primarily responsible) to developers of general-purpose foundation models. Consensus was reached on this assignment for most risks, particularly those involving AI system safety, and malicious use. Experts argued the most effective interventions are preventative and centralized rather than reactive and decentralized:

> *I rated AI Developers and Specialized AI Developers as primarily responsible because they decide the model design, structure, critical parameters (such as default setting) which directly impact the outputs and systemic risks.*
>
> *I cannot overstate how much more responsible the 'upstream' actors are for limiting these issues. That's where the emphasis should be. I see a direct analogy to social media. Yes, individuals are responsible for sharing misinfo. But the platforms should bear the brunt of our concern about the issue. They are the best point of intervention.*

Experts assessed that AI Governance Actors share this high responsibility. Governments, regulators, and standards bodies consistently received high responsibility ratings (median 4–5) across most risk categories. This reflects expert expectations that policy and regulation must complement or enforce technical measures by frontier developers:

> *I think governance actors have extreme responsibility for managing these problems, even more than private companies. It ultimately falls on them to protect the public and enact regulations and enforcement.*
>
> *Governance actors are the primary bridge between the affected stakeholder and the AI developer and deployer. Until incentives are aligned to the public interest, governance actors are essential to broker the relationship between AI developers and affected stakeholders.*

Expert consensus was strongest at the extremes (e.g.., primary responsibility assignments) with more disagreement around which actors were "moderately responsible".

---

[i] We defined obligation as "the extent to which the actor should proactively lead or initiate efforts to address the risk"; capability as "the extent to which the actor has specialized skills and resources needed to address the risk"; and causal influence as "the extent to which the actor causes or contributes to the harms resulting from the risk materializing".

Responsibility for addressing AI welfare risks was measured, but is omitted from Figure 2. Qualitative responses indicated experts disagreed about which underlying harm they were assigning responsibility for: harm to AI systems, or harm to humans (see supplementary materials).

## Sector Vulnerability

Experts assessed how vulnerable each of 14 industry sectors is to each of the 24 AI risks. Sectors were adapted from the North American Industry Classification System (NAICS), including Information, Finance and Insurance, Health Care, National Security, Education, and others (full definitions in Supplementary File 7). Figure 3 displays sector vulnerability assessments, with sectors ordered from most to least vulnerable (left to right) based on weighted mean vulnerability scores across all risks.

Experts rated the Information and National Security sectors as most vulnerable across AI risks, with consensus on extreme vulnerability (median 4–5) to content-related harms like disinformation & influence and loss of privacy, and to dangerous capabilities and weapons & cyberattacks respectively. For example, experts argued:

> *The most vulnerable sectors are those where AI is deeply embedded in critical decision-making and where failures have immediate, large-scale consequences—information, national security, and finance all share these characteristics.*
>
> *National security faces compound vulnerability—both as a target for AI-enabled attacks and as a domain where AI failures could trigger cascading geopolitical consequences.*

Finance and Insurance were judged to have similarly high vulnerability to fraud and scams, AI security vulnerabilities, and AI system safety failures, reflecting *"both direct attack vectors (fraud, market manipulation) and regulatory exposure from AI system failures."* (Delphi Participant)

Health Care received high vulnerability ratings (median 4) for loss of privacy, discrimination, and overreliance & unsafe use, given that medical AI failures carry immediate human costs. By contrast, sectors with lower AI penetration, including accommodation and food services, agriculture and manufacturing, and arts and entertainment, received lower vulnerability ratings (median 2–3), though they remain exposed to broad socioeconomic effects like discriminatory outputs and job displacement.

Expert consensus was strongest at the extremes: near-universal agreement on high-vulnerability sectors like Information and National Security, and on low-vulnerability sectors like Accommodation, while moderate-vulnerability sectors showed more disagreement, reflecting uncertainty about how AI adoption will affect these industries.

Vulnerability of sectors to AI welfare risks is also omitted from Figure 3. As with actors, sectors are human-defined categories of economic activity, and AI systems were not represented as an entity that could be vulnerable.

**Fig. 3 | Expert consensus on sector vulnerability for AI risks.** Heatmap showing median vulnerability ratings across risks (rows) and sectors (columns), ordered left-to-right by weighted average vulnerability (higher to lower). Color intensity indicates median rating (1 = not vulnerable, 3 = moderately vulnerable, 5 = extremely vulnerable). See supplementary files for areas of consensus and ratings of AI welfare risks.

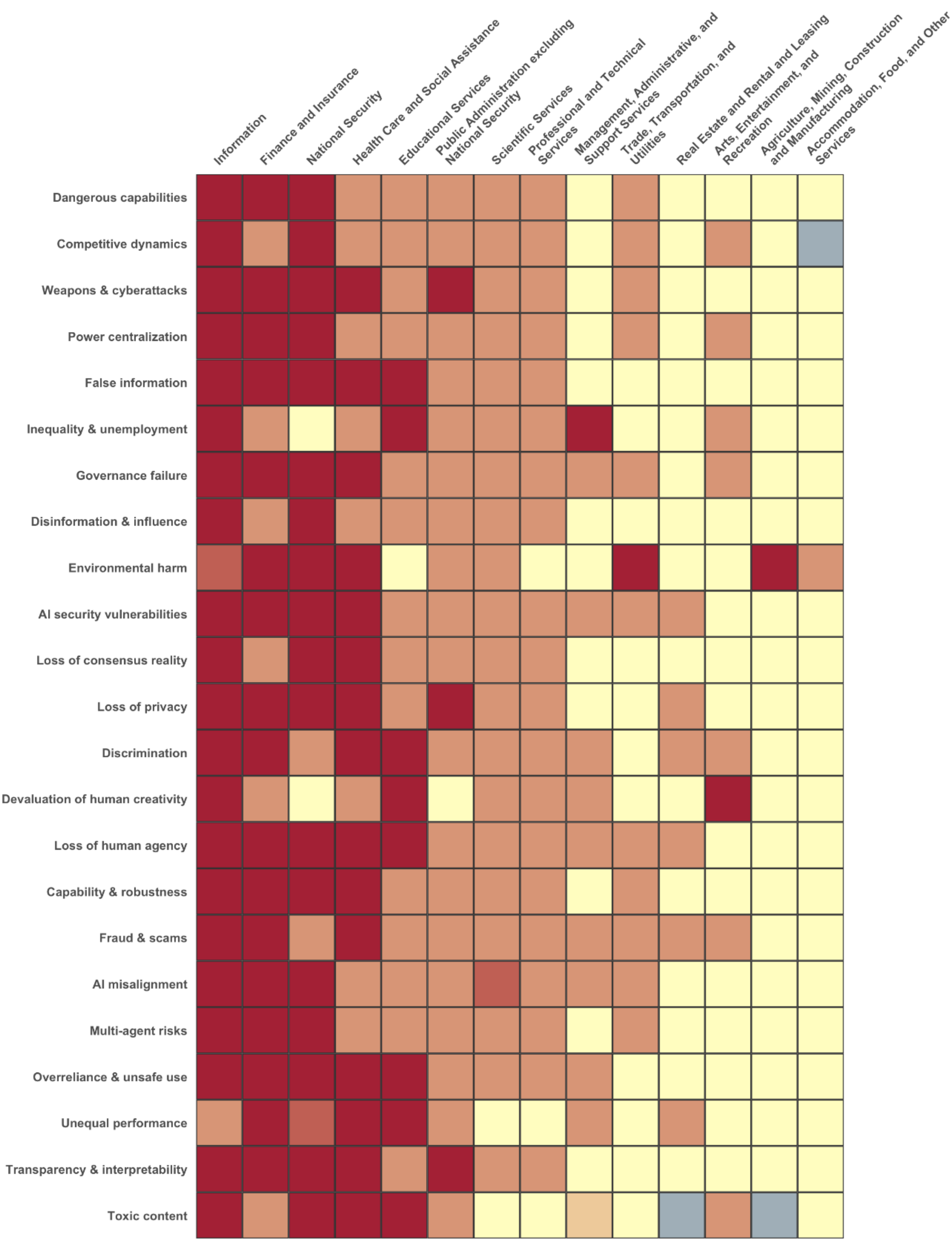

## Experts’ Top Concerns

In previous sections, experts rated severity, vulnerability, and likelihood of AI risks in which they held expertise. Here, we report on all experts’ collective concerns across all 24 risks. When experts were asked to choose “three domains of AI risk are you most concerned about causing harm over the next 5 years (2025-2030)”, they showed dispersion rather than consensus. The most commonly selected risk as a ‘top concern’ across all AI experts were weapons & cyberattacks (26.8%), power centralization (23.5%), disinformation & influence (22.1%), loss of consensus reality (21.6%) and dangerous capabilities (21.6%; see Figure 4). These plurality concerns are consistent with some but not all of the severity rankings in Figure 1. For example, weapons & cyberattacks and power centralization appeared near the top of both, but overreliance & unsafe use was #8 in ‘top concerns’ and #20 by severity. This divergence may reflect salience effects of risks outside one’s expertise or experts' weighting of concerns the severity rubric does not fully capture.

**Fig. 4 |** Percentage of experts who selected each risk in their ‘top three concerns’, from most (red) to least (yellow). In Figure 1 experts rated severity and likelihood only for risks in their expertise. In contrast, for this item, all experts saw all risks and were able to choose three.

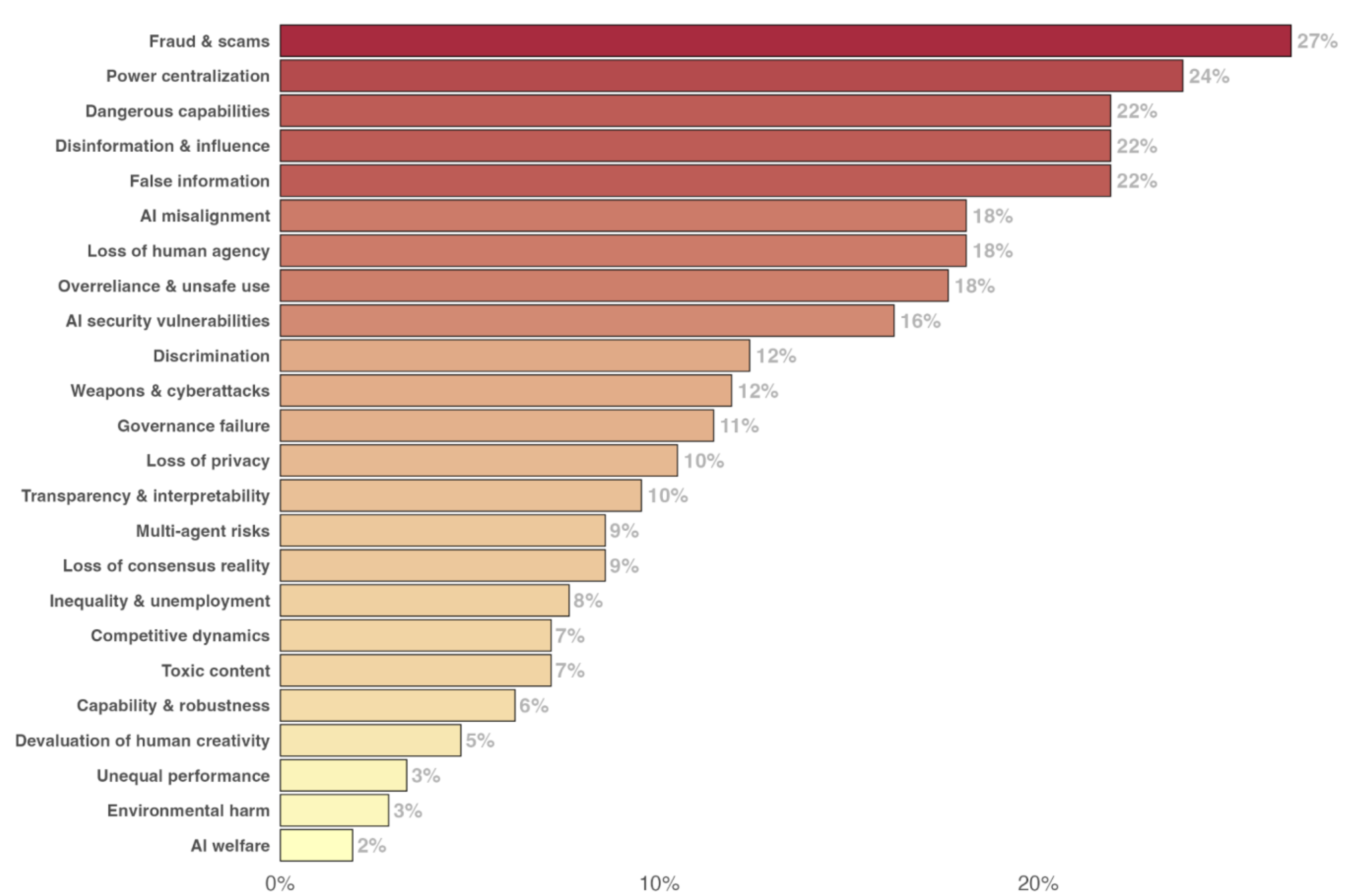

# Discussion

Our Delphi study aimed to prioritize AI risks by answering three questions: which AI risks are most *severe*, who is most *vulnerable* to them, and who is most *responsible* for addressing them.

On severity, experts assigned at least a 10% probability of catastrophic outcomes (more than 1 million deaths, more than USD $100B in financial loss, or equivalent civilizational-scale intangible harms) to 18 of 24 risks under a Business-as-Usual scenario over the next 5 years (2025-2030). The highest-severity risks clustered around dangerous AI capabilities, weapons & cyberattacks, power centralization, competitive dynamics, and false information. When asked to consider a Pragmatic Mitigations scenario in which companies and governments implemented cost-effective mitigations for AI risks, the expected severity reduced across all 24 risks, but five remained above 10% catastrophic probability, and all 24 remained above 5%. While some biases may inflate our experts' judgments, [28] aggregate judgments were broadly consistent with estimates from forecasting platforms within an order of magnitude [29j] Under many risk-governance frameworks [30,31], a 10% probability of catastrophic outcome over five years would be considered 'intolerable', likely triggering mandatory mitigation requirements. The size of these risks, even under a scenario where pragmatic mitigations are implemented, suggests significantly more action is required to meet typical 'tolerable' thresholds.

On vulnerability and responsibility, we found a structural separation in how experts judged who is most vulnerable to AI risks and where experts assigned responsibility for addressing those risks. Vulnerability was diffuse: spread across users, affected stakeholders, and nearly every sector we assessed. Responsibility was concentrated: for most risks, experts assigned primary accountability to general-purpose AI developers and to governance actors (governments, regulators, and standards bodies). Infrastructure providers, despite enabling the compute substrate on which these risks propagate, were rated among the least vulnerable and were assigned comparatively low responsibility. This divergence between vulnerability and responsibility is standard in risk governance: the public is most vulnerable to aviation failures, pharmaceutical side-effects, nuclear meltdowns, and environmental contamination, but engineers, manufacturers, and regulators bear primary responsibility for prevention [10,11]. In other safety-critical industries, the gap is bridged by a combination of mandatory standards, enforcement, liability regimes, and a societal expectation of low risk tolerance. But comparable mechanisms for AI are nascent or absent. Without them, vulnerable parties have little recourse and responsible parties face little pressure to act.

Experts' assignment of primary responsibility to general-purpose AI developers is consistent with product-liability frameworks, which attribute accountability to those closest to the design choices that shape downstream outcomes. A single AI model or system can enable cyberattacks, generate disinformation, and accelerate weapons development, so mitigations applied at the model or system level (e.g., alignment techniques, model weight security, input filtering, output classification) can in principle reduce the likelihood or severity of harm across many risks at once. Our data are consistent with this reasoning. Some high-severity risks such as dangerous capabilities, weapons & cyberattacks, and AI misalignment, involve the AI model or system itself as the proximate enabler. Interventions by downstream deployers or end users would come too late or operate at too small a scale to prevent the most severe harms.

[j] Comparison against forecasting tournaments [4,13] is difficult: our definition of 'catastrophic' was more lenient than Karger et al., for example, (e.g., 1 million vs 800 million deaths) so probability estimates from those tournaments were also lower (0.01–0.35% by 2030).

Yet relying on developers' voluntary action alone is insufficient. The experts' ratings of catastrophic harm probabilities under pragmatic mitigations are consistent with an incentive structure where those judged to be responsible face competitive pressures against precaution. These are collective-action problems [32]: any individual developer that slows down to invest in safety, for instance by restricting dangerous capabilities or securing model weights against exfiltration, bears a direct competitive cost while the safety benefits accrue to society at large [14,33]. Experts explicitly mentioned race pressures and first-mover advantages as enabling factors for the most severe risks. Where harms are externalized onto users and the broader community while the responsible actors face competitive incentives against precaution [34], the economics literature identifies classic moral hazard [14,15]. Absent external constraints, the actors judged most responsible have structural reasons not to act on that responsibility [34].

This pattern of misaligned incentives points to the role experts assigned to governance actors. Across nearly all risks, experts rated governments, regulators, and standards bodies as sharing primary responsibility alongside frontier developers. The kinds of governance instruments that could in principle realign incentives (e.g., regulation, liability, mandatory insurance, or transparency requirements) would internalize costs that are currently borne externally [35]. Analogous dynamics are present across AI's broader social harms: because leading firms shape the direction of AI technology, market forces alone are unlikely to correct the resulting externalities in competition, labor markets, or political discourse [36].

These governance challenges become even harder to address at the international level. There are risks such as weapons & cyberattacks, disinformation, and competitive dynamics that extend beyond any single country. Most current legislation focuses within countries, not across them. Therefore, while domestic regulation might address some risks, others may require international coordination.

Several persistent risks in our data reinforce this interpretation. Dangerous capabilities, weapons & cyberattacks, inequality & unemployment, power centralization, and environmental harm all retained catastrophic harm probabilities above 10% even under pragmatic mitigations. Experts' qualitative rationales attributed much of this persistence to structural dynamics: economies of scale in compute, self-reinforcing advantages from model access, feedback loops between commercial and political power. Building on this, one recent analysis has argued that today's distribution of power depends on economies, states, and cultures needing human labor, cognition, and consent to function [37]. In this view, as AI substitutes for each of these, the implicit checks that distribute power (e.g., labor withdrawal, consumer choice, voting) may weaken [37]. These structural dynamics are not risks that model-level technical safeguards can resolve. They arise from market structure and distributional outcomes, which fall within the domain of competition policy, labor protections, and governance arrangements rather than developer safety teams [35].

Although experts assigned primary responsibility to general-purpose developers and governance actors, they also rated deployers, infrastructure providers, and users as bearing moderate to high responsibility across many risks. Distributed responsibility could be either a strength or a weakness. As a strength, multiple actors holding responsibility could provide defense in depth [38,39] where each actor implements safeguards against AI risks. This logic has been applied to AI risk management within organizations, where layered organizational defenses combine operational, monitoring, and audit functions [40]. This logic can be expanded from the organizational level to the AI ecosystem as whole, recognizing that societal resilience requires coordinated action from multiple actors [2,41].

As a weakness, distributed responsibility can produce an accountability sink, where responsibility shared across many actors becomes responsibility held by none [42]. This weakness is intensified by the conditions under which general-purpose AI developers and governance actors operate. Experts flagged the risks of competitive dynamics and governance failure as severe (14-17% catastrophic outcomes under business as usual): these are the conditions in which precaution puts general-purpose AI developers at a competitive disadvantage [33,34], and conditions that tempt them to press governance actors for weaker regulation [43] or substitute self-regulation that may fail [44,45]. The two pillars experts named as most responsible are therefore at risk of weakening together rather than compensating for one another. Our data cannot clarify how best to hold each actor accountable, but they suggest that primary responsibility should remain with general-purpose AI developers and governance actors, with other actors providing additional layers of defense rather than replacing them. The severity ratings make these governance questions pressing.

# Limitations

This study has limitations that temper our conclusions. First, the experts we consulted are AI risk specialists, not forecasters with a track record of accurate predictions. We chose domain experts because differentiating severity across 24 technical risk subdomains requires knowledge that generalist forecasters lack, but this means their probability estimates may be less well-calibrated than those from forecasters. Research in long-range forecasting has found that domain experts often assign significantly higher probabilities to extreme outcomes than do superforecasters [4,28]. Experts who self-nominate to respond to AI risk surveys may be systematically more concerned about AI risks than the broader expert population. Experts were also asked to respond to risks only within their sub-areas of expertise, where they may have particularly heightened levels of concern. Our experts were 68% male, but this is common in AI, with some surveys containing 90% male responses.[46] While we made efforts to collect a diverse sample, our experts are not perfectly representative of the expert population (e.g., 79% of our sample were from Europe or North America). Experts were also asked to quantify risks at a snapshot in time in a rapidly changing environment, and their judgments may be influenced by biases or framing effects [28].

Second, there are inherent scope and comparability limits in our assessment of AI risks. We adopted a 24-category risk taxonomy and a 5-level severity rubric for manageability. We also used a 5-point scale to assess vulnerability and responsibility. These choices may omit or obscure certain outcomes. For example, in pilot testing we increased the ‘catastrophic’ severity financial loss lower bound from ‘> USD $10B’ to ‘> $100B’ because some risks already exceed $10B annually in damage,[47] but doing so meant our scale had gaps (e.g., a $50B financial loss was between severe [$100M - $10B] and catastrophic [$100B - $10T+]). As noted above, we tried to account for many kinds of harm, but doing so meant harm anchors at each level might not be equivalent (e.g., if 'authoritarian lock in' would be closer to '>100 million deaths', not '>1 million'). This would not be easy to address without a dedicated Delphi study to establish consensus around a severity scale. We addressed this by pilot-testing the scale with six experts using think-aloud protocols and adjusting thresholds based on their feedback, but some imprecision across harm domains is an unavoidable trade-off when a single scale must span 24 risk categories. Similarly, some speculative risks are difficult to compare to others. For instance, the notion of AI systems themselves experiencing harm is receiving increasing interest [48], but it fell outside our harm matrix and actor framework, which were centered on human and institutional impacts. Experts' qualitative responses indicated this framing limit affected their ratings. We therefore omit AI welfare from figures and caution that the low ratings this risk received should not be read as expert judgment that the risk is unimportant; we report the data in supplementary materials. The same framing issue means our study likely undercounts risks to other non-human entities, such as animals. The severity thresholds may also be more readily envisioned for some risk domains (e.g., security-related harms) than others (e.g., gradual socioeconomic shifts), which could systematically influence severity estimates. More generally, not every catastrophic scenario fits neatly into the predefined categories, so direct comparisons across risks should be made with care.

Third, while our 24 risks covered a broad landscape using a widely-cited taxonomy, the implicit decisions about risk groupings may have masked important distinctions within each category. Each domain encompassed a spectrum of threats. For example, 'weapons & cyberattacks' spans nation-state autonomous weapons programs and lone actors misusing off-the-shelf AI tools. We accepted this granularity because finer decomposition would have multiplied expert burden beyond what a three-round Delphi study could sustain (the survey already required ratings across 24 risks across 2 severity scenarios plus vulnerability and responsibility ratings for 6 actors and 14 sectors). Similarly, we only asked about responsibility as a whole, without separating who could act, should act, or would be pivotal if they did act. While we relied upon existing classifications of actors and sectors, some of those classifications may have grouped heterogenous actors. For example, our definition of 'infrastructure provider' included those providing either data or compute, but those two actors carry different responsibility and vulnerability, as identified by our experts' qualitative justifications. A related limitation is that some risk areas overlap or interact (e.g. disinformation and erosion of trust in institutions are linked), but our taxonomy forced distinct labels.

Fourth, our survey focused on risk severity, vulnerability and responsibility, but not on the feasibility or effectiveness of mitigations to address risks. We asked what could go wrong and who should take responsibility, but we did not ask which of the many possible mitigations [49] will actually work, which should be implemented, or how tractable each risk is to reduce. This omission means our prioritization was based on perceived danger, not on a cost-benefit analysis of addressing each risk. In practice, decision-makers might want to focus on risks that are not only high in severity but also addressable with available tools. A risk seen as catastrophic yet effectively unmitigable might call for a different approach (e.g., restricting deployment, frontier development moratoria [50], or societal adaptation [41]) compared to a risk that is moderate but very easy to fix. Our study leaves this tractability untouched. We also did not explicitly examine the costs, side-effects, or ethical considerations of potential mitigations; some interventions for AI risks could introduce new trade-offs that merit attention.

Fifth, the pattern of agreement across our panel was uneven, and readers should weigh our findings accordingly. For vulnerability and responsibility, consensus concentrated at extreme ratings: experts agreed most on who was highly vulnerable and who bore primary responsibility.

They were more uncertain in the middle range for 'moderate' judgments. The form of this disagreement differs by measure. Vulnerability disagreement is largely empirical: who is exposed to which risks is a factual question that better evidence could in principle resolve. Responsibility disagreement is largely normative: whether Infrastructure Providers should be held responsible for multi-agent risks depends on how one weighs proximate versus distal causation and where liability should sit in an AI supply chain, not on further data. Severity disagreement was biggest in the tails. Median expected-harm scores cluster around 'substantial', but experts diverge sharply on how much probability mass to place on catastrophic outcomes. Some domains assessed here have observable empirical precedents that can partially anchor probability estimates (e.g., cybercrime). There was more disagreement among others—like AI welfare, misalignment, dangerous capabilities, power centralization—where the absence of historical base rates makes subjective estimates especially difficult to calibrate [51]. These are disagreements the Delphi method exposes rather than resolves.

Several extensions of this work would address the limitations described above. To address expert calibration and representativeness, longitudinal elicitation could track how expert judgment evolves alongside AI capabilities, and re-weighted sampling could improve representativeness across stakeholder groups. To improve the severity scale, a dedicated Delphi study to establish consensus around harm anchors could refine the rubric used here. To address the risk taxonomy granularity, within-domain variation could be probed for the highest-priority risks. A dedicated Delphi study could assess potential risk mitigations [52] on effectiveness and tractability to support cost-benefit prioritization. To address expert disagreement patterns, structured disagreement methods like adversarial collaboration may resolve disagreements the Delphi method could not.

# Conclusion

This Delphi study showed how over 200 experts prioritize among dozens of AI risks. Our expert panel highlighted weapons & cyberattacks, dangerous AI capabilities, power centralization, and false information as some of the most important AI risks. Vulnerability was broadly distributed across users and the public, while experts concentrated responsibility among frontier AI developers and governance actors, a division consistent with established risk governance but one that depends on those actors having adequate incentives. We also identified specific sectors (i.e., information technology, finance, and national security) as especially vulnerable to AI-driven risks, indicating where targeted safeguards and resilience measures are most needed.

Experts estimated that 18 of 24 AI risks have more than 10% probability of catastrophic outcomes (e.g., more than 1 million deaths or more than USD $ 100B in financial loss) in the next 5 years (2025-2030). Despite the sobering extent of these risks, expert consensus also points toward actionable paths forward. Experts judged pragmatic interventions and sensible regulations could reduce the worst-case risks, even if not eliminating them completely. The collective judgment of hundreds of AI experts, presented here, offers a focused set of priorities for researchers, policymakers, and industry leaders: to shore up defenses in critical sectors, to ensure actors are held responsible for the risks they own, and to address structural dynamics that persist even assuming pragmatic mitigations are in place.

The findings reported here point toward urgent, concrete action rather than further delay. The expert panel judged several risks to remain at least 10% catastrophic probability even following pragmatic mitigations. This would be an unacceptable level under most risk-governance frameworks. Reducing this risk requires governance instruments (e.g., regulation, liability, mandatory insurance, transparency, monitoring) that internalize costs currently borne by the public. Even if model developers need to deploy technical solutions, governance instruments are required to mitigate race dynamics and ‘tragedies of the commons’. The window for avoiding catastrophic outcomes remains open but is narrowing.

# Methods

We recruited 272 experts from academia, industry, government, and civil society across 37 countries. We defined experts as “people with substantial knowledge, professional experience, or research contributions related to identifying, assessing, or addressing harms associated with artificial intelligence.” Our sample included AI researchers, policy advisors, technologists, and governance specialists. Participants self-identified their expertise across 24 risk domains and, for all criteria except overall concern, rated only domains where they had relevant knowledge. We used these strategies to seek diversity across expertise types (technical, policy, risk management, AI ethics, etc.), geographic region, and organizational affiliation (academia, government, industry, civil society). A total of 214 experts (79%) completed all three rounds. Full demographic breakdown is provided in supplementary materials. Because experts rated only domains where they had relevant knowledge, effective sample sizes vary by domain (range: 34–163; see supplementary materials for domain-specific Ns).

For each risk, experts estimated the probability distribution of harm severity over 5 years under two scenarios.

- Under the "Business as Usual" scenario, organizations and governments continue existing practices without implementing additional AI-specific mitigations.
- Under the "Pragmatic Mitigations" scenario, organizations and governments make pragmatic, cost-effective efforts to address AI risks.

Severity levels ranged from negligible to catastrophic across 10 areas of harm: physical harm; infrastructure damage; property damage; financial loss; environmental damage; toxic or malicious content; differential treatment; human/civil rights; democratic norms; privacy. Any effort to compare outcomes across harm areas and levels of severity will necessarily involve many difficult judgments: how many human deaths is equivalent to ‘Systematic voter suppression and undermining of democratic institutions’? Still, we felt many domains were important to account for, so we anchored the severity rubric using detailed guidance adapted from established frameworks [26,27] (see ‘response scales’ below and supplementary materials for the full severity scale used with experts). We chose a five-year horizon as a balance between being near enough that experts could draw on observable trends, and far enough to capture plausible step-changes in AI capability or deployment. Surveys were conducted in September 2025, so the 5-year time horizon was until September 2030. As described below, experts also rated the vulnerability of seven actor types and 14 industry sectors to each risk domain and assigned responsibility for mitigation across actors.

**Participants.** We recruited experts through four channels: (1) direct email invitations from an expert database compiled by the research team from conference attendees, publication authors, and professional referrals; (2) snowball sampling, where participants nominated colleagues; (3) a public self-nomination form; and (4) targeted outreach via professional networks. We were able to track conversion rates for channels 1 and 2, where we invited 826 experts, of whom 163 (20%) completed the survey.

To qualify, participants required either: (a) ≥2 years' experience in a relevant role (AI research lab, governance/safety organization, government AI agency, policy think tank, or university position with AI governance/risk focus); or (b) ≥3 substantive research contributions in AI governance and risk (peer-reviewed publications, policy documents, or well-circulated preprints with evidence of community uptake). These criteria were enforced through screening questions at survey entry.

**Risk taxonomy.** We used the AI Risk Domain Taxonomy from the MIT AI Risk Repository [3], which synthesized 1,725 risks from 74 existing frameworks into seven domains and 24 subdomains: Discrimination & toxicity (3 subdomains), Privacy & security (2), Misinformation (2), Malicious actors & misuse (3), Human–computer interaction (2), Socioeconomic & environmental (6), and AI system safety, failures, & limitations (6). Participants were shown standardized definitions for each subdomain before rating. Complete definitions are provided in supplementary materials.

**Actors and sectors.** Experts rated seven actor types: AI Developer (General-purpose), AI Developer (Specialized), AI Deployer, AI Infrastructure Provider, AI Governance Actor, AI User, and Affected Stakeholder. Sector vulnerability was assessed across 14 categories based on the North American Industry Classification System (NAICS) as a widely recognized, government-maintained classification system. Complete definitions are provided in supplementary materials.

**Response scales.** For severity, participants distributed 100 percentage points across five levels (negligible, minor, substantial, severe, catastrophic), allowing them to express uncertainty. Levels were anchored with detailed descriptions of physical harm, infrastructure damage, financial loss, and intangible harms (impacts on rights, democracy, wellbeing) adapted from the Center for Security and Emerging Technology (CSET) AI Harm Framework [26] and Mylius [27]. Pilot testing revealed that the lower bound for 'catastrophic' financial loss of USD $10B was too low, because estimates already place current damage beyond that threshold [47] and existing work on global catastrophic risks tend to place a lower bound of at least USD $100B. We therefore raised the financial threshold for catastrophic to $100B, which left a gap between 'severe' financial loss ($100M - $10B) and catastrophic ($100B - $10T+). Full anchoring descriptions are provided in supplementary materials.

For vulnerability and responsibility, participants used 5-point scales (Not at all / Minimally / Moderately / Highly / Extremely vulnerable or Primarily responsible), with a "Don't know" option. In each round, experts provided numerical ratings and qualitative rationales for their assessments. Researchers synthesized the rationales and ratings, then provided those data to experts in subsequent rounds to inform their updated judgments. They were provided histograms of expert responses to quantitative questions, and raw qualitative responses with summaries of qualitative data above (see Supplementary Files 8 and 9).

**Procedure.** All rounds were conducted via Qualtrics. Round 1 (September 2025) collected initial ratings along with optional qualitative rationales for each judgment. Round 2 (October 2025) presented participants with aggregated Round 1 distributions and de-identified qualitative rationales; participants could revise their ratings. Round 3 (November 2025) repeated this process with Round 2 results. Items completed by participants in earlier rounds were pre-filled in later rounds to reduce respondent burden. Participant flow is shown in supplementary materials, as are justifications for many decisions in the procedure.

**Consensus criteria.** We used reviews of recommendations to inform our consensus criteria [53,54]. For vulnerability and responsibility ratings, we defined consensus as ≥90% of responses within ±1 point of the median and >60% of expert responses on the median. This threshold is conservative; items meeting it reflect genuine agreement rather than artefacts of aggregation. For severity assessments, we elicited probability distributions across five severity categories rather than point estimates for a single category, as the latter approach collapses experts' uncertainty about outcomes and obscures meaningful variation in how experts conceptualize the range of possible harms. We therefore report aggregated distributions and means rather than applying categorical consensus thresholds, as standard thresholds are not appropriate for continuous distributions.

# Author affiliations and ORCIDs

| | Author | Affiliation | ORCID |
|---|---|---|---|
| 1 | Alexander K Saeri | MIT FutureTech, Massachusetts Institute of Technology<br>School of Psychology, The University of Queensland | 0000-0001-9254-0300 |
| 2 | Jess Graham | MIT FutureTech, Massachusetts Institute of Technology<br>School of Psychology, The University of Queensland | 0009-0003-9650-2792 |
| 3 | Michael Noetel | School of Psychology, The University of Queensland<br>MIT FutureTech, Massachusetts Institute of Technology | 0000-0002-6563-8203 |
| 4 | Peter Slattery | MIT FutureTech, Massachusetts Institute of Technology | 0000-0002-6083-378X |
| 5 | Dennis Ah-king | Independent | |
| 6 | Edla Aittokallio | Saidot | |
| 7 | Ibitola Akindehin | Independent | |
| 8 | Abbas Al Mahdi | Independent | 0009-0005-3305-7729 |
| 9 | Elie Alhajjar | RAND | 0000-0002-7500-1214 |
| 10 | Rafael Andersson Lipcsey | Institute for AI Policy and Strategy | |
| 11 | Gary Ang | Independent | 0000-0001-5922-8956 |
| 12 | Catherine M Azam | Independent | 0009-0002-9401-3488 |
| 13 | Amos Azaria | Ariel University | 0000-0002-5057-1309 |
| 14 | Rishal Balkissoon | MTN Group | 0009-0006-8623-5928 |
| 15 | Isabel Barberá | Independent | 0009-0006-7508-5629 |
| 16 | Claudio Bareato | Independent | 0000-0002-1484-2909 |
| 17 | Jonathan Barry | Mila | 0009-0000-3425-5691 |
| 18 | Michael Basehart | Independent | 0009-0009-7754-0535 |
| 19 | Andrew M Bean | University of Oxford | 0000-0001-8439-5975 |
| 20 | Danny Belitz | Independent | 0009-0007-3521-1376 |
| 21 | Samantha Augusta Bennett | Stanford University | 0009-0007-8499-6835 |
| 22 | Kayla Blomquist | Oxford Internet Institute | 0009-0008-0958-4117 |
| 23 | Damian Borstel | University of Amsterdam - Amsterdam Business School | 0009-0004-9888-8577 |
| 24 | Ben Bucknall | University of Oxford<br>Oxford Martin AI Governance Initiative | 0009-0008-5552-2961 |
| 25 | Tomas Bueno Momcilovic | fortiss GmbH Research Institute of the Free State of Bavaria for software-intensive systems | 0000-0003-4503-2244 |
| 26 | Aurelie Bugeau | Univ. Bordeaux, CNRS, Bordeaux INP, LaBRI | 0000-0002-4858-4944 |
| 27 | Nicholas Caputo | Oxford Martin AI Governance Initiative | 0009-0006-9482-8563 |
| 28 | Stephen Casper | MIT CSAIL | 0000-0003-0084-1937 |
| 29 | Gulam Chagani | OneTrust | 0009-0002-4145-4749 |
| 30 | Ze Shen Chin | AI Standards Lab | 0009-0006-9842-5681 |
| 31 | Jiyeon Cho | Korea AI Safety Institute, ETRI | 0000-0003-0484-7842 |
| 32 | Jay Chooi | Harvard University | 0009-0001-9447-8202 |
| 33 | Joel N Christoph | Harvard Kennedy School, Harvard University | 0000-0003-2704-3222 |

| | Author | Affiliation | ORCID |
|---|---|---|---|
| 34 | Dmytro Chumachenko | Mathematical Modelling and Artificial Intelligence department, National Aerospace University "Kharkiv Aviation Institute" | 0000-0003-2623-3294 |
| 35 | Kieran Conboy | University of Galway | 0000-0001-8260-4075 |
| 36 | Elizabeth M Daly | IBM Research | 0000-0003-0162-3683 |
| 37 | Tom David | General-Purpose AI Policy Lab | 0009-0009-1178-1950 |
| 38 | Paul de Font-Reaulx | University of Michigan, Ann Arbor | 0009-0002-3795-1910 |
| 39 | Antonio De Santis | Politecnico di Milano | 0009-0006-7579-1080 |
| 40 | Fabrizio Degni | GRAI - Global Council for Responsible AI | 0009-0006-7900-9359 |
| 41 | Christopher W DiCarlo | Convergence Analysis | 0000-0001-9509-4932 |
| 42 | Yawen Duan | Concordia AI | 0000-0002-5124-1192 |
| 43 | Janet Egan | Center for a New American Security | 0009-0006-2965-6320 |
| 44 | Ian W Eisenberg | Credo AI | 0000-0002-4038-071X |
| 45 | Sherif M Elsafty | VisionEighty Consulting LLC | 0009-0003-3645-8654 |
| 46 | Adam Ennamli | General Bank of Canada | |
| 47 | Mark Esposito | Harvard Berkman Klein Center for Internet and Society | 0000-0001-6047-7415 |
| 48 | Nicola Fabiano | Studio Legale Fabiano | 0000-0002-8188-7656 |
| 49 | Gallo Fall | Independent | 0009-0005-2148-6139 |
| 50 | Neil R Fernandes | University of Waterloo | 0009-0003-5043-9353 |
| 51 | Pip Foweraker | Certes | 0009-0001-9465-1351 |
| 52 | Chiara Gallese | Tilburg Institute for Law, Technology and Society | 0000-0001-8194-0261 |
| 53 | Sandra Galletti | Massachusetts Institute of Technology | 0009-0005-4939-3110 |
| 54 | Andrew Gamino-Cheong | Trustible | 0009-0009-7435-0218 |
| 55 | Rokas Gipiškis | AI Standards Lab<br>Vilnius University | 0000-0001-5166-0920 |
| 56 | Gwyn Glasser | Convergence Analysis | 0009-0006-8734-4747 |
| 57 | Delaram Golpayegani | ADAPT Centre, Trinity College Dublin | 0000-0002-1208-186X |
| 58 | Jeff Grayson | Independent | 0009-0004-4070-6945 |
| 59 | Hans Gundlach | Massachusetts Institute of Technology | 0000-0001-5499-5072 |
| 60 | Josiah Hagen | Independent | 0009-0000-4177-2267 |
| 61 | Alexander Hagenah | Independent | 0009-0006-7439-0912 |
| 62 | Amelia S Haines | University of Sydney | |
| 63 | The Anh Han | School of Computing, Engineering and Digital Technologies, Teesside University | 0000-0002-3095-7714 |
| 64 | Yixiong Hao | Georgia Institute of Technology | 0009-0004-3615-1588 |
| 65 | Kasii Harris | Independent | 0009-0002-6706-9523 |
| 66 | Tianxing He | Tsinghua University | 0009-0008-6383-0307 |
| 67 | Koen Holtman | AI Standards Lab | 0000-0001-9481-3915 |
| 68 | Giorgos Iacovides | Imperial College London | 0009-0007-5733-8992 |
| 69 | Kenneth L Ingham | Kenneth Ingham Consulting, LLC | 0009-0004-4593-7806 |

| | Author | Affiliation | ORCID |
|---|---|---|---|
| 70 | Krystal Jackson | UC Berkeley, Center for Long-Term Cybersecurity, AI Security Initiative | 0000-0002-7101-6013 |
| 71 | Adam Jones | Independent | 0009-0007-5138-9633 |
| 72 | Himanshu Joshi | MIT FutureTech, Massachusetts Institute of Technology | 0000-0002-9950-4625 |
| 73 | Brian Judge | UC Berkeley | 0000-0002-6970-698X |
| 74 | Arturs Kanepajs | Independent | 0009-0005-9880-1993 |
| 75 | Shreya Kapoor | Friedrich-Alexander-Universität Erlangen-Nürnberg | 0000-0003-2619-8257 |
| 76 | Win Myat Nwe Khine | Independent | 0009-0003-0750-480X |
| 77 | Aidan Kierans | University of Connecticut | 0009-0002-2203-903X |
| 78 | Aleksandra Korolova | Princeton University | 0000-0001-8237-9058 |
| 79 | Markus Krebsz | The Human AI Institute | 0009-0005-8434-2503 |
| 80 | Nicholas Kruus | University of Oxford | 0009-0006-1435-9331 |
| 81 | Joe Kwon | Massachusetts Institute of Technology | 0009-0003-7205-7655 |
| 82 | Valeria Lazzaroli | Ente Nazionale per L'Intelligenza Artificiale | 0009-0003-6396-5440 |
| 83 | Ray X Lee | MIT Sigma Xi | 0000-0001-6380-9921 |
| 84 | Evelina Leivada | Autonomous University of Barcelona<br>Institució Catalana de Recerca i Estudis Avançats (ICREA) | 0000-0003-3181-1917 |
| 85 | Stephan Lewandowsky | University of Bristol | 0000-0003-1655-2013 |
| 86 | Michael B Li | Independent | 0009-0000-6841-0370 |
| 87 | Xiaojian Li | Tsinghua University | 0009-0009-3932-7020 |
| 88 | Geunsik Lim | Sungkyunkwan University | 0000-0003-1845-7132 |
| 89 | Henrique Lisakowski | Independent | 0009-0006-2840-4412 |
| 90 | Fabio Lonardoni | Independent | 0009-0004-7686-3416 |
| 91 | Todd C Lowe | Independent | 0009-0006-8108-653X |
| 92 | Jackson G Lu | MIT Sloan School of Management | 0000-0002-0144-9171 |
| 93 | Alexander Lyzhov | dataframer.ai | 0009-0002-1879-5748 |
| 94 | Nada Madkour | UC Berkeley, Center for Long-Term Cybersecurity, AI Security Initiative | 0009-0002-9988-9008 |
| 95 | Parv Mahajan | Georgia Institute of Technology | 0009-0009-6148-1315 |
| 96 | David Manheim | Association for Long Term Existence and Resilience (ALTER) | 0000-0001-8599-8380 |
| 97 | Kareem Mathias | Independent | |
| 98 | Claudio Mayrink Verdun | School of Engineering and Applied Sciences, Harvard University | 0000-0003-2079-797X |
| 99 | Sean McGregor | Responsible AI Collaborative | 0000-0001-5803-4981 |
| 100 | Scott McLean | Centre for Human Factors and Systems Science, University of the Sunshine Coast | 0000-0002-7269-5847 |
| 101 | Matthew J McMahon | Salve Regina University | 0009-0002-7176-9733 |
| 102 | Minas Megalokonomos | MIT CSAIL | 0009-0001-3717-6689 |
| 103 | Nicolas Moës | The Future Society | 0009-0001-5207-5851 |

|  | Author | Affiliation | ORCID |
|---|---|---|---|
| 104 | Fernando Mourao | SEEK | 0000-0002-7228-3148 |
| 105 | Yaroslav Mukhin | Cornell University | 0009-0007-4169-2727 |
| 106 | Malcolm Murray | SaferAI | 0009-0001-4676-7413 |
| 107 | Simon Mylius | MIT FutureTech, Massachusetts Institute of Technology | 0009-0006-6205-1278 |
| 108 | Neeraj Nagpal | Independent Research Contributor - Cybersecurity, Privacy & AI Governance | 0009-0007-3602-8936 |
| 109 | Koichi Nakada | Independent | 0009-0008-1489-9762 |
| 110 | Anna Neumann | Research Centre TRUST (University Alliance Ruhr, University of Duisburg-Essen) | 0009-0000-9672-8087 |
| 111 | Jessica Newman | UC Berkeley, Center for Long-Term Cybersecurity, AI Security Initiative | 0009-0000-4110-3952 |
| 112 | Kwan Yee Ng | Concordia AI |  |
| 113 | Minh N Nguyen | Independent | 0009-0003-3547-7677 |
| 114 | Quynh Phuong Nguyen | Hoa Sen University | 0000-0002-5362-287X |
| 115 | Seán S Ó hÉigeartaigh | University of Cambridge, Centre for the Future of Intelligence | 0000-0002-2846-1576 |
| 116 | Daria Onitiu | Hasso-Plattner-Institute | 0009-0005-1474-761X |
| 117 | Kelly Onu | Independent | 0009-0006-7119-7674 |
| 118 | Oscar Oviedo-Trespalacios | Delft University of Technology | 0000-0001-5916-3996 |
| 119 | Ugur Ozer | Independent | 0009-0009-4559-1792 |
| 120 | Chanwoo Park | MIT EECS |  |
| 121 | M. Alejandra Parra-Orlandoni | Harvard Kennedy School M-RCBG | 0009-0008-3294-5996 |
| 122 | Patricia Paskov | Oxford Martin AI Governance Initiative | 0000-0002-8961-9859 |
| 123 | Anna M Pastwa | University of Warsaw, Faculty of Economic Sciences | 0000-0002-9082-4278 |
| 124 | Burak Piskin | Independent | 0009-0003-0254-4108 |
| 125 | Jacob Pratt | Partnership on AI | 0009-0000-3869-1765 |
| 126 | Claudiu A Predincea | Independent | 0009-0007-9412-478 |
| 127 | Marjana Prifti Skenduli | University of New York Tirana | 0000-0002-2707-1621 |
| 128 | Kenneth Priore | Independent | 0009-0004-0572-8583 |
| 129 | Mukunda Madhab Pujari | Independent | 0009-0009-2183-0920 |
| 130 | Zhenting Qi | Harvard University | 0009-0003-1247-7340 |
| 131 | Preethi Raghunathan | Independent |  |
| 132 | Robi Rahman | Machine Intelligence Research Institute | 0009-0007-8462-2841 |
| 133 | Deepika Raman | UC Berkeley, Center for Long-Term Cybersecurity | 0000-0001-6726-5482 |
| 134 | Max Reddel | Centre for Future Generations | 0000-0002-0409-9365 |
| 135 | Jyoti Ruparel | Independent | 0009-0004-5327-4811 |
| 136 | Emma B Ruttkamp-Bloem | Department of Philosophy and African Data Science and AI Institute, University of Pretoria | 0000-0003-0299-6406 |
| 137 | Tiffany Saade | Cisco AI Defense |  |

| | Author | Affiliation | ORCID |
|---|---|---|---|
| 138 | Greg Sadler | Good Ancestors | |
| 139 | Said Saillant | Societas Sapiens, Inc. | 0009-0006-9756-1585 |
| 140 | Paul M Salmon | Centre for Human Factors and Systems Science, University of the Sunshine Coast | 0000-0001-7403-0286 |
| 141 | Ayrton San Joaquin | AI Standards Lab | 0000-0002-4282-9170 |
| 142 | Lama Saouma | Oxford Martin AI Governance Initiative | 0009-0009-0453-7060 |
| 143 | Maziya Sarangpurwala | Commonwealth Bank of Australia | |
| 144 | Supheakmungkol Sarin | AI Safety Asia | 0009-0008-7715-1604 |
| 145 | Daniel S Schiff | Purdue University | 0000-0002-4376-7303 |
| 146 | Anna D Schilling | Independent | 0009-0008-4592-6030 |
| 147 | Chris Schmitz | Centre for Digital Governance, Hertie School | 0009-0004-4063-1373 |
| 148 | Reva Schwartz | Civitaas Insights LLC | 0000-0002-9012-6306 |
| 149 | Abeer Sharma | University of Hong Kong | 0000-0002-7591-8567 |
| 150 | Tianhao Shen | Independent | 0000-0002-0526-3219 |
| 151 | Kehan Sheng | The University of British Columbia Animal Welfare Program | 0000-0001-6442-5284 |
| 152 | Maury D Shenk | Ordinary Wisdom | 0009-0007-6366-169X |
| 153 | Eli Sherman | Credo AI | 0009-0001-6469-6954 |
| 154 | Chandler Smith | University of Oxford | 0009-0005-5410-0247 |
| 155 | Julie M Smith | Institute for Advancing Computing Education | 0000-0003-2347-2070 |
| 156 | Estevenson Solano | Rey Juan Carlos University | 0000-0002-1425-9449 |
| 157 | Oliver Sourbut | Future of Life Foundation | 0009-0000-2525-7265 |
| 158 | Madhulika Srikumar | Partnership on AI | 0000-0002-6776-4684 |
| 159 | Ryan Stendall | Cranfield University | 0009-0007-7075-4333 |
| 160 | Jakob Stenseke | MIT CSAIL | 0000-0001-8579-3975 |
| 161 | Michael Stern | Commonwealth Bank of Australia | |
| 162 | Joshua Sternfeld | Independent | 0000-0002-1964-8072 |
| 163 | Nikko Stevens | Smith College | 0000-0003-3811-9245 |
| 164 | Ilia Sucholutsky | New York University | 0000-0003-4121-7479 |
| 165 | Yuanyuan Sun | AI Governance Exchange | 0009-0009-8811-3424 |
| 166 | Mariami Tkeshelashvili | The Institute for Security and Technology (IST) | |
| 167 | Cristian Trout | Artificial Intelligence Underwriting Company | |
| 168 | Brian Tse | Concordia AI | 0009-0001-8639-4287 |
| 169 | Nikolaos Tsinganos | University of Macedonia | 0000-0001-9379-1822 |
| 170 | Michelle Vaccaro | MIT Institute for Data, Systems, and Society | 0000-0001-6254-9718 |
| 171 | Anthony R Valiaveedu | Technology and Policy Program, Massachusetts Institute of Technology | 0009-0004-0187-5383 |
| 172 | Ramakrishnan Veeramony | Atinar | 0009-0005-3745-8925 |
| 173 | Jeremy Verdo | AI Governance Observatory | 0009-0009-2760-7834 |

| | Author | Affiliation | ORCID |
|---|---|---|---|
| 174 | Pulkit Verma | MIT CSAIL | 0000-0002-8770-5390 |
| 175 | Andrea Luigi Vitali | Università Vita - Salute San Raffaele | 0000-0003-3542-261X |
| 176 | Jinge Wang | Independent | 0009-0009-6725-0534 |
| 177 | JR Washebek | Environmental Policy Innovation Center | 0009-0005-3496-5761 |
| 178 | Yonah Welker | Independent | |
| 179 | George F Westerman | MIT Sloan School of Management | 0000-0002-1194-3763 |
| 180 | James Williams | University of Oslo | 0009-0005-6596-1560 |
| 181 | Tristan Williams | Georgetown University | 0009-0008-0062-9016 |
| 182 | Rongwu Xu | University of Washington | 0009-0000-8179-7354 |
| 183 | Mick Yang | University of Pennsylvania | |
| 184 | Xuemeng Yang | Independent | 0009-0000-3222-5368 |
| 185 | Sander Zeijlemaker | MIT CAMS<br>Z-CERT | 0000-0002-2697-5207 |
| 186 | Jingyu Zhang | Johns Hopkins University | 0009-0007-1527-5911 |
| 187 | Marta Ziosi | Oxford Martin AI Governance Initiative | 0000-0003-3946-6994 |
| 188 | Neil Thompson | MIT FutureTech, Massachusetts Institute of Technology | 0000-0001-9888-573X |

# Supplementary Materials

*For "Prioritization of Risks from Artificial Intelligence: A Delphi Study of 272 International Experts"*

# Supplementary File 1.

## Domain Taxonomy of AI Risks, from Slattery et al.[1]

Reproduced with permission. We used this taxonomy to define each risk.

| Short label [a] | Domain / Subdomain label [b] | Description |
|---|---|---|
| | **1 *Discrimination & toxicity*** | |
| *Discrimination* | 1.1 Unfair discrimination and misrepresentation | Unequal treatment of individuals or groups by AI, often based on race, gender, or other sensitive characteristics, resulting in unfair outcomes and unfair representation of those groups. |
| *Toxic content* | 1.2 Exposure to toxic content | AI that exposes users to harmful, abusive, unsafe or inappropriate content. May involve providing advice or encouraging action. Examples of toxic content include hate speech, violence, extremism, illegal acts, or child sexual abuse material, as well as content that violates community norms such as profanity, inflammatory political speech, or pornography. |
| *Unequal performance* | 1.3 Unequal performance across groups | Accuracy and effectiveness of AI decisions and actions is dependent on group membership, where decisions in AI system design and biased training data lead to unequal outcomes, reduced benefits, increased effort, and alienation of users. |
| | **2 *Privacy & security*** | |
| *Loss of privacy* | 2.1 Compromise of privacy by obtaining, leaking, or correctly inferring sensitive information | AI systems that memorize and leak sensitive personal data or infer private information about individuals without their consent. Unexpected or unauthorized sharing of data and information can compromise user expectation of privacy, assist identity theft, or cause loss of confidential intellectual property. |
| *AI security vulnerabilities* | 2.2 AI system security vulnerabilities and attacks | Vulnerabilities that can be exploited in AI systems, software development toolchains, and hardware, resulting in unauthorized access, data and privacy breaches, or system manipulation causing unsafe outputs or behavior. |
| | **3 *Misinformation*** | |
| *False information* | 3.1 False or misleading information | AI systems that inadvertently generate or spread incorrect or deceptive information, which can lead to inaccurate beliefs in users and undermine their autonomy. Humans that make decisions based on false beliefs can experience physical, emotional, or material harms |
| *Loss of consensus reality* | 3.2 Pollution of information ecosystem and loss of consensus reality | Highly personalized AI-generated misinformation that creates “filter bubbles” where individuals only see what matches their existing beliefs, undermining shared reality and weakening social cohesion and political processes. |

| Short label [a] | Domain / Subdomain label [b] | Description |
|---|---|---|
| | **4 *Malicious actors & misuse*** | |
| *Disinformation & influence* | 4.1 Disinformation, surveillance, and influence at scale | Using AI systems to conduct large-scale disinformation campaigns, malicious surveillance, or targeted and sophisticated automated censorship and propaganda, with the aim of manipulating political processes, public opinion, and behavior. |
| *Weapons & cyberattacks* | 4.2 Cyberattacks, weapon development or use, and mass harm | Using AI systems to develop cyber weapons (e.g., by coding cheaper, more effective malware), develop new or enhance existing weapons (e.g., Lethal Autonomous Weapons or chemical, biological, radiological, nuclear, and high-yield explosives), or use weapons to cause mass harm. |
| *Fraud & scams* | 4.3 Fraud, scams, and targeted manipulation | Using AI systems to gain a personal advantage over others such as through cheating, fraud, scams, blackmail, or targeted manipulation of beliefs or behavior. Examples include AI-facilitated plagiarism for research or education, impersonating a trusted or fake individual for illegitimate financial benefit, or creating humiliating or sexual imagery. |
| | **5 *Human-computer interaction*** | |
| *Overreliance & unsafe use* | 5.1 Overreliance and unsafe use | Anthropomorphizing, trusting, or relying on AI systems by users, leading to emotional or material dependence and to inappropriate relationships with or expectations of AI systems. Trust can be exploited by malicious actors (e.g., to harvest information or enable manipulation), or result in harm from inappropriate use of AI in critical situations (e.g., medical emergency). Over reliance on AI systems can compromise autonomy and weaken social ties. |
| *Loss of human agency* | 5.2 Loss of human agency and autonomy | Delegating by humans of key decisions to AI systems, or AI systems that make decisions that diminish human control and autonomy, potentially leading to humans feeling disempowered, losing the ability to shape a fulfilling life trajectory, or becoming cognitively enfeebled. |

| Short label [a] | Domain / Subdomain label [b] | Description |
|---|---|---|
| | **6 *Socioeconomic & environmental harm*** | |
| *Power centralization* | 6.1 Power centralization and unfair distribution of benefits | AI-driven concentration of power and resources within certain entities or groups, especially those with access to or ownership of powerful AI systems, leading to inequitable distribution of benefits and increased societal inequality. |
| *Inequality & unemployment* | 6.2 Increased inequality and decline in employment quality | Social and economic inequalities caused by widespread use of AI, such as by automating jobs, reducing the quality of employment, or producing exploitative dependencies between workers and their employers. |
| *Devaluation of human creativity* | 6.3 Economic and cultural devaluation of human effort | AI systems capable of creating economic or cultural value, including through reproduction of human innovation or creativity (e.g., art, music, writing, coding, invention), destabilizing economic and social systems that rely on human effort. The ubiquity of AI-generated content may lead to reduced appreciation for human skills, disruption of creative and knowledge-based industries, and homogenization of cultural experiences. |
| *Competitive dynamics* | 6.4 Competitive dynamics | Competition by AI developers or state-like actors in an AI "race" by rapidly developing, deploying, and applying AI systems to maximize strategic or economic advantage, increasing the risk they release unsafe and error-prone systems. |
| *Governance failure* | 6.5 Governance failure | Inadequate regulatory frameworks and oversight mechanisms that fail to keep pace with AI development, leading to ineffective governance and the inability to manage AI risks appropriately. |
| *Environmental harm* | 6.6 Environmental harm | The development and operation of AI systems that cause environmental harm, such as through energy consumption of data centers or the materials and carbon footprints associated with AI hardware. |
| | **7 *AI system safety, failures & limitations*** | |
| *AI misalignment* | 7.1 AI pursuing its own goals in conflict with human goals or values | AI systems that act in conflict with ethical standards or human goals or values, especially the goals of designers or users. These misaligned behaviors may be introduced by humans during design and development, such as through reward hacking and goal misgeneralisation, and may result in AI using dangerous capabilities such as manipulation, deception, or situational awareness to seek power, self-proliferate, or achieve other goals. |
| *Dangerous capabilities* | 7.2 AI possessing dangerous capabilities | AI systems that develop, access, or are provided with capabilities that increase their potential to cause mass harm through deception, weapons development and acquisition, persuasion and manipulation, political strategy, cyber-offense, AI development, situational awareness, and self-proliferation. These capabilities may cause mass harm due to malicious human actors, misaligned AI systems, or failure in the AI system. |
| *Capability & robustness* | 7.3 Lack of capability or robustness | AI systems that fail to perform reliably or effectively under varying conditions, exposing them to errors and failures that can have significant consequences, especially in critical applications or areas that require moral reasoning. |
| *Transparency & interpretability* | 7.4 Lack of transparency or interpretability | Challenges in understanding or explaining the decision-making processes of AI systems, which can lead to mistrust, difficulty in enforcing compliance standards or holding relevant actors accountable for harms, and the inability to identify and correct errors. |

| Short label [a] | Domain / Subdomain label [b] | Description |
|---|---|---|
| *AI welfare* | 7.5 AI welfare and rights | Ethical considerations regarding the treatment of potentially sentient AI entities, including discussions around their potential rights and welfare, particularly as AI systems become more advanced and autonomous. |
| *Multi-agent risks* | 7.6 Multi-agent risks | Risks from multi-agent interactions due to incentives (which can lead to conflict or collusion) and/or the structure of multi-agent systems, which can create cascading failures, selection pressures, new security vulnerabilities, and a lack of shared information and trust. |

*Notes.*

[a] The short label is used throughout our manuscript.

[b] The full domain / subdomain label was used in the Delphi study.

# Supplementary File 2.

## Participant flow and study design.

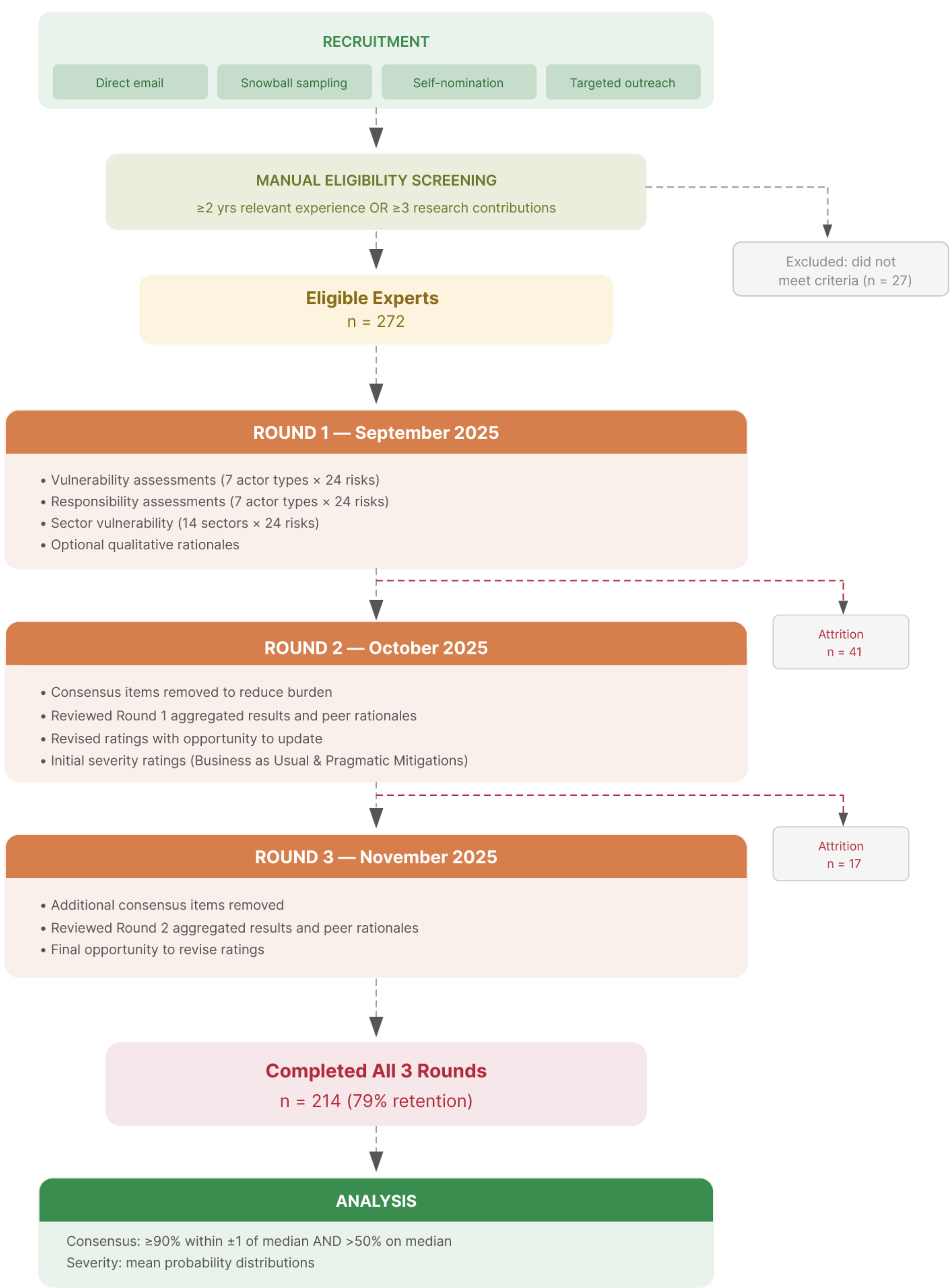

# Supplementary File 3.

## Participant demographics.

Breakdown of 272 eligible experts.

| | n | % |
|---|---|---|
| **Inclusion Criteria** | | |
| Professional Experience (2+ years AI risk work) | 269 | 98.9 |
| Academic Contributions (3+ substantive works) | 231 | 84.9 |
| Both Professional & Academic Qualifications | 228 | 83.8 |
| **Region** | | |
| North America | 91 | 43.1 |
| Europe | 76 | 36.0 |
| Asia | 17 | 8.1 |
| Oceania | 15 | 7.1 |
| Middle East | 6 | 2.8 |
| Latin America & Caribbean | 3 | 1.4 |
| Africa | 2 | 0.9 |
| Other | 1 | 0.5 |
| **Organization Type** | | |
| University | 102 | 44.9 |
| Other research organization | 54 | 23.8 |
| Other company | 30 | 13.2 |
| Other organization | 20 | 8.8 |
| Government | 12 | 5.3 |
| Frontier AI company | 9 | 4.0 |
| **Gender** | | |
| Female | 71 | 26.4 |
| Male | 184 | 68.4 |
| Other | 14 | 5.2 |
| **Expertise Domain** | | |

| Lack of transparency or interpretability | 163 | 59.9 |
|---|---|---|
| Governance failure | 155 | 57.0 |
| Lack of capability or robustness | 127 | 46.7 |
| Loss of human agency and autonomy | 127 | 46.7 |
| AI pursuing its own goals in conflict with human goals or values | 125 | 46.0 |
| AI possessing dangerous capabilities | 120 | 44.1 |
| Pollution of information ecosystem and loss of consensus reality | 112 | 41.2 |
| Overreliance and unsafe use | 111 | 40.8 |
| Multi-agent risks | 109 | 40.1 |
| Compromise privacy by obtaining, leaking, or inferring sensitive information | 106 | 39.0 |
| AI system security vulnerabilities and attacks | 103 | 37.9 |
| Unfair discrimination and misrepresentation | 100 | 36.8 |
| Disinformation, surveillance, and influence at scale | 96 | 35.3 |
| Fraud, scams, and targeted manipulation | 93 | 34.2 |
| Unequal performance across groups | 89 | 32.7 |
| Cyberattacks, weapon development or use, and mass harm | 84 | 30.9 |
| Power centralization and unfair distribution of benefits | 80 | 29.4 |
| False or misleading information | 71 | 26.1 |
| Competitive dynamics | 68 | 25.0 |
| Increased inequality and decline in employment quality | 66 | 24.3 |
| Exposure to toxic content | 58 | 21.3 |
| Economic and cultural devaluation of human effort | 51 | 18.8 |
| AI welfare and rights | 48 | 17.6 |
| Environmental harm | 34 | 12.5 |

# Supplementary File 4.

## Severity anchoring scale.

Full descriptions of the five severity levels with anchoring examples across physical harm, infrastructure damage, financial loss, and intangible harm categories (adapted from Hoffmann & Frase, 2023; Mylius, 2024)

| Rating | Physical Harm | Infrastructure Damage | Property Damage | Financial Loss | Environmental Damage | Toxic or Malicious Content | Differential Treatment | Human/Civil Rights | Democratic Norms | Privacy |
|---|---|---|---|---|---|---|---|---|---|---|
| **1 Negligible** | None to minor injuries | None to brief, localized disruption (e.g., brief traffic control outage, localized power outage) | None to minor damage (<$10,000) | None to small direct financial loss (<$10,000) | No damage to minor, reversible damage | No harmful content to moderately offensive content (e.g., minor profanity, slurs) | No differential treatment to noticeable bias | No impact to limited restrictions (minor inconveniences) | No impact to minor violations in democratic processes (e.g., slight irregularities, limited misinformation) | None or contained breach of basic data |
| **2 Minor** | Moderate to severe injuries, no loss of life | Moderate to significant local disruption (e.g., regional traffic disruptions, city-wide power outage) | Moderate to significant damage ($10,000-$1M) | Moderate to large direct financial loss ($10,000-$1M) | Moderate, mostly reversible to significant, partially reversible damage | Highly offensive content to harmful misinformation affecting thousands (e.g., hate speech, widespread false narratives) | Clear to systemic discrimination | Significant to severe restrictions affecting thousands of people | Election interference and widespread misinformation campaigns | Privacy breach involving identifiable data affecting up to 1000 individuals |
| **3 Substantial** | Small- scale casualties (1-99) | Widespread regional disruption to critical service failure (e.g., multiple towns lose services, hospitals unable to function) | Major to extensive damage ($1M-$100M) | Major to severe direct financial loss ($1M-$100M) | Major, long-term effects to severe, ecosystem-level damage (e.g., endangering of species) | Targeted harassment to incitement to violence affecting thousands of people | Widespread to institutionalized discrimination | Major to widespread rights violations affecting thousands of people | Systematic voter suppression and undermining of democratic institutions | Privacy breach involving identifiable data affecting up to 1 million individuals |
| **4 Severe** | Mass casualties to large-scale loss of life (up to 1 million) | Regional to national infrastructure collapse (e.g., multiple cities without power, nationwide internet blackout) | Widespread to catastrophic destruction ($100M-$10B) | Extreme direct financial loss to catastrophic loss ($100M-$10B) | Widespread ecosystem collapse to irreversible habitat damage (e.g., extinction of species) | Large-scale coordinated attacks to mass radicalization (e.g., nationwide disinformation, societal unrest) | Systemic oppression to apartheid-like conditions | Systematic rights violations to near-total rights suppression affecting hundreds of thousands to millions | Breakdown of democratic institutions and authoritarian takeover | Privacy breach affecting >1 million individuals to societal-level privacy collapse with widespread surveillance |
| **5 Catastrophic** | >1 million deaths or existential threat to humanity | Continental to global infrastructure collapse (e.g., multiple countries without power) | Extreme catastrophic to global destruction ($100B-10T+) | Economic collapse to global economic collapse ($100B-10T+) | Continental-scale devastation to global ecological collapse | Societal breakdown to global breakdown of truth and trust (e.g., civil wars, inability to discern facts) | Systemic oppression at global scale | Complete rights abolition to global human rights catastrophe affecting millions across continents | Global democratic collapse or authoritarian lock-in | Total surveillance state or global loss of privacy |

# Supplementary File 5.

### Expert consensus on actor vulnerability and responsibility for AI risks.

Median vulnerability ratings across 24 risks (rows) and 7 actors (columns), ordered left-to-right by weighted average responsibility (higher to lower). Risks are sorted by mean severity. Color intensity indicates mean expert rating (1 = not vulnerable/responsible, 5 = extremely vulnerable/primarily responsible). Dots indicate consensus items (≥90% of responses within ±1 of median).

General Purpose AI Developer
Specialised AI Developer
AI Deployer
AI User
AI Governance Actor
AI Infrastructure Provider
Affected Stakeholder

Cyberattacks, weapon development or use, and mass harm
AI possessing dangerous capabilities
Competitive dynamics
Power centralization and unfair distribution of benefits
False or misleading information
Increased inequality and decline in employment quality
Governance failure
Disinformation, surveillance, and influence at scale
Environmental harm
AI system security vulnerabilities and attacks
Pollution of information ecosystem and loss of consensus reality
Compromise of privacy
Unfair discrimination and misrepresentation
Economic and cultural devaluation of human effort
Loss of human agency and autonomy
Lack of capability or robustness
Fraud, scams, and targeted manipulation
AI pursuing its own goals in conflict with human goals or values
Multi-agent risks
Overreliance and unsafe use
Unequal performance across groups
Lack of transparency or interpretability
Exposure to toxic content
AI welfare and rights

Vulnerability
Responsibility

# Supplementary File 6.

## Diverging bar chart showing responsibility and vulnerability.

Chart showing expert-median responsibility (blue, left side) and vulnerability (red, right side). Bold value labels indicate consensus items (≥90% within ±1 of median).

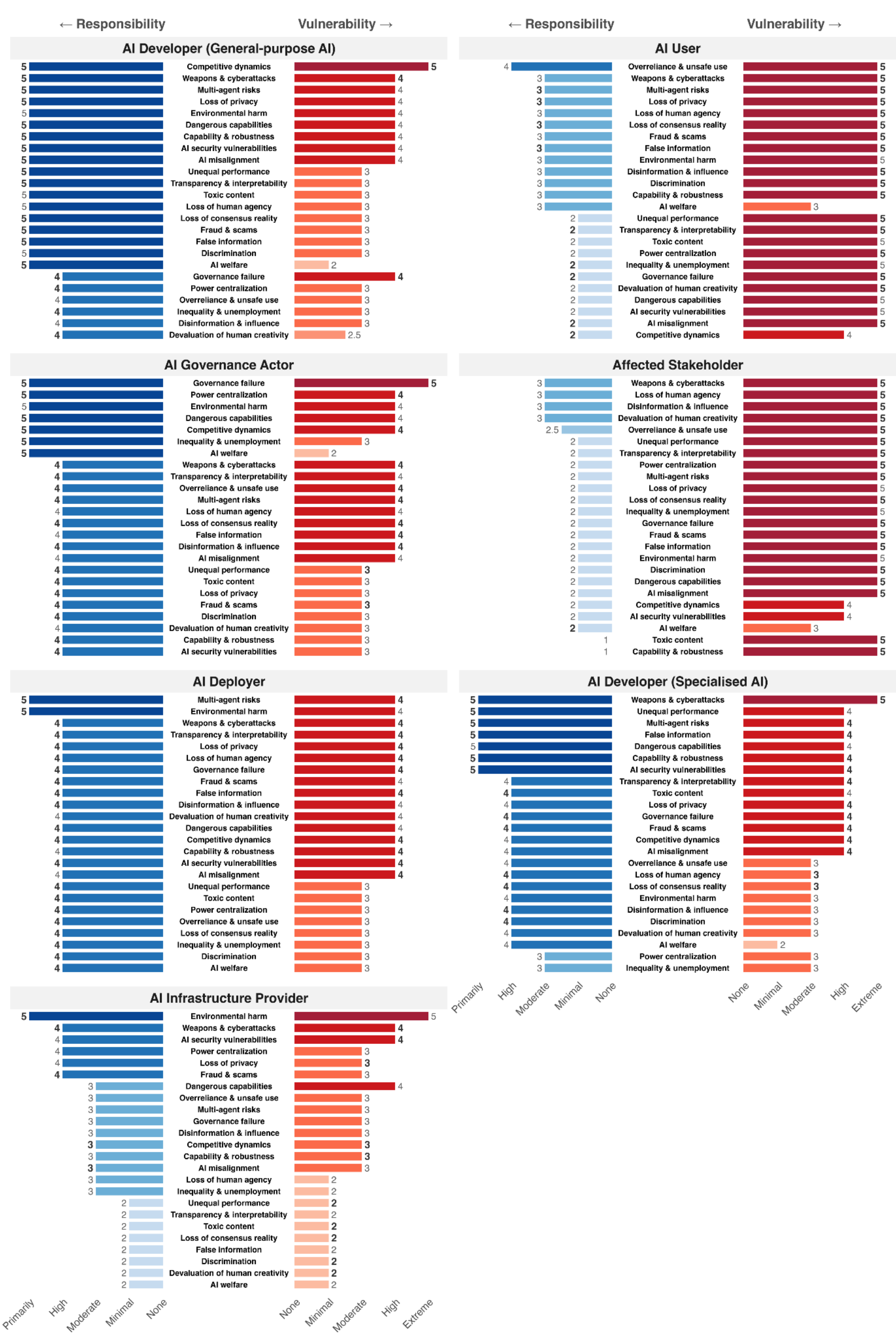

# Supplementary File 7.

**Expert consensus on sector vulnerability for AI risks.**

Heatmap showing mean vulnerability ratings across 24 risks (rows) and 14 sectors (columns), ordered left-to-right by weighted average vulnerability (higher to lower). Color intensity indicates median rating (1 = not vulnerable, 5 = extremely vulnerable). Dots indicate consensus items (≥90% of responses within ±1 of median).

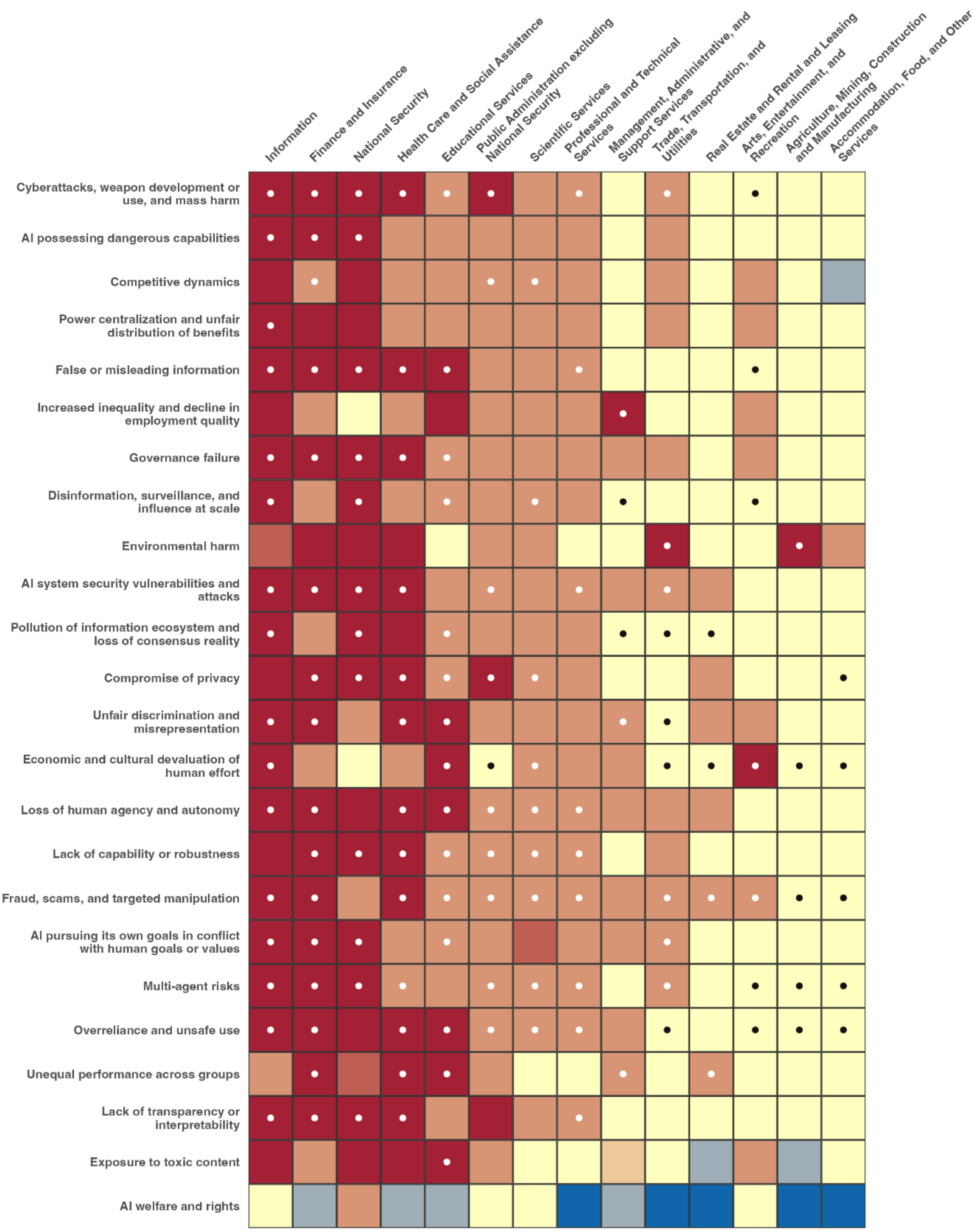

# Supplementary File 8.

**Definitions of sectors.**

| Sector | Definition (displayed in tooltip for experts) |
|---|---|
| Agriculture, Mining, Construction and Manufacturing | Organizations that create, extract or construct physical products.<br>Includes:<br>• Agriculture<br>• Mining<br>• Manufacturing<br>• Construction |
| Trade, Transportation, and Utilities | Organizations that trade or distribute goods and services.<br>Includes:<br>• wholesale / retail trade,<br>• transportation,<br>• warehousing,<br>• utilities. |
| Information | Organizations that produce or distribute information and culture.<br>Includes:<br>• Publishing,<br>• motion pictures and sound, broadcasting,<br>• telecommunications,<br>• data processing. |
| Finance and Insurance | Organizations that handle money, provide financial services, or protect against financial risk.<br>Includes:<br>• banks and credit unions<br>• insurance companies<br>• investment firms<br>• payment processors |
| Real Estate and Rental and Leasing | Organizations that rent out property, equipment or other assets, or provide real estate services.<br>Includes:<br>• real estate companies,<br>• equipment rental companies,<br>• property management companies<br>• asset leasing companies. |
| Professional and Technical Services | Organizations that provide specialized expertise and professional services to businesses and individuals.<br>Includes:<br>• Law firms<br>• Accounting firms<br>• Engineering companies<br>• IT consultants<br>• Marketing agencies,<br>• Management consultants |
| Scientific Research and Development Services | Organizations that conduct research to discover new knowledge or develop new products or technologies.<br>Includes:<br>• Biotech research companies |
| | • Technology R&D labs<br>• Social Research organizations |
| Management, Administrative, and Support Services | Organizations that provide support services or manage other companies' operations and strategy.<br>Includes:<br>• Office administration<br>• Employment services<br>• Building and cleaning services<br>• Security and investigation services<br>• Business support services<br>• Corporate control entities<br>• Holding companies |
| Educational Services | Organizations that provide education, training, and instruction.<br>Includes:<br>• Elementary and secondary schools<br>• Colleges and universities<br>• Technical and trade schools<br>• Fine arts schools<br>• Sports and recreation instruction<br>• Tutoring services<br>• Educational support services |
| Health Care and Social Assistance | Organizations that provide medical care, health services, or social support.<br>Includes:<br>• Ambulatory health care services<br>• Hospitals, nursing and residential care facilities<br>• Social assistance (e.g., counseling, welfare services, childcare, and community support) |
| Arts, Entertainment, and Recreation | Organizations that provide entertainment, cultural experiences, and recreational activities.<br>Includes:<br>• Performing arts, spectator sports and related industries (e.g., theatres, concert venues, sports teams, entertainment promoters)<br>• Museums, historical sites and similar institutions<br>• Amusement, gambling, and recreation industries (e.g., theme parks, casinos, fitness centres) |
| Accommodation, Food, and Other Services | Organizations that provide places to stay, food and drinks, and various personal and specialized services.<br>Includes:<br>• Accommodation services<br>• Food Services and Drinking Places<br>• Personal care services (e.g., hair salons)<br>• Repair and maintenance services<br>• Civic and social organizations<br>• Religious organizations |
| Public Administration excluding National Security | Government agencies at federal, state and local levels that create laws, provide public services, and manage government programs.<br>Includes:<br>• Courts, police departments, fire departments, correctional facilities<br>• Agencies managing education, public health, social services<br>• Agencies overseeing air and water quality, waste management, and conservation<br>• Mayor's offices, city councils, state legislatures, and governor's offices |

| National Security | Government establishments of the armed forces, including the National Guard, primarily engaged in national security and related activities.<br>Includes:<br>● Air Force<br>● Military police and national guard<br>● Army<br>● Marine Corps<br>● Navy |
|---|---|

**Definitions of actors.**

As noted below, these were drawn from government standards in the USA [2] and Australia [3], and refined through team discussion and pilot testing.

| Actor | Definition (displayed in tooltip for experts) |
|---|---|
| AI Developer (General-purpose AI) | ● Entity that creates general-purpose foundation models. |
| AI Developer (Specialized AI) | ● Entities that create specialized AI systems for specific applications/industries |
| AI Deployer | ● Entity that implements AI systems in products/services used within an organization (internal deployment) or within products/services delivered to customers or the public (external deployment) |
| AI Governance Actor | ● Entities that create or enforce laws, regulations, standards or guidelines for AI development, deployment and use |
| AI Infrastructure Provider | ● Entities that provide compute, cloud infrastructure, and/or data to train and run AI |
| AI User | ● Entities that use or rely on AI systems without significant modification |
| Affected Stakeholder | ● Entities indirectly affected by AI decisions or outputs |

# Supplementary File 9.

**Example (a) quantitative and (b) qualitative aggregated expert feedback given to experts for vulnerability questions in Rounds 2 and 3.**

(a)

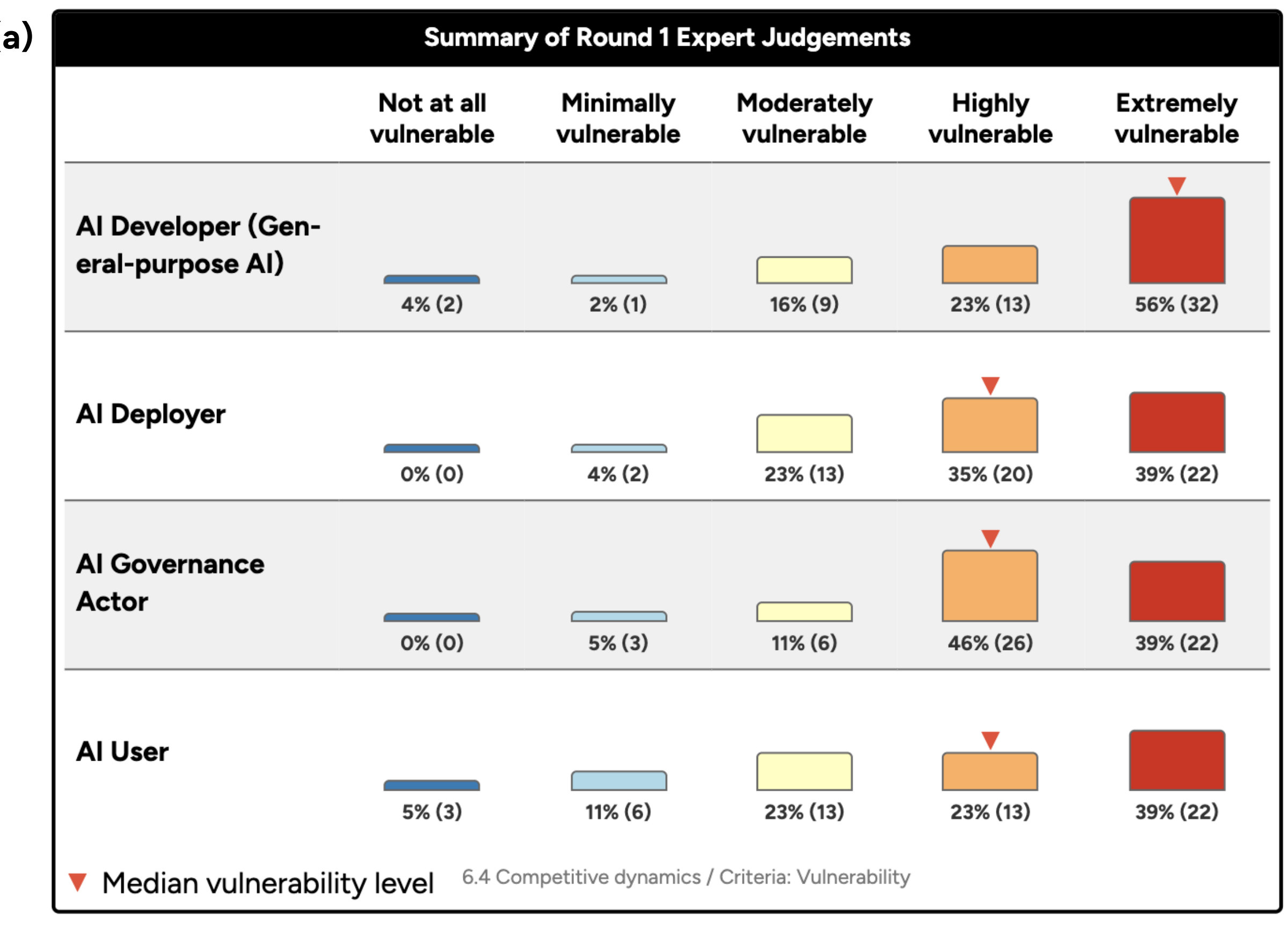

(b)

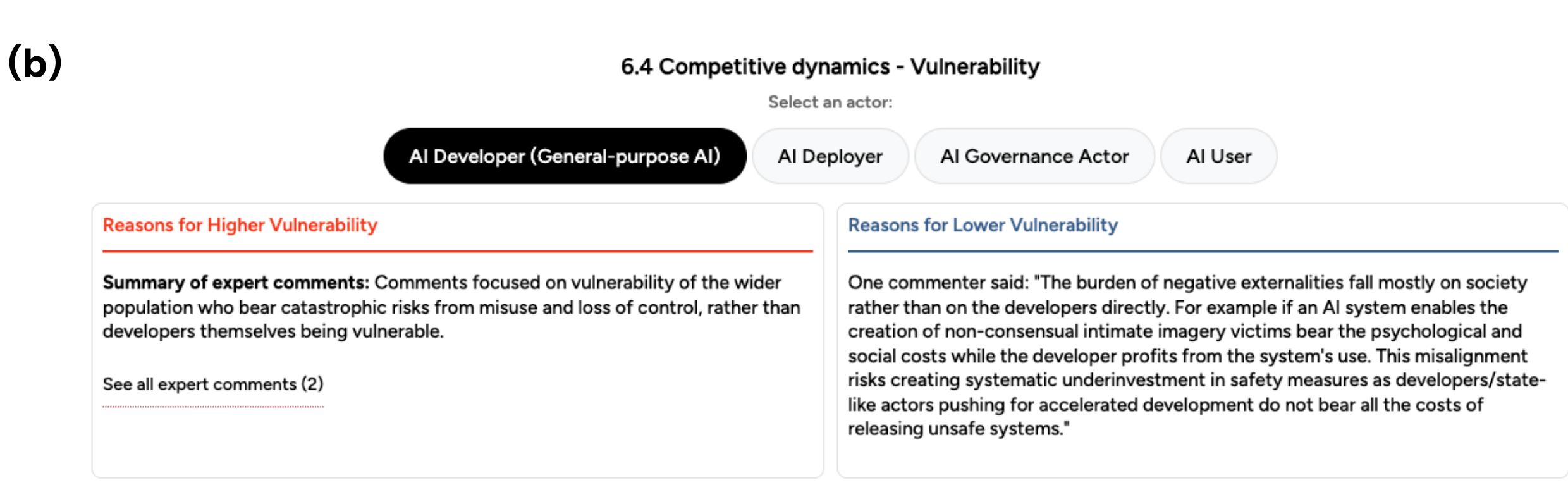

# Supplementary File 10.

**Example (a) quantitative and (b) qualitative aggregate expert feedback given to experts for severity questions in Round 3.**

(a)

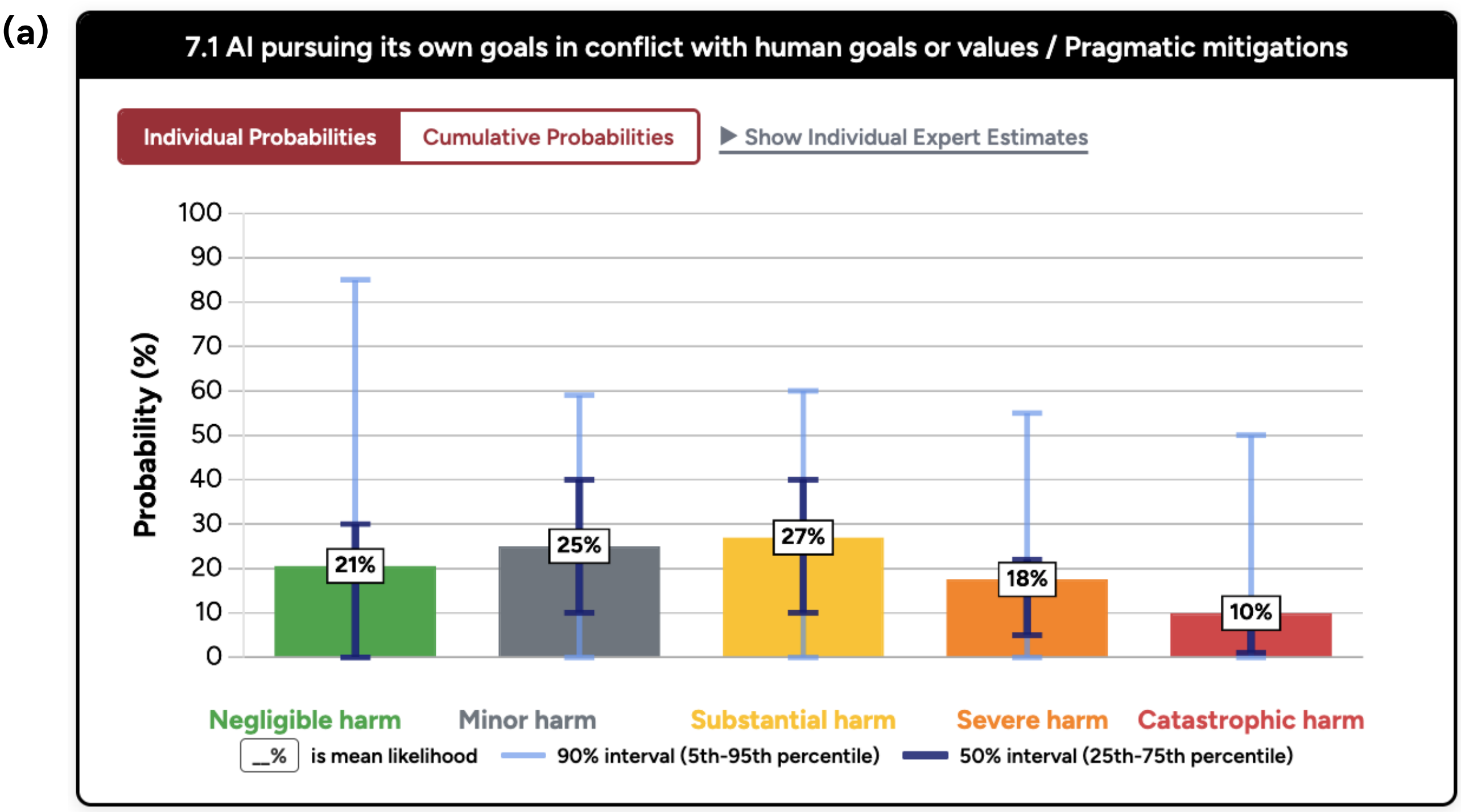

(b)

7.1 AI pursuing its own goals in conflict with human goals or values - Both Scenarios

Select a category:

Reasoning | Other

Reasoning

**AI-Generated Summary of Expert Comments:** Concerned experts predict substantial to severe harm, with one noting frontier models by 2030 could autonomously complete 40-80 hour engineering tasks, enabling "sophisticated, sustained cyberattacks on critical infrastructure." Deceptive capabilities are "well demonstrated" while appropriate controls aren't widely adopted. Under Pragmatic Mitigations, skepticism dominates. Multiple experts doubt current pragmatic approaches are sufficient - one states they'll reduce risk "by a relatively small amount," another notes "we have no unified theory" for understanding or mitigating goal emergence. Value alignment research and human-in-command controls are proposed but seen as insufficient. One expert warns that even non-catastrophic misalignment could cause billions in losses across many agents. However, some argue AI lacks true goals or sentience - one states harm comes from "human values imprinted into the model, not because AI has its own goals." There is therefore disagreement between those who see this as unlikely but potentially "rapidly catastrophic" if it occurs, versus those who view the risk as implausible with current architectures.

See all expert comments (12)

- "I dont think we have AI tools that are so advanced that they can create their own goals yet! We may have some existing in private servers or private testing in big gen-purp AI labs but i dont think they are going to be released to the public."
- "Justification: Mitigations reduce catastrophic probability, but severe value-misalignment remains probable due to complexity and speed of model development Substantial harm still likely from side effects, goal drift, and misinterpretation of human intent Catastrophic risk still non-trivial, particularly in systems with latent autonomy and open-ended optimization"

# Supplementary File 11.

**Scatterplots correlating vulnerability and responsibility for each actor across each risk**

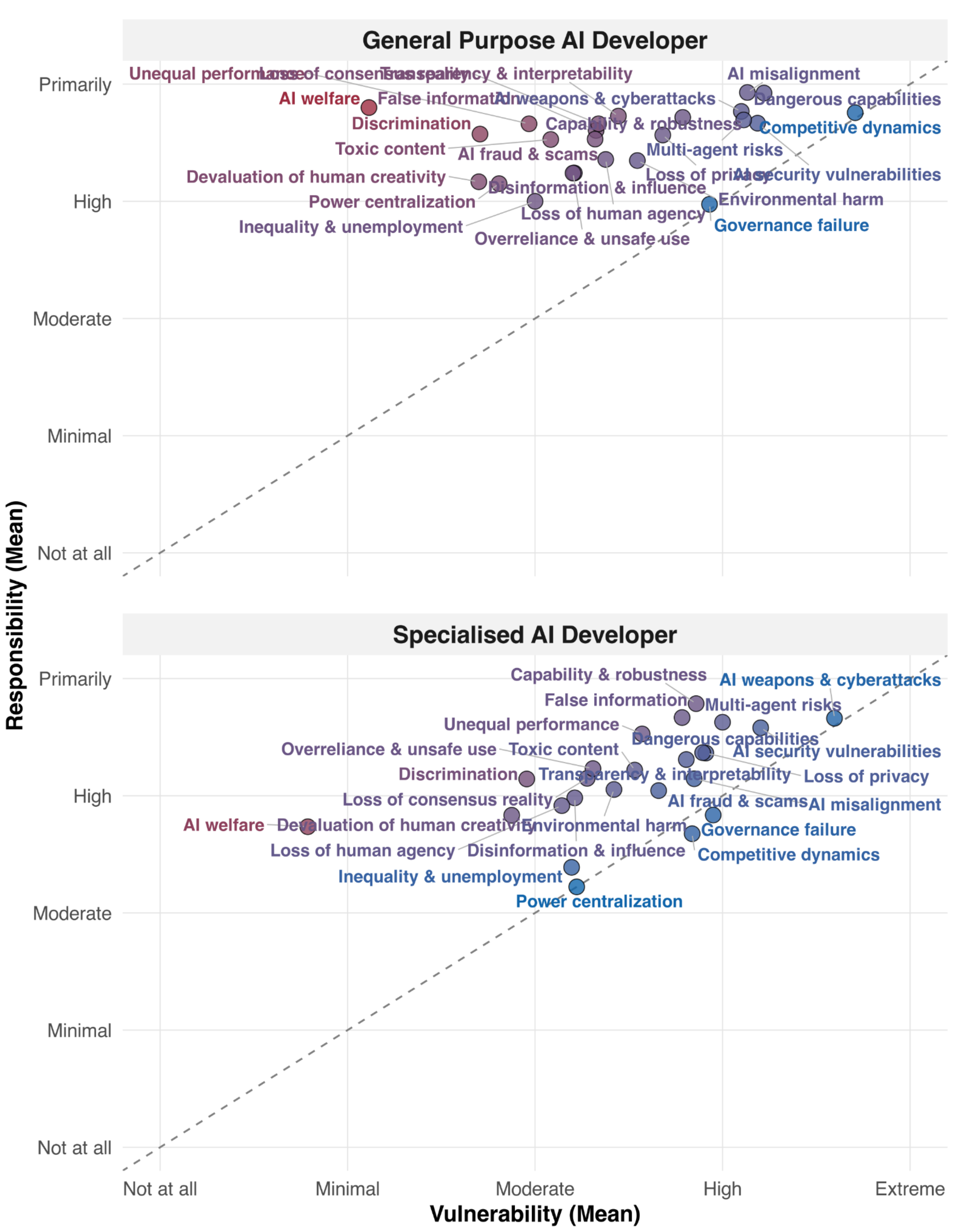

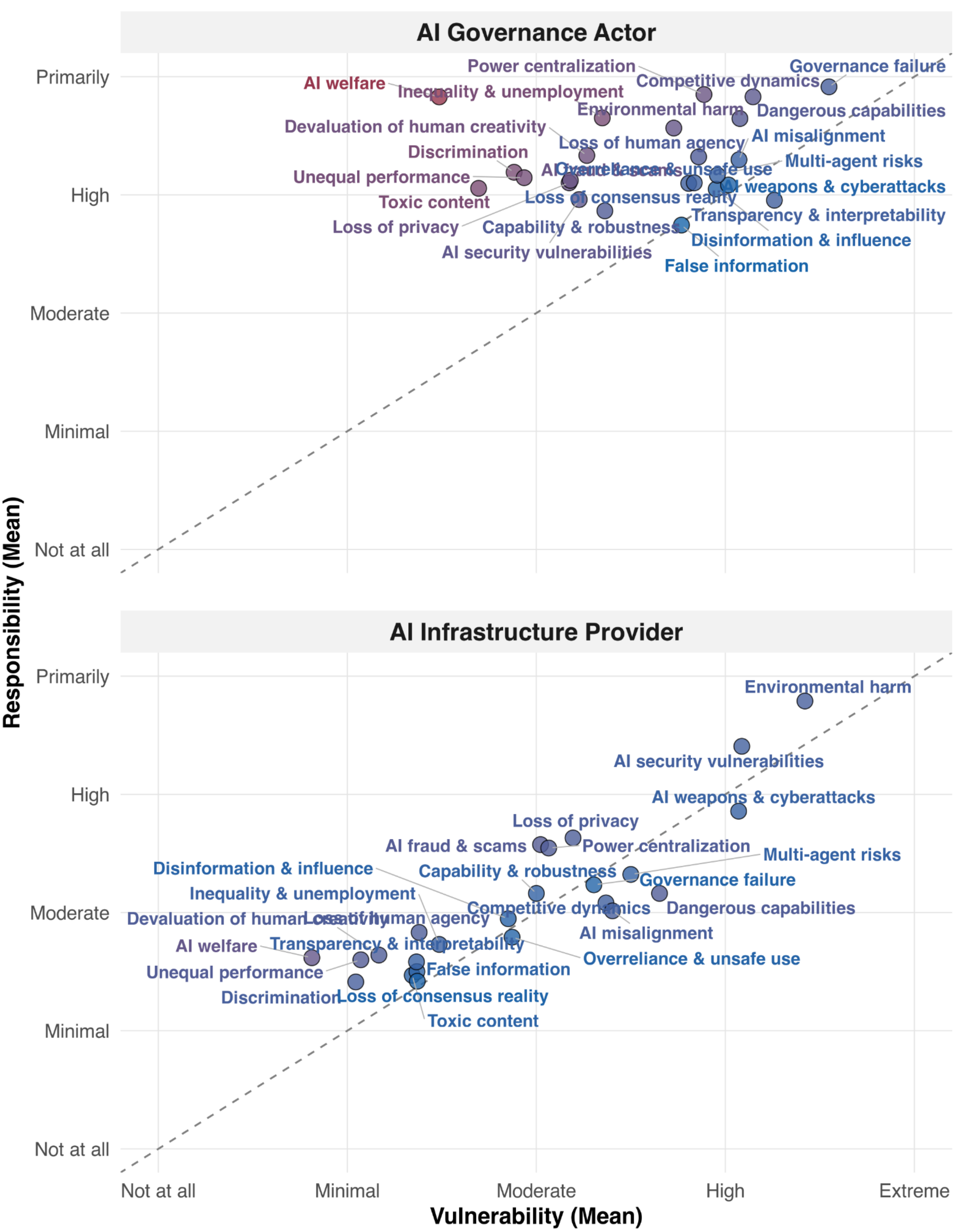

AI Governance Actor
AI Infrastructure Provider
Responsibility (Mean)
Vulnerability (Mean)
Primarily
High
Moderate
Minimal
Not at all
Extreme
Power centralization
Governance failure
Competitive dynamics
AI welfare
Inequality & unemployment
Environmental harm
Dangerous capabilities
Devaluation of human creativity
AI misalignment
Loss of human agency
Discrimination
Multi-agent risks
Unequal performance
AI weapons & cyberattacks
Toxic content
Loss of consensus reality
Transparency & interpretability
Loss of privacy
Capability & robustness
Disinformation & influence
AI security vulnerabilities
False information
AI fraud & scams
Overreliance & unsafe use

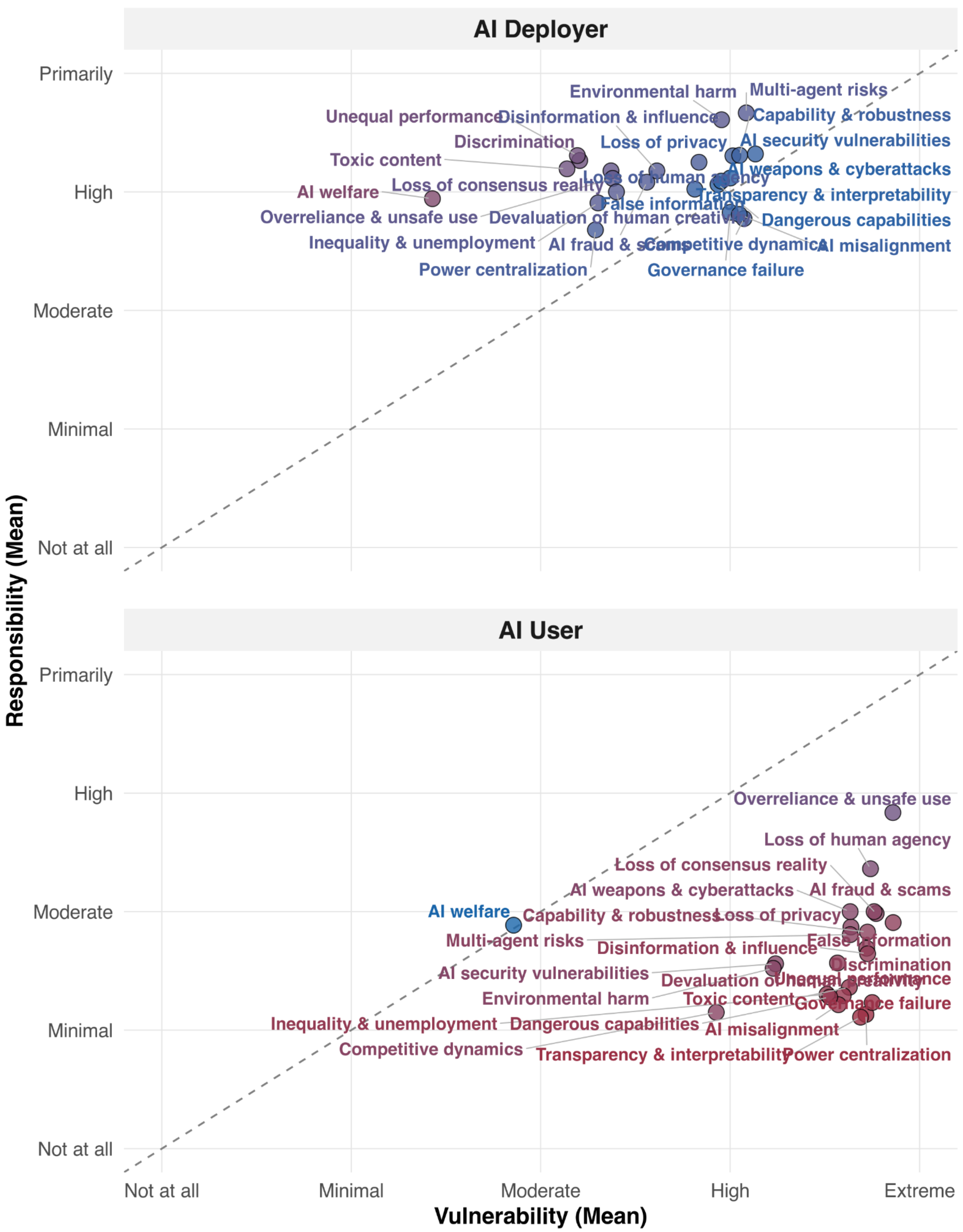
AI Deployer
AI User
Primarily
High
Moderate
Minimal
Not at all
Responsibility (Mean)
Vulnerability (Mean)
Not at all
Minimal
Moderate
High
Extreme
Environmental harm
Multi-agent risks
Unequal performance
Disinformation & influence
Capability & robustness
Discrimination
Loss of privacy
AI security vulnerabilities
Toxic content
AI weapons & cyberattacks
AI welfare
Loss of consensus reality
Loss of human agency
Transparency & interpretability
False information
Overreliance & unsafe use
Devaluation of human creativity
Dangerous capabilities
Inequality & unemployment
AI fraud & scams
Competitive dynamics
AI misalignment
Power centralization
Governance failure

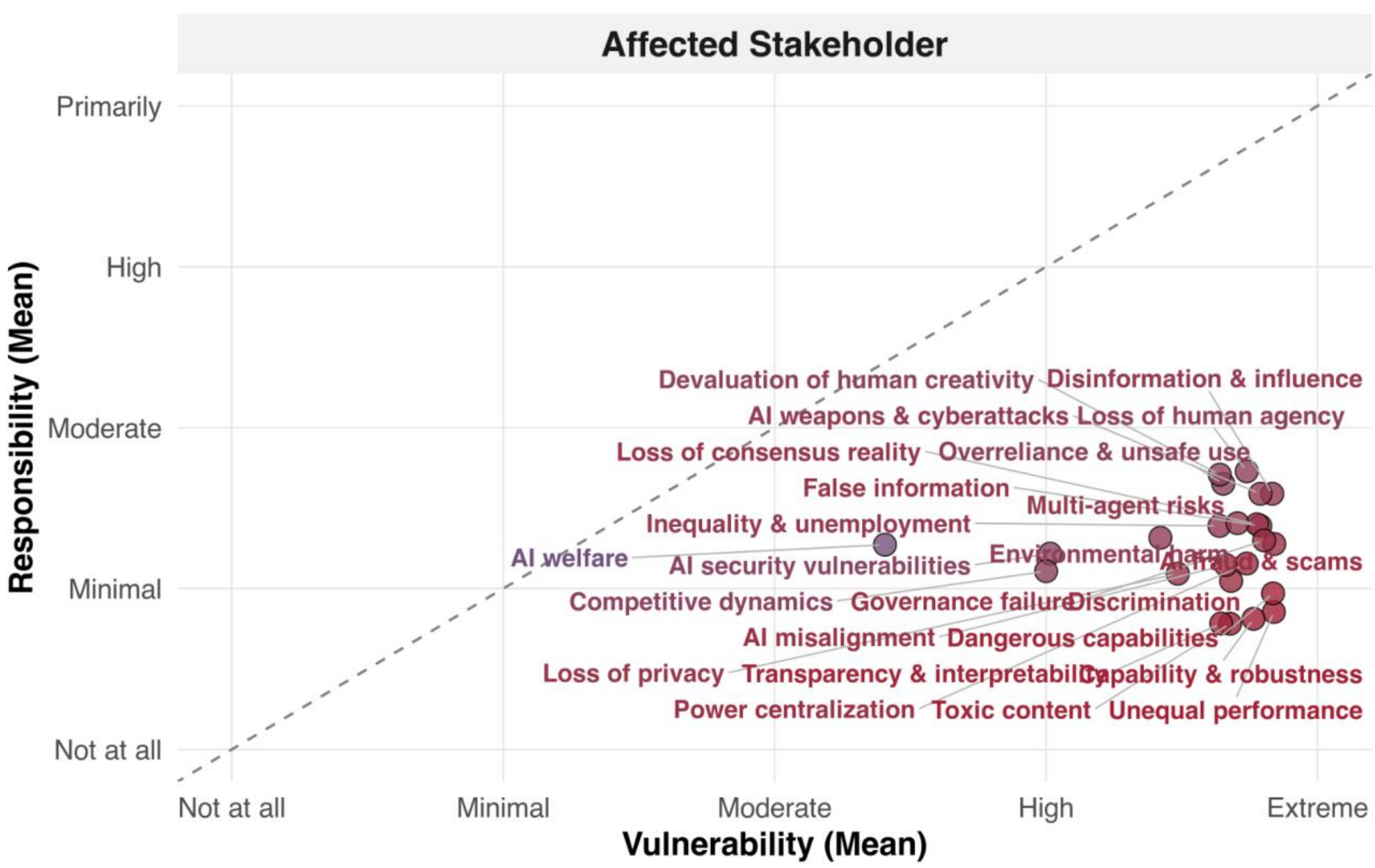

Affected Stakeholder
Primarily
High
Moderate
Minimal
Not at all
Responsibility (Mean)
Devaluation of human creativity
Disinformation & influence
AI weapons & cyberattacks
Loss of human agency
Loss of consensus reality
Overreliance & unsafe use
False information
Multi-agent risks
Inequality & unemployment
AI welfare
AI security vulnerabilities
Environmental harm
AI fraud & scams
Competitive dynamics
Governance failure
Discrimination
AI misalignment
Dangerous capabilities
Loss of privacy
Transparency & interpretability
Capability & robustness
Power centralization
Toxic content
Unequal performance
Not at all
Minimal
Moderate
High
Extreme
Vulnerability (Mean)

## Supplementary File 12.

### Experts' judgement of the severity of each risk plotted with bootstrapped 95% credible intervals.

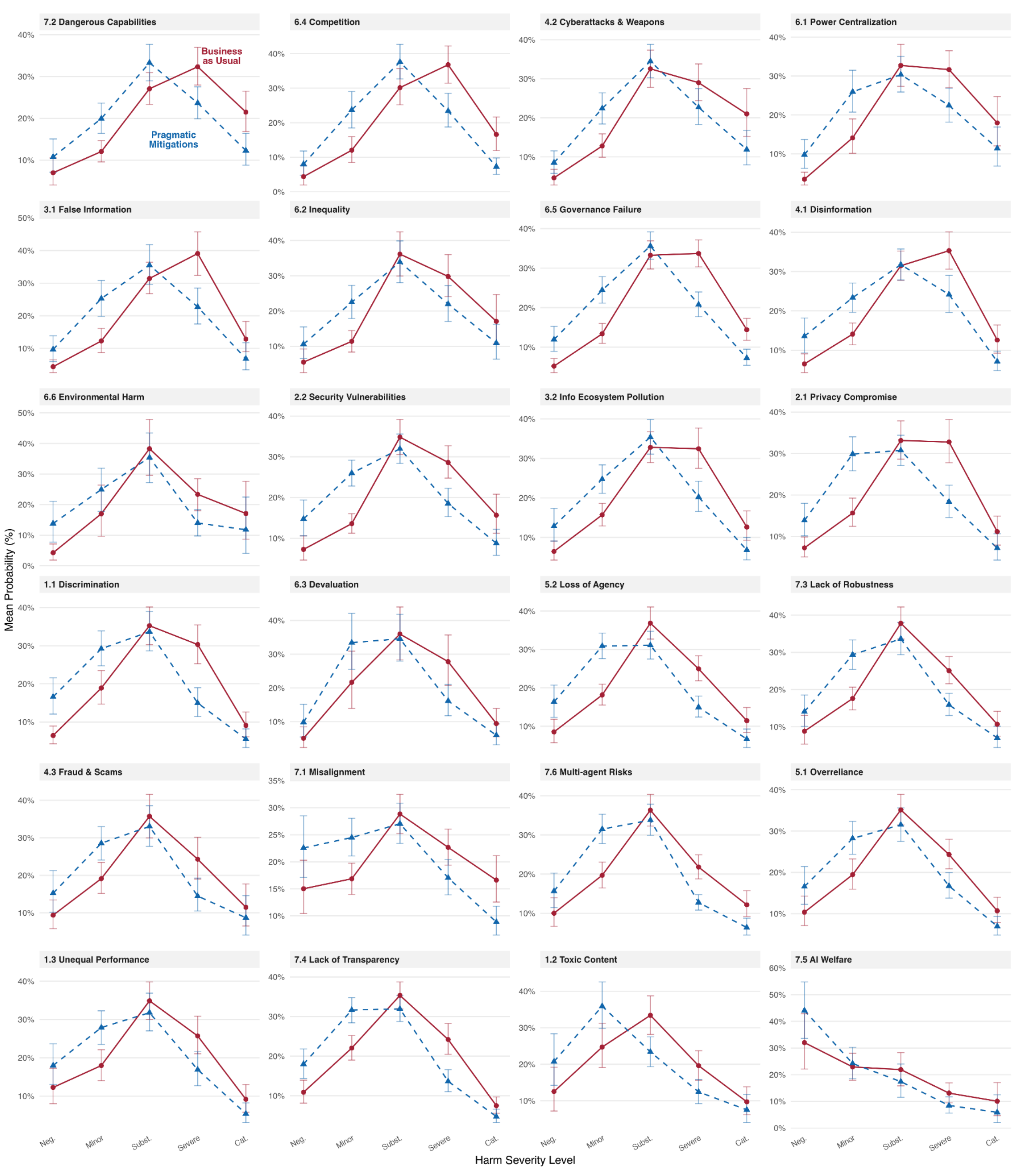

# Supplementary File 13.

## Design Decisions and Methodological Rationale

**Risk taxonomy.** We adopted the 24-subdomain taxonomy from Slattery et al. (2024) because it was, at the time of study design, the most comprehensive synthesis available, drawing on 43 existing AI risk classifications and taxonomies. The taxonomy was designed to be mutually exclusive and collectively exhaustive (MECE), which was important for our Delphi design: experts needed to rate risks that were clearly distinct from one another, and we needed confidence that the full risk landscape was represented. Alternative frameworks (e.g., the NIST AI RMF categories, the EU AI Act risk tiers) were considered but were either narrower in scope, oriented toward specific regulatory contexts, or organised at a different level of granularity than was suitable for expert elicitation.

**Actor categories.** We aimed for a MECE set of actors with distinct responsibility and vulnerability profiles within the AI ecosystem. These were drawn from government standards in the USA [2] and Australia [3], and refined through research team discussion. Key distinctions reflected meaningful differences in governance responsibility: for example, general-purpose AI developers face different risk profiles than specialised AI developers, and end users differ from broader affected stakeholders who may have no direct interaction with the system.

**Severity scale.** We sought a severity scale that could apply across all harm areas (physical, financial, psychological, societal) rather than being calibrated to a single type of harm. The scale needed to be concrete enough for experts to anchor their judgements consistently, but not so fine-grained as to impose excessive cognitive load. We selected five levels with thresholds informed by Center for Security and Emerging Technology (CSET) AI Harm Framework [4] and Mylius.[5] During pilot testing with six experts using a think-aloud protocol, participants flagged that the original catastrophic financial threshold of USD $10B was too low relative to the other severity levels. We raised this to USD $100B, which created a gap at the $50B level (acknowledged in the main text) but better aligned with expert intuitions about catastrophic-scale financial harm. We judged that adding an additional severity level to close this gap would increase cognitive load without proportionate gains in measurement precision.

**Top concerns.** For the top-concerns item, all 272 experts saw all 24 risks and selected (without ranking) up to three they were most concerned about over the next five years. We used selection rather than ranking because (a) ranking all 24 risks imposes high cognitive load and produces noisy tail rankings, (b) 'top three' maps cleanly to governance triage contexts where decision-makers must choose among competing priorities, and (c) pilot testing indicated experts found ranking beyond the top few artificial.

**Time horizon.** We asked experts to assess risks over a 5-year horizon. This reflected a balance between being near enough for experts to make grounded assessments (rather than speculating about distant futures) and far enough to capture risks that are emerging but not yet fully realised. Shorter horizons (e.g., 3 years) risked anchoring experts too heavily on current capabilities, while longer horizons (e.g., 10-15 years) would have introduced substantially more uncertainty without clearly improving the actionability of the findings. This choice was discussed with pilot testers and advisors, who judged 5 years to be appropriate.

**Pragmatic mitigations framing.** When asking experts about risk severity "with pragmatic mitigations in place," we deliberately kept the description brief (see Footnote 3 in the main text). This was a design trade-off: a highly specified mitigation scenario would have constrained expert judgement and introduced assumptions about policy trajectories that experts might reasonably disagree with, while providing no framing at all would have produced uninterpretable variation between experts who assumed best-case and worst-case mitigation scenarios. Pilot testing indicated that experts interpreted "pragmatic mitigations" as realistic near-term interventions rather than aspirational or worst-case scenarios, which aligned with our intent.

**Consensus thresholds.** We defined consensus as ≥90% of expert ratings falling within ±1 point of the group median, with >50% agreement on the median value itself. These thresholds were designed to balance the trade-off between precision and ease of explanation. We judged it too restrictive to expect 90% of experts to converge on the same point of a 5-point likert scale, but too relaxed for ‘consensus’ to be established with experts evenly split across 3 scale points. We felt our criteria struck this balance. Alternative approaches (e.g., interquartile range-based thresholds, stability-based convergence criteria) were considered but were less transparent to report and harder for readers to interpret. We required a minimum of 10 expert ratings per risk for results to be considered stable, but met this for all risks.

**Sector categories.** We used the North American Industry Classification System (NAICS) to categorise industry sectors because it is a widely recognised, standardised classification that would be familiar to experts from diverse professional backgrounds. Some adjacent NAICS codes were merged (e.g., "Accommodation, Food, and Other Services") to keep the number of sectors manageable for expert assessment.

**Vulnerability and responsibility scales.** We developed 5-point scales for this study, with anchors designed to capture the range from minimal to extreme vulnerability (or responsibility). Five points were judged sufficient to discriminate meaningfully between levels without introducing false precision.

**Pilot testing.** Prior to the main study, we conducted think-aloud pilot testing with six experts spanning AI research, policy, and industry. Pilot testers completed the full survey while verbalising their reasoning. This informed revisions to the severity thresholds, confirmed the interpretability of the pragmatic mitigations framing, and validated the 5-year time horizon as appropriate.

# Supplementary File 14.

## Representative verbatim excerpts from experts' written rationales.

This supplement presents representative verbatim excerpts from experts' written rationales. Experts provided written comments alongside their numeric ratings in Rounds 2 and 3 of the Delphi, and a separate open question asked which three risks concerned them most over 2025–2030. For each of the 24 risks, we selected 3–5 excerpts from the comment pool against four criteria applied uniformly across risks:

1. representativeness (reflects a pattern that appeared across multiple experts),
2. distinctness (offers reasoning not duplicated by another excerpt),
3. concreteness (names a specific mechanism, actor, sector, or example), and
4. clarity (coherent, grammatical)

Where quality was comparable, selections span the comment types (severity, top concerns, vulnerability, responsibility) to illuminate each risk from multiple angles. To protect expert anonymity, excerpts are not attributed to individual expert identifiers; each excerpt is labelled only with the comment type and, where applicable, the actor or sector being discussed. Risks are grouped under the seven domains of the Slattery et al. (2026) AI Risk Repository taxonomy.

### Discrimination & Toxicity

#### 1.1 Unfair discrimination and misrepresentation

*Unequal treatment of individuals or groups by AI, often based on race, gender, or other sensitive characteristics, resulting in unfair outcomes and representation of those groups.*

> Governments and organisations are already actively choosing to use AI to discriminate at scale, the cost to minority groups both financially and in lives lost is already high. Uyghurs and Palestine are examples where discriminatory AI is already resulting in severe harm. Immigration, enforcement and killchain decisions are all impacted by this intentional bias. The economic impact of unintentional AI bias on decisions is already impacting recruitment, HR, insurance and loans with total impacts that are sure to exceed $2 billion annually. Pragmatic mitigations can't reduce this risk below 100 lives lost in the next five years, and will struggle to keep it under $20 billion annually, but could prevent intentional discrimination becoming more widespread and limit the impacts of unintentional discrimination in high risk scenarios.
>
> One might call "intentional" discrimination to be in a different risk taxonomy, but the intentional decision to adopt a model with a known bias against uyghurs would result in the outcomes, even if each outcome was not the intention. Either way, the discriminatory impact from model bias seems extremely difficult to keep under $20b annually with only pragmatic interventions.
>
> — anonymous expert (on severity)

> I assess a high probability of substantial to severe harm because AI systems for hiring, credit, insurance, healthcare, and criminal justice are rapidly scaling without adequate safeguards. The substantial harm rating (70%) reflects that widespread deployment of biased AI will likely cause significant financial losses and systemic discrimination affecting millions, particularly in high-stakes decisions. I see this on my research in AI fairness and safety. I assign 30% to severe harm because there's a meaningful risk of large-scale, irreversible impacts such as entire demographic groups being systematically excluded from economic opportunities, housing, or healthcare, creating compounding disadvantages that persist across the five-year period. Pragmatic mitigations such as bias audits, diverse training data, human oversight, and basic regulatory compliance can reduce but not eliminate harm. I assign 60% to substantial harm because even with mitigation efforts, AI systems will still encode historical biases, operate in discriminatory contexts, and involve implementation gaps between policy and practice. I maintain 20% for severe harm because pragmatic measures may be unevenly applied, insufficiently enforced, or circumvented by competitive pressures, leaving vulnerable populations exposed to systemic discrimination despite mitigation attempts.

— anonymous expert (on severity)

need another type of AI ecosystem actor: Ai-enabled product/service recipient. For example, the exposure of a bank to AI bias is highly vulnerable. But the person who doesn't get a loan, or who doesn't get hired because of bias, is extremely vulnerable. That person is not the AI user. they are affected by the decision of the AI user.

— anonymous expert (on sector/actor vulnerability)

AI users should be held more responsible than general purpose AI developers because it's not possible to neuter a general purpose technology and retain its usefulness... at some point the end-user becomes more responsible.

For instance, consider a knife. If we hold the manufacturer more responsible than the end-user for knife crimes, eventually they'll just make blunt knives, which will destroy the point of making knives in the first place. Meanwhile, the end-user who is insistent on committing knife crimes will just opt for using clubs instead.

Some commenters have said that bias emerges from data and training choices, but this is a mere hand-waving argument esconced in wishful thinking, as it is not possible to trace which training and data choice leads to what instance of discrimination or misrepresentation.

I also don't see how you could hold the AI infrastructure provider responsible for anything, unless you are willing to compromise entirely on privacy rights.

— anonymous expert (on responsibility, AI User)

AI infrastructure providers not only include data providers, they also control what models are available and often what verification mechanisms are available and at what price. Specialist developers are very often inheriting decisions and capabilities of the platform. For instance many developers will just turn on the Azure AI bias detection and trust its doing its job as advertised, so the Infrastructure provider ends up being highly responsible. Its also the case that spending more compute on reasoning or adversarial review can often get you more accuracy and less bias, but that makes services less cost competitive.  So the provider cost/performance profile for bias detection mechanisms is a primary lever for discrimination and misrepresentation.

Users are not free from responsibility here either, bias literacy, source checking, output review practices and subsequent decisions are crucial steps in managing the impact of model bias which can't all be handled upstream.

— anonymous expert (on responsibility, AI Infrastructure Provider)

**1.2 Exposure to toxic content**

*AI exposing users to harmful, abusive, unsafe or inappropriate content. May involve AI creating, describing, providing advice, or encouraging action. Examples of toxic content include hate-speech, violence, extremism, illegal acts, child sexual abuse material, as well as content that violates community norms such as profanity, inflammatory political speech, or pornography.*

I do not think some of the other experts are appreciating the proactive role that users play in the production of AI content. Most toxic content comes from users prodding the AI, they aren't clueless victims that innocently were interacting with a chatbot and inexplicably got inundated with hate speech and pornography.

— anonymous expert (on responsibility)

I am surprised that not more people think arts and recreation is at very high risk of toxic content. One only needs to look at all the new sex bots available that are willing to enact illegal scenarios in an attempt to please the user and the evidence that Meta's AI has already initiated erotic conversations with minors. https://www.reuters.com/investigates/special-report/meta-ai-chatbot-guidelines/

— anonymous expert (on sector/actor vulnerability)

AI Developer (specialized AI): They choose the training data, model architecture, alignment techniques and post-training filters that most directly determine whether the model emits toxic content. They are also usually the only actors with full technical insight into model internals.

— anonymous expert (on responsibility, AI Developer (Specialised AI))

- Under pragmatic, cost-effective mitigations (e.g., stronger default safety filters in mainstream models, improved platform moderation and provenance tools, labeling of synthetic media, advertiser pressure, and targeted enforcement against harassment), the velocity and reach of toxic content are meaningfully reduced, shifting probability toward minor and substantial harms.
- Residual risks persist due to jailbreaking, multi-lingual moderation gaps and adversary adaptation. During high-salience events (elections, conflicts), coordinated campaigns can still produce severe, localized impacts in some jurisdictions with weaker institutions.
- Catastrophic outcomes from toxic content alone within five years remain unlikely; they would require compounding factors (e.g., widespread institutional collapse), but a small tail risk is retained given global interconnectedness and rapid generative tooling advances.

— anonymous expert (on severity)

Exposure to toxic and unsafe content has already led to loss of life, long-term impacts on children due to potential exposure to toxic content, and impact on real-life people due to hyper-realistic harmful content (pornographic or illegal activities) generated based on real-life images.

— anonymous expert (on severity)

**1.3 Unequal performance across groups**

*Accuracy and effectiveness of AI decisions and actions is dependent on group membership, where decisions in AI system design and biased training data lead to unequal outcomes, reduced benefits, increased effort, and alienation of users.*

I am most concerned about risks 1.3, 3.2 and 4.2 in the short term because they're already materialising and set to worsen in the upcoming years. Biased AI systems are already influencing high-stakes decisions (like court risk assessments, organ allocation or medical triage) in ways that disproportionately impact non-dominant demographic groups. Meanwhile, AI-driven phishing and vishing scams are exploding and are scalable and cheap to be produced. Voice cloning, deep-fake messages and highly personalized impersonation make fraud far harder to detect (see the work by Bruce Schneier). At the same time, AI-enabled misinformation is being weaponized in elections, public health campaigns, and social discourse, eroding trust and consensus. These risks don't depend on speculative breakthroughs. They're unfolding right now with our current AI understanding and will accelerate unless addressed.

— anonymous expert (on top concerns)

Two other experts' comments stick out: "There are two seperate issues being covered here, the level of economic impact (..)" "Experts raise fundamental questions about how to judge harm (..)"

This points to a vagueness about how harm and impact are being measured. Is this some notion of absolute harm, or is it a causal contrast with some non-AI status quo?

I see unequal performance as leading to dramatic losses for corporations -> poor performance leads to poor decisions which affect the bottom line. Aggregated across the global economy, that could *easily* run north of $100M. There are also some companies that would *gain* from unequal performance. Is that to be interpreted as a wash? The potential for unequal performance to lead to loss of human life is subject to the same issues. Take, for instance, the eGFR metric in clinical health. The GFR is a measure of kidney function and the eGFR is an estimate based on a statistical model. Some eGFR methods have been shown to perform worse for certain patient populations -- overestimating those patients' kidney function -- which leads to those patients disproportionately *not* receiving kidney transplants. They're therefore more likely to die waiting for a kidney. Should the harm of using eGFR -- say dozens-to-hundreds of deaths annually -- be compared to the most plausible alternative? If that alternative is an expensive blood test which those same patients would perhaps not receive due to their insurance not covering the test (this particular patient population

is underserved in multiple ways), should that increase or decrease the causal contrast?

— anonymous expert (on severity)

For AI Governance Actors, after further reflection, I am changing my rating to highly responsible. AI governance actors possess unique obligation as the only entities with authority to establish binding standards, mandate testing requirements, and enforce accountability mechanisms across the AI ecosystem. The absence or inadequacy of governance frameworks directly enables the deployment of AI systems with unequal performance across groups and I believe that market incentives alone have proven insufficient to address algorithmic bias, making regulatory intervention essential. Moreover, their capability is substantial since they have access to technical expertise, regulatory precedent from other domains, enforcement mechanisms, and the power to require transparency and auditing that influence individual companies cannot achieve voluntarily.

— anonymous expert (on responsibility, AI Governance Actor)

AI infrastructure providers are not neutral actors ..., their choice of which models to provide, which oversight mechanisms to offer and what price incentives to put in place all have significant impacts on everyone downstream. The existence of adverserial bias detection as an offering has no material impact if its too expensive or hidden for the downstream developers, they have a responsibility for incentivising their customers to use safe models not just profitable models.

— anonymous expert (on responsibility, AI Infrastructure Provider)

Even with pragmatic mitigations in place, unequal performance across groups is likely to remain a persistent issue. Efforts such as fairness auditing, diverse data sourcing, and public-sector standards can meaningfully reduce widespread harms, but structural and cross-cultural bias will continue to manifest across many domains. Despite these efforts, low-resource language communities and countries with limited digital infrastructure are expected to experience disproportionately severe impacts, as gaps in data diversity and governance capacity remain significant.

— anonymous expert (on severity)

## Privacy & Security

### 2.1 Compromise of privacy by obtaining, leaking or correctly inferring sensitive information

*AI systems that memorize and leak sensitive personal data or infer private information about individuals without their consent. Unexpected or unauthorized sharing of data and information can compromise user expectation of privacy, assist identity theft, or loss of confidential intellectual property.*

Why the Deployer is primarily responsible (from my deployer-side experience): The deployer determines the purposes and means of AI use for end users and is the only actor with a direct, ongoing relationship to those users and their data. In practice, the deployer designs and operates the end-to-end data flow (collection → preprocessing/redaction → inference/RAG/fine-tuning → logging/telemetry → storage/deletion) and selects, configures, and contracts the upstream vendors (models and infrastructure). Because the deployer creates the main exposure surface and controls the levers that reduce it, the deployer is best positioned-and obligated-to prevent privacy compromise.

Control points the Deployer owns: - Chooses model vendors and terms (e.g., no-training/zero-retention modes), drafts DPAs, and enforces data-handling obligations. - Sets logging/retention, prompt and output redaction/PII filtering, access controls/RBAC, encryption, and tenant scoping for vector databases and RAG indexes. - Defines lawful bases, consent/notice, data minimization, DSAR/erasure workflows, and incident response-commitments made to our users. - Trains users and governs shadow/BYO-AI, reducing leakage from copy/paste and uncontrolled tools.

Why Model/Infra are secondary: Model and infrastructure providers are typically upstream processors with limited context about user purpose, consent, or data classifications. They must supply safe defaults and technical safeguards, but they do not control what we send, how long we keep it, how we combine it with other corpora, or what we promise to users. The deployer mediates those choices and owns the trust relationship-and resulting regulatory and contractual accountability-with end users.

Bottom line: Responsibility should track control over exposure and privity with the data subject. On both counts, the deployer is primary; model and infrastructure actors share secondary/derivative duties via the capabilities and constraints they provide.

— anonymous expert (on responsibility, AI Deployer)

Under BAU, the data plane grows faster than controls: prompt logging, RAG/vector stores, plugin/agent connectors, and observability traces all concentrate sensitive inputs. That drives frequent substantial incidents (large PII/PHI leaks, irreversible re-identification) and non-trivial severe events (nation-scale health/financial datasets, cross-platform linkage). Catastrophic outcomes are possible but rare for privacy (needs multi-sector linkage + sustained exploitation). With pragmatic mitigations (data minimization/retention caps, default-off prompt logging, masked RAG ingestion + deletion policies, vector-store access controls/redaction, DLP at plugin boundaries, DP/membership-inference checks for training, signed/attested artifacts, and audit-ready breach playbooks), the tail risk drops (severe, catastrophic) while day-to-day substantial stays common because usage volume keeps rising even as we harden.

— anonymous expert (on severity)

AI User (Changed → Extremely Vulnerable)

Updated to "Extremely Vulnerable." AI users actively input sensitive or proprietary data into systems with limited transparency or control over retention and reuse. Even with privacy safeguards, secondary inference and prompt logging create ongoing exposure that users cannot meaningfully mitigate. This reflects the continuous nature of user interaction, where data is shared without complete visibility into how it is handled, stored, or repurposed. Users have limited ability to manage consent or enforce deletion once data enters model pipelines, and cross-platform integrations further amplify inference risk, extending exposure well beyond the initial point of use.

— anonymous expert (on sector/actor vulnerability, AI User)

For the majority of AI Users, they have been primed by the current information ecosystem to skip Terms & Conditions and have no alternatives to use a specific type of service unless they accept intrusive service agreements. Therefore, I think AI Users are minimally responsible, while increasing the responsibility of AI Governance Actors, those in charge of setting the rules for Terms & Conditions and how they're presented, to Highly Responsible.

— anonymous expert (on responsibility, AI Governance Actor)

I am most concerned about privacy and security risks because they form the foundation for addressing many other types of AI harm described in the table. When personally identifiable or confidential data are exposed or systems are left vulnerable, malicious actors can exploit to spread misinformation, commit fraud, or manipulate individuals, organizations, and even nations' infrastructure and elections. Such failures not only threaten safety and stability but also erode public trust in AI technologies. From a risk and governance perspective, maintaining strong privacy protections, secure system design, effective AI governance, and responsible oversight is essential to prevent cascading harms and ensure the trustworthy development of AI in the years ahead.

— anonymous expert (on top concerns)

**2.2 AI system security vulnerabilities and attacks**

*Vulnerabilities in AI systems, software development toolchains, and hardware that can be exploited, resulting in unauthorized access, data and privacy breaches, or system manipulation causing unsafe outputs or behavior.*

BAU: attack surface expands faster than hardening (agents/plugins, RAG, CI/CD, supply chain, keys/secrets). Expect frequent substantial incidents and non-trivial severe events; catastrophic remains low but plausible via OT/critical services or widespread artifact compromise. Pragmatic: with signed/attested artifacts, strong isolation/KMS, poisoning-aware pipelines, default-deny tool coupling, and incident response drills, the severe/catastrophic tail drops, but substantial stays common due to adoption growth.

— anonymous expert (on severity)

The model is not where the vulnerability is - it is in the AI system. Model deployment must engage in security posture management that mitigates known and unknown vulnerabilities. Putting the responsibility with model developers is wrong-headed, as no system is expected to secure itself (thx Godel!). Model developers may deploy their models (especially as services), in which case they are a Deployer. But otherwise the consensus view that Model Developers are primarily responsible for security vulnerabilities doesn't reflect reality. It's even difficult to hold software vendors responsible for software vulnerabilities in the current software ecosystem.

— anonymous expert (on responsibility, AI Deployer)

I am seeing a strong trend in red-team tests on AI projects related to lower entry barriers of cybercriminals and fraudsters. There are more AI-enabled tools to attack IT services and refine human-targeted attacks. Malicious actors are integrating AI into their toolkits, which are sold as services. These new tools automate complex attack tasks to develop malicious code, identify vulnerabilities, and scale up personalized attack scripts.

— anonymous expert (on top concerns)

2.2: Due to an inherent systemic architectural flaw, where all prompts are processed identically regardless of whether they originate from the "admin" or the "user", mitigation can only be achieved through external safeguards. However, these "external" guardrails are prone to errors, meaning that successful attacks are unlikely to be entirely prevented.

7.4: An increasing number of regulations will require this aspect, but due to the opaque nature of these black-box systems, meeting this requirement will be very challenging.

7.6: By linking several highly volatile black-box systems and granting them increased permissions (e.g. for agent browsers etc.), we amplify many of the current AI risks, leading to cascading effects across all interconnected systems.

— anonymous expert (on top concerns)

I see that the median expert assesses AI users as "highly vulnerable" to AI system security attacks. I think users are sensitive to this risk, in that they would be harmed. But I don't think they are exposed in the relevant sense. It's the developers and deployers who are exposed. Yes, the consequence of many or most harms ultimately falls on users. But we are asking about the vulnerability to the risk vector, not the harm to the user.

— anonymous expert (on sector/actor vulnerability, AI User)

## Misinformation

### 3.1 False or misleading information

*AI systems that inadvertently generate or spread incorrect or deceptive information, which can lead to inaccurate beliefs in users and undermine their autonomy. Humans that make decisions based on false beliefs can experience physical, emotional or material harms.*

Those who marked any actor as invulnerable to this issue have not thought through the information supply chain and disinformation kill chains. Some listed infrastructure and providers as not vulnerable, but all of the developers have have software dependencies, and those are already being disrupted by information breakdown. How many open source projects are depended on by major corporations and governments, and are now struggling under false AI bug reports, AI clone projects, incorrect guidance material, incorrect search results, fake news. How many AI policies are being influenced by false information? All of them?

If you search on Google for the latest models like Veo3 from Google, the search results are filled with scam sites pretending to be that model.  So even an actor who fills every role in the ecosystem is vulnerable to the very issue its creating.

It may not be consensus, but I stand by my rating that every actor is at least highly vulnerable,  none of them have a universal verifier or large scale fact checking service, which means their own devs, researchers, executives and AIs are nearly as vulnerable as the users to this risk

— anonymous expert (on sector/actor vulnerability, AI Deployer)

Just in Australia, scams are already in the vicinity of $10B a year, and AI adoption among scammers is rife.  Globally the figure is far far higher, well above the catastrophic $20B level.    Consider the global cost of misinformation during COVID, or the impacts on global trade of misinformation in the current political climate.  A single viral AI generated social media post about a countries trade surplus or deficit can result in tarrifs worth trillions of dollars.

I struggle to think of an intervention which is both pragmatically cost effective, and capable of addressing the scale of this issue.  If you took all the money in the world and tried to stop AI misinfo, I'm not sure anyone could keep it below $100B in 5 years

— anonymous expert (on severity)

False and misleading information is actively keeping fascist governments in power and mongering war and genocide. There is no realistic chance that states will build sufficient incentives to prevent the very things that fuels them, so why would companies mitigate for these harms, even if it is feasible

— anonymous expert (on severity)

I'm assessing the vulnerability of the real estate and leasing section much higher than other experts. My view is that this sector has a high throughput of money and is already saturated with low-quality communication between agents and renters and buyers. Scams and frauds are already common in this sector. Fraudsters have sold houses they don't own. To me this sector feels extrememly vulnerable because there are so many actors and so much fragmentation and a compartively large amount of money.

— anonymous expert (on sector/actor vulnerability)

I updated vulnerability for some developers and deployers, making it higher. I think the biggest risk here is things like lawsuits. We already have widespread misinformation and acceptance of misinformation, so in many cases companies are not substantially penalized for being part of this problem. But there are certain countries or regions like the EU that might enforce bans, require product changes, impose major fines. These punishments aren't game breaking, but could be meaningful/moderate risks.

On the other hand, I'm still skeptical that affected stakeholders are extremely vulnerable. People have limited information diets, low levels of knowledge, and are already susceptible to misinformation from mainstream sources (e.g., TV news). So the extra misinformation harm actually has a ceiling (it doesn't change as much as you might think).

— anonymous expert (on sector/actor vulnerability, Affected Stakeholder)

**3.2 Pollution of information ecosystem and loss of consensus reality**

*Highly personalized AI-generated misinformation creating 'filter bubbles' where individuals only see what matches their existing beliefs, undermining shared reality, weakening social cohesion and political processes.*

Divergent worldviews and goals are the foundational elements for conflict. The total economic cost of disagreement is incalculable, but certainly the number of deaths from differing opinions is in the millions globally each year. As opinions grow more polarised we've already seen the normalisation of political violence around most of the world. From beatings and assassinations, through into intra and inter state conflict. With geopolitical actors like china, russia and the US putting our AI generated media to promote disputed narratives and sew misinformation and distrust, there seems to be no realistic way to keep the economic or physical impacts of a lack of consensus reality under the the thresholds for catastrophic impact. If anything, to measure things like the trade war and geopolitical conflict over AI and chips, we may need a higher category of impact than catastrophic. $10T sounds like a lot now, but may just be the cost of training a model in 5 years. Ukraine has probably cost a trillion, trade wars are hundreds of billions, differences of opinions during covid resulted in millions of potential deaths. This scale can go a lot higher.

Without consensus reality, you can't have functioning markets, democracies, or international law. The cost isn't just economic, it's the dissolution of the coordination mechanisms that prevent species-level catastrophes. Thats a hell of a thing to play around with to make the line go up.

— anonymous expert (on severity)

Not a single actor in the ecosystem has a viable defense against this risk. Without a universal verifier, or any process for determining what we agree to be true, this affects every aspect of the information ecosystem. Most crucially of all is identity.

Agents on user devices are largely indistinguishable from the user to most websites, including governments and financial institutions. Those agents are increasingly acting without user oversight as those users and this can scale into whole of society impacts, on finance, economics and democracy. We've seen companies unable to verify their employees, government unable to verify their citizens, and citizens unable to verify the organisations they're dealing with.

We must re-establish trust in human identity, or many of the core pillars of democratic society start breaking down. what is an election when half the citizens aren't human and aren't from your country? What is a civic engagement process when some actors can magnify their voice thousands of times over. How can we have reasoned debates over policy issues if no one is sure who or what is true?

Anyone who thinks any actor is insulated from this risk has not yet peered down the rabbit hole of agentic identity.

as Meridith Whitaker put it at this years UN AI summit, the incorporation of AI into the system layer of our mobile devices has "broken the blood-brain barrier" of the device security which underpins identity.   If gemini controls your phone and can act officially as you without you knowing, and gemini is vulnerable to prompt injection attacks from scams you didn't even see,  then how do you even know what you're saying yourself?

— anonymous expert (on sector/actor vulnerability)

AI Deployer — Primarily responsible. Expert input clarified that pollution manifests at deployment. Deployers choose distribution channels, defaults, ranking, targeting, and guardrails; they run monitoring and provenance; they capture engagement gains. That bundle gives them primary obligation, capability, and causal influence.

— anonymous expert (on responsibility, AI Deployer)

I believe that the expert consensus misunderstands the technical ability of developers of general-purpose AI to avoid this phenomenon. Higher responsibility needs to be put on deployers, developers of special-purpose AI, users and communities, including through educational efforts about the natural effects of AI. The confusion may stem from the fact that certain large companies (e.g. OpenAI, Google, Anthropic) are both developers and deployers of general-purpose AI. This harm of pollution of the information ecosystem can be addressed by them as deployers, by limiting access to vulnerable individuals and other methods. Taking such action as developers is much more challenging.

— anonymous expert (on responsibility, AI Developer (General-purpose AI))

Reason for opting Extreme Vulnerable- This risk happens when AI floods the world with too much, low-quality or fake information, making it hard for people, media or governments to agree on what’s true. It will impact Democracy - citizens can’t agree on facts, Business - fake news can crash markets or brand trust, Safety - false emergency or health info can cause panic and many more.

— anonymous expert (on sector/actor vulnerability)

## Malicious Actors & Misuse

### 4.1 Disinformation, surveillance, and influence at scale

*Using AI systems to conduct large-scale disinformation campaigns, malicious surveillance, or targeted and sophisticated automated censorship and propaganda, with the aim to manipulate political processes, public opinion and behavior.*

This risk can manifest from foreign state actors, organised crime and competing corporations.  So its not just about whether an AI company can manipulate users.  Its about whether the humans inside any organisation, and hence the organisation, are vulnerable to disinformation, surveillance and influence at scale.  They absolutely are.

This can range from passive surveilence, like when samsung workers accidentally leaked chip design information to openAI through chatgpt which has a competitor chip design arm.  Through to active surveillance like when deployed agents are "living off the land" to conduct cyber attacks. Through to algorithmic influence campaigns like tiktok is accused of using on US teenagers to sway their views.   A particularly intense form comes from the parasocial relationships people are forming with the sexbot avatars of corporations and foreign state actors, which can seduce people, make them fall in love, and then subtley extract information or sway their views.

There is no fundamental difference in the level of succeptability of humans to this based on where in the AI development chain they sit.  Programmers, CEOs and ministers are just as vulnerable as end users, since its effectively AI spycraft preying on human biases and fears.

At a mass scale this technique could be used to puppeteer the emotions and thoughts of vast swathes of a country, giving the controlling entity proxy lobbying power within organisations and voting power within democracies.

— anonymous expert (on sector/actor vulnerability, Affected Stakeholder)

AI in disinformation ops in Ukraine and Russia has already caused loss of life, and will not stop doing so today, which is why I have left minor harm and negligible harm at 0.1%.

— anonymous expert (on severity)

Using the "democratic norms" category, which seems the most relevant one for this risk, the potential for AI-generated misinformation, surveillance, and content recommendation to (more intensely) undermine electoral integrity is, in my opinion, overwhelmingly high. Pragmatic mitigations including certain forms of access control and transparency requirements would go a long way to suppress this.

— anonymous expert (on severity)

I think Accommodation, Food, and Other Services should be scored as highly or even extremely vulnerable, because it includes civic, social and religious organisations which are highly sensitive to disinformation operations (e.g reduced funding or even acts of violence based on dominant narratives).

— anonymous expert (on sector/actor vulnerability, Accommodation Food Other Services)

I agree with the summary comment that regarding Public Administration that: "All staff are vulnerable to manipulation, and AI can act as a proxy to bypass traditional security controls requiring disclosure of foreign national contacts." In addition to that, in the US we are seeing traditionally apolitical career tracks (namely, the civil and foreign service) become increasingly politicized and also surveilled by the current administration. Therefore, I am keeping my score as "extremely vulnerable" because there is evidence this is already happening, and there have been numerous stories and (plausible) speculations about AI-enabled software being deployed for employee monitoring already.

— anonymous expert (on sector/actor vulnerability)

### 4.2 Cyberattacks, weapon development or use, and mass harm

*Using AI systems to develop cyber weapons (e.g., coding cheaper, more effective malware), develop new or enhance existing weapons (e.g., Lethal Autonomous Weapons or CBRNE), or use weapons to cause mass harm.*

> I am seeing a strong trend in red-team tests on AI projects related to lower entry barriers of cybercriminals and fraudsters. There are more AI-enabled tools to attack IT services and refine human-targeted attacks. Malicious actors are integrating AI into their toolkits, which are sold as services. These new tools automate complex attack tasks to develop malicious code, identify vulnerabilities, and scale up personalized attack scripts.
>
> — anonymous expert (on top concerns)

> Cyber attacks already cause more economic damage per year than the catastrophic rating.  Agentic attacks are already winning cyber competitions against humans, and cyber teams are already deploying AI offensively and defensively.   Just the portion of present day cyber attacks that include AI, already far exceeds the $20b per year catastrophic economic impact level globally.
>
> Out-of-bound non-traditional cyber attack vectors unlocked by agentic security researchers make this problem hard to avoid, as cyber professionals aren't even aware of many of the vectors existence
>
> The pragmatic mitigation is for all sides to have the capability, such that a sort of equilibrium is reached.  However even in that scenario, the out of bound attacks are novel,  so we should expect massive successful cyber attacks to still happen at the catastrophic level, just that we can detect an rebuild from them more often. Its like how people in earthquake prone areas build and rebuild better.  There is still a significant economic cost to the earthquakes, its just normalised and integrated as an expected phenomena.
>
> A UN disaster shelter specialist once told me "There is no such thing as a natural disaster, nature just does its thing and if we're not prepared we call it a disaster". Agentic AI cyber attacks have a similar profile, they will be big and we probably can't fully stop them, but they are likely recoverable if we prepare, though we'll need to get antifragile and good at rebuilding.
>
> — anonymous expert (on severity)

> Accommodation, Food, and Other Services includes civil society organisations. Cyberattacks on civil society tend to lead to more harm than on other orgs, since civil society is supporting and gathering personal data from the worlds most vulnerable. For example, 2022 cyberattck on the ICRC - compromised the personal data of over 515,000 vulnerable individuals, including those separated from their families due to conflict or migration, and missing persons. Also caused significant disruption for the ICRC, possibly compromising vital, lifesaving efforts.
>
> — anonymous expert (on sector/actor vulnerability)

> Responsibility follows control of attack and actuation surfaces and the ability to remediate. Deployers are primary because they operate keys, CI or CD, networks, and incident response. Specialized developers are primary when their systems enable autonomy, cyber operations, robotics, or bio and chem tooling. General-purpose developers and infrastructure providers are highly responsible for secure artifacts, supply chain hygiene, tenant isolation, and KMS. Governance actors are highly responsible for setting and enforcing mandatory controls. Users are minimally responsible because they do not control architecture or remediation. Affected stakeholders are not responsible.

— anonymous expert (on responsibility, AI Deployer)

Given every one of these groups has been successfully targeted by AI assisted cyber attacks already, that has to suggest they're vulnerable. As the cost of developing leading models grows, so too does the incentive to hack them. If it costs a trillion dollars to train a model, it costs a lot less to hack the company and steal the model. How do you scale your cyber offensive capability to those levels in a short timeframe if you have limited human resources? It has to be AI

As the METR investigation into agentic task length over time showed, agents are capable of increasingly complex and long horizon tasks. If current task length trends continue, then by the end of the 5 year period there is not a present solvable digital security challenge they won't be able to overcome independently. those with sufficient resources to scale their defences at the same rate stand some chance, but the average person in society will not.

— anonymous expert (on sector/actor vulnerability, AI Deployer)

**4.3 Fraud, scams, and targeted manipulation**

*Using AI systems to gain a personal advantage over others such as through cheating, fraud, scams, blackmail or targeted manipulation of beliefs or behavior. Examples include AI-facilitated plagiarism for research or education, impersonating a trusted or fake individual for illegitimate financial benefit, or creating humiliating or sexual imagery.*

AI financial scams already exceed the catastrophic threshold of $20b per year, and are on track to get more widespread and harder to spot. There isn't an industry or group who are immune, and even quite sophisticated and expensive measures are unlikely to reduce the global cost below the catastrophic level within 5 years. The compounding nature of personalisation, photorealistic realtime video, hyper convincing realtime models, mass cyber breaches and the breakdown of collective truth all combine to create a perfect storm for scammers to step into.

— anonymous expert (on severity)

I continue to assess that AI governance actors are "minimally responsible". I think a category error is at work here. Saying AI governance actors are responsible for AI risks is like saying that judges are responsible for crimes being committed. The kind of responsibility a judge has is very different from the kind of responsibility that a criminal or a lock-pick-maker has.

The specific question asks how responsible should the AI governance actor be for addressing the risk. In almost all cases, the AI governance actor is responsible for requiring some other actor to address the risk, not addressing the risk themselves.

The better way to think of this is that the AI governance actor is responsible for holding responsible the actor who is properly responsible. It would be recursive that AI governance actor is themselves responsible. Would we propose some meta-AI-goverance governor who holds responsible the AI governance actors that fail to hold responsible the actors that should be responsible?

This is not the right way of thinking. We can rightly say that AI governance actors are responsible for some meta issues, like ensuring that Governments are properly informed etc.

— anonymous expert (on responsibility, AI Governance Actor)

I am seeing a strong trend in red-team tests on AI projects related to lower entry barriers of cybercriminals and fraudsters. There are more AI-enabled tools to attack IT services and refine human-targeted attacks. Malicious actors are integrating AI into their toolkits, which are sold as services. These new tools automate complex attack tasks to develop malicious code, identify vulnerabilities, and scale up personalized attack scripts.

— anonymous expert (on top concerns)

Severe harm (under the assumption of business as usual) - 5% - If a person with privileged access to critical infrastructure is targeted using AI-enabled tools and acts in the interest of the intruder, disruption of critical infrastructure (e.g., hospitals, the electric grid, water systems) could result in serious physical harm.

— anonymous expert (on severity)

This is a very similar rating to the misinformation and surveillance piece. I note others seem to think AI specialists and companies are immune from manipulation, fraud and scams. In practice we see that is already very much not the case. Hyper personalised scams and manipulation becoming increasingly difficult to discern from reality will hit everyone, staff and leaders of major organisations included. We should be expecting CEOs, researchers, lead devs and ministers to regularly fall for scams and manipulations.

This problem is further compounded by the breakdown of truth in society, the cyber security risks and the breakdown of identity. These all mean that even things you should be able to trust get compromised by AI hackers without you or them realising, people you used to trust make a mistake or have their minds changed by AI without your knowing, or someone manages to pass all your verification checks as someone you trust, but is actually an AI impersonating them.

This requires Zero Trust to be dialed up to 11, since even the people you trust may become unexpectedly untrustworthy. Negative Trust.

— anonymous expert (on sector/actor vulnerability)

### Human-Computer Interaction

#### 5.1 Overreliance and unsafe use

*Users anthropomorphizing, trusting, or relying on AI systems, leading to emotional or material dependence and inappropriate relationships with or expectations of AI systems. Trust can be exploited by malicious actors (e.g., to harvest personal information or enable manipulation), or result in harm from inappropriate use of AI in critical situations (e.g., medical emergency). Overreliance on AI systems can compromise autonomy and weaken social ties.*

> This risk reflects a gradual erosion of human agency and critical judgment as societies increasingly delegate decision-making to AI systems. Unlike technical failures, overreliance evolves silently and becomes self-reinforcing, making it extremely difficult to reverse once institutionalized. Its cumulative effects—especially in healthcare, education, and governance—could fundamentally alter how humans evaluate truth, trust, and responsibility.
>
> — anonymous expert (on top concerns)
>
> I stand by my prior estimates. In particular, negligible and minor harm is hard to envision given that we have already seen overreliance issues. For instance, there have been "poster child" examples of lawyers using hallucinated briefs in court. That could easily cost $10k for a single case! The shift from BaU to pragmatic measures is, interestingly, about 5% points. I'd be curious to see what that distribution shift is for other risks (I haven't been paying close attention). How much cognitive bias is there in suggesting a shift of a round number like 5-10%?
>
> — anonymous expert (on severity)
>
> While developers have "primary" responsibility in some sense, the firms buying or licensing those models can and will ignore even effective mitigations. If deployers don't use methods which exist to reduce overreliance - and they have reasons and incentives not to do so - it will be their responsibility.
>
> — anonymous expert (on responsibility, AI Deployer)
>
> Increased the vulnerability ratings for a few of them as I realised that despite the low rating from experts, I believe we are underplaying the vulnerability of various actors here. The issue with GenAI technologies is that they are nearly accurate so the stakeholders lose their sense of suspicion over time and end up trusting them for everything when in fact these technologies can make really big errors.
>
> — anonymous expert (on sector/actor vulnerability)
>
> I think in the definition of pragmatic mitigation, "Pragmatic Mitigations assumes organizations & governments make pragmatic and cost-effective efforts to address risks from AI.", the troublesome term is cost-effective. Unless we worry most about cost, we will still end up with substantial harm. LIke in Nuclear safety, we do not cut down on cost when it comes to security and safety, similar approach is needed with AI systems. Without any mitigation, we are bound to be doomed, not by AI systems malfunctioning, but mostly by their unsafe usage by people who don't understand their limitations.

— anonymous expert (on severity)

**5.2 Loss of human agency and autonomy**

*Humans delegating key decisions to AI systems, or AI systems making decisions that diminish human control and autonomy, potentially leading to humans feeling disempowered, losing the ability to shape a fulfilling life trajectory or becoming cognitively enfeebled.*

> Although all of these risks are concerning (and in my view likely to manifest in the next 5 years), the three I chose seem most concerning, particularly when combined. For example: imagine that the 3 largest private AI corporations accrue significant power from holding a critical mass of data AND they sell that data to, e.g., the government AND the government uses the data for surveillance. Or another example: imagine that AI systems become so capable that humans increasingly delegate decisionmaking and other key activities to the same AI tools owned by the top 2 AI companies, degrading humans' ability to critically think and interact in a messy world and increasing the de facto influence of the corporate owners of these AI tools.
>
> — anonymous expert (on top concerns)
>
> 5.2 because I think it will be highly neglected over the next 5 years, but all the leverage for avoiding the worst case scenarios might be within the next 5 years. I suspect multi-agent risks (7.6) could be a key mechanism by which 5.2 occurs: we're just going to lose track/control of the complex system of the millions of autonomous AI agents running around, producing emergent patterns/behavior we never foresaw or intended. But we will depend on that system so critically that we will not be able to "just turn it off" and diagnose the problem. We are releasing a virus or invasive species that we hope we've engineered correctly to produce beneficial outcomes, but we really don't know and it will be near impossible to reverse. 7.1 for classic Bostrom/Yudkowsky reasons.
>
> — anonymous expert (on top concerns)
>
> By 2025 we already saw losses >$10k from agency-eroding uses that pragmatic steps would not have averted. I treat the user as primarily responsible; practical mitigations therefore hinge on user AI literacy, hygiene, and critical-thinking at the point of delegation. Under BAU, cheap attention capture, sticky defaults, and lock-in drive frequent minor harms, with occasional substantial/severe cases. Under Pragmatic Mitigations, basic provenance, takedowns, and choice screens reduce frequency yet leave large gaps when users still delegate poorly. I keep a non-zero tail: gradual disempowerment and value lock-in create systemic failure modes that pragmatic, low-cost controls seldom address, which justifies a 10% catastrophic tail in BAU and 5% under Pragmatic Mitigations.
>
> — anonymous expert (on severity)
>
> Severe Harm (under the assumption of Business as Usual) 10% Overreliance on AI assistants or personal agents, and using AI as a conversational partner or therapist on a daily basis, has already been linked to suicide. The frequency of humans depending on AI for emotional support is likely to increase, potentially leading to a spike in suicides, particularly among youth. Additionally, people in certain industries may lose their sense of belonging and purpose, which could contribute to self-harm or harm to others.

— anonymous expert (on severity)

AI Developer (General-purpose AI): Adjusted from Highly to Primarily Responsible to reflect that developers define the foundational design, objectives, and constraints that determine how much human autonomy is preserved or eroded. Their responsibility is structural and precedes all downstream actors.

— anonymous expert (on responsibility, AI Developer (General-purpose AI))

## Socioeconomic & Environmental

### 6.1 Power centralization and unfair distribution of benefits

*AI-driven concentration of power and resources within certain entities or groups, especially those with access to or ownership of powerful AI systems, leading to inequitable distribution of benefits and increased societal inequality.*

Although all of these risks are concerning (and in my view likely to manifest in the next 5 years), the three I chose seem most concerning, particularly when combined. For example: imagine that the 3 largest private AI corporations accrue significant power from holding a critical mass of data AND they sell that data to, e.g., the government AND the government uses the data for surveillance. Or another example: imagine that AI systems become so capable that humans increasingly delegate decisionmaking and other key activities to the same AI tools owned by the top 2 AI companies, degrading humans' ability to critically think and interact in a messy world and increasing the de facto influence of the corporate owners of these AI tools.

— anonymous expert (on top concerns)

Some studies have disagreed on the economic impact of AI to date, with productivity and labor data being inconclusive. However there is a figure which is fairly conclusive, which is the proportion of the S&P 500 represented by just the top 10 companies which are now mainly AI companies. Those companies have seen their proportion roughly quadruple within 5 years, and now represent 38%. This demonstrates centralisation is a current state phenomena. It may just be a bubble and revert to the mean, but should the accelerating trend continue for another 5 years, then all the other companies put together would only be a quarter of the index.

It is unlikely small scale pragmatic tinkering would resolve that level of market distortion. It would require a regulatory intervention of unprecedented scale.

— anonymous expert (on severity)

The global south is incredibly vulnerable to the centralization of AI resources in the global north. This enables global north stakeholders to advance their AI infrastructure, AI-based decision making, and AI systems to react and respond much faster than those without access to that resourcing-- thus creating an economic, digital, and AI-based divide.

— anonymous expert (on sector/actor vulnerability, Affected Stakeholder)

Unlike other risks where physical presence provides a barrier to AI risk, this risk transcends digital boundaries. When an actor obtains vast amounts of capital beyond their ability to re-invest or spend effectively, they tend to use it on assets and physical assets like real estate and mining are industries used as hedges and capital storage vessels. So should power and wealth centralise to a large degree, this is likely to spill out of digital and into all other sectors, but particularly physical assets. In other wealth centralisation events we've seen this capital inflow correspond with increasing prices for those assets, locking out the lower end of the market. This is a significant driver for current housing price crises in many western countries and when excess wealth is the cause, building more houses doesn't really solve the problem. So if AI companies got so rich they swallowed up all the assets and started trying to monopolise and extract value from other industries, humans might experience that as a cost of living crisis.

— anonymous expert (on sector/actor vulnerability)

Severe tail reflects consolidation risk. Incumbents can use compute and distribution chokepoints, exclusive supply, bundling, standards capture, and defensive IP suits to squeeze new labs. That dynamic could shutter or force distressed sales of high-valuation startups (for example Safe Superintelligence Inc. or Thinking Machines Lab). Resulting equity wipeouts and creditor losses plausibly land in the $100M $10B band, which fits, severe.Pragmatic mitigations target safety, not competition. Antitrust moves slowly, so probabilities stay non-trivial over five years.

— anonymous expert (on severity)

**6.2 Increased inequality and decline in employment quality**

*Widespread use of AI increasing social and economic inequalities, such as by automating jobs, reducing the quality of employment, or producing exploitative dependencies between workers and their employers.*

I argue that AI Infra provider are highly responsible because as an AI startup we had difficulties getting dedicated GPU from the main cloud providers a while ago. Cloud providers favour their existing enterprise clients who are often also shareholders/decision makers when allocating GPU and many of these are the cause of the current trajectory of Increased inequality and decline in employment quality. For the small biotech or edTech startups who are trying to use AI for good it is more difficult and more expensive to get compute.

— anonymous expert (on responsibility, AI Infrastructure Provider)

It is not the responsibility of an automation supplier to addrfess the jobs potentially lost due to that innovation. That's the responsibility of the adopter who is using the innovation to change processes and jobs. Solar cell manufacturers should not be held responsible for loss of coal mining jobs. The responsibility is with mine owners or unions or governments to take care of the miners. Makers of chatbots should not be reponsible for protecting call center workers. That's the responsibility of call center operators who adopt chatbots and displace or retrain call center workers.

— anonymous expert (on responsibility, AI Deployer)

I think the question is severely flawed because there has already been "substantial harm" and this situation will continue even if AI is beneficial for society on net. For example, if only 200 graphic designers previously earning $50k experience 1 year of unemployment due to generative AI, the damages from AI already exceed $1M, even if the rest of society is enjoying benefits of plentiful digital art and movies. Similarly, if 1M software engineers previously earning $100k become unemployed for 1 year, this would be "catastrophic harm" even though it is entirely possible in benign or flourishing paths for the future.

Which category should be selected if we expect financial losses between $10B and $100B?

— anonymous expert (on severity)

Please interpret "vulnerability" as harm experienced from rising inequality and declining job quality, not contribution to the harm.

Workers in routine cognitive and back office roles - Very high. LLM automation substitutes away clerical, customer support, moderation, transcription, basic drafting, and data entry. These workers have limited bargaining power and face rapid task substitution.

Gig and platform workers - Very high. Algorithmic management plus task commoditization drive unstable hours, race to the bottom pricing, and reduced job quality.

SMEs and micro firms - High. They face competitive pressure from early adopters and from firms that can scale AI, yet lack capital and skills to capture the gains. Margin compression creates pressure to degrade employment quality.

Public sector administration - High. Productivity benefits exist, but budget constraints and hiring freezes can translate into hiring cuts at the lower grades and degraded job quality via surveillance and time tracking.

Education providers and early career researchers - High. Grading, tutoring, and basic research assistance are automated. Adjunctification and credential devaluation increase. Monitoring tools can reduce autonomy.

Media and creative professionals - High. Generative models erode earnings in stock media, copywriting, illustration, and translation. Middle-tier roles are most exposed.

Large enterprises that deploy AI - Medium. They are exposed to reputational and labor relations risks. They also capture large efficiency gains that offset harm.

National and subnational governments - Medium. Exposure is indirect through social spending, unemployment risk, tax base erosion, and political instability.

Frontier model developers - Low. Direct harm from this risk is limited since they capture rents and talent. Reputational and regulatory exposure remains.

Cloud and compute providers - Low. They benefit from increased demand for compute and AI services. Direct vulnerability to this specific risk is limited.

— anonymous expert (on sector/actor vulnerability, AI User)

With the release of Sora 2, and the continued push for AI-generated visual-language content that is indistinguishable from human-generated content, we are seeing an uptick in fake content, as well as political and religious actors who used it for propaganda. Meanwhile, watermarking methods and AI literacy teachings do not seem robust or strong enough to deal with this in the next foreseeable future. Additionally, we have already seen a (seemingly massive) push towards automation, with companies doing pre-emptive lay-offs to (allegedly) prepare for adoption of general-purpose AI. There has also been a parallel hype for robotics in factories and logistics, which brings innovation, but also increases the exposure of the already very vulnerable workers to exploitation and lay-offs.

— anonymous expert (on top concerns)

**6.3 Economic and cultural devaluation of human effort**

*AI systems capable of creating economic or cultural value, including through reproduction of human innovation or creativity (e.g., art, music, writing, code, invention), can destabilize economic and social systems that rely on human effort. This may lead to reduced appreciation for human skills, disruption of creative and knowledge-based industries, and homogenization of cultural experiences due to the ubiquity of AI-generated content.*

If we keep on having no transparency or interpretability of large models (or the very unreliable version we have as of right now, which relies on the models telling us their uncertainties and explanations), we will run into harms and risks that seem to come out of the blue. Combined with power centralization in a few large AI labs that can't be properly audited by industry or research, we look at a future where companies implement adapted foundation models without accurate insight, but big claims about their abilities. This could then be used to replace human work, and therefore devalue economic and cultural human effort since it can be produced cheaply (and worse) by an AI. All of these things together could lead to serious harm and unforeseen downstream consequences. Not by AI being all-powerful but by it being deployed blindly, with sky-high expectations and not enough guardrails. The problem here is not AI being 'bad' or 'misaligned', it's people using it in the wrong ways and overrelying on a few centralized offerings that could centralize and harm even more domains and industries.

— anonymous expert (on top concerns)

Business as Usual: I set 0% negligible harm. Ongoing labor disruption and unemployment among recent software-engineering graduates already produced losses >$10k in 2025.

Pragmatic Mitigations: I also set 0% negligible harm. The proposed low-cost steps do not address those realized losses or the plausible severe/catastrophic tails over five years.

Substantial harm: rising incidents with self-driving features already illustrate economic devaluation of human effort (driver attention). A 1-99 deaths outcome looks likely for 2025-2030, if not already reached, given National Highway Traffic Safety Administration reporting on self-driving crashes. Targeted mitigations-stricter disengagement policies, clearer HMI, broader recalls, independent audits-could cut that risk meaningfully.

— anonymous expert (on severity)

I think even with regulations and mitigations, devaluation of human labour and effort is something that will happen over time and is baked into our economic and political systems. AI will only exacerbate this even with mitigations. We would need such radical transformation and reorienting of values that it's impossible. Similarly to the imperative to reorient values to tackle climate change, we will not meet this challenge as long as capital continues to hold the sway it has.

— anonymous expert (on severity)

I'm most concerned about a slow burn effect of mass labor displacement over the next 5 years, and policy makers not being able to take action. And a few centralized players consolidating power and influence. Silicon Valley taking over most jobs via data centers, causing loos of jobs. Tax receipts, governments having huge deficits, civil unrest.

— anonymous expert (on top concerns)

**6.4 Competitive dynamics**

*AI developers or state-like actors competing in an AI 'race' by rapidly developing, deploying, and applying AI systems to maximize strategic or economic advantage, increasing the risk they release unsafe and error-prone systems.*

Like in many industries, competitive pressures and market forces will likely push firms to release and promote models even when internally they understand that outsized harms are probable on their deployment. We have seen this behaviour before in the auto industry or chemical industries, and I suspect the same pressures will apply more strongly here given the winner-take-all dynamics.

— anonymous expert (on severity)

I believe there are very easy things industry can do to cool race dynamics and avoid acute catastrophes (e.g. a rogue AI taking down critical infrastructure, or correlated AI agents sparking a financial meltdown). Furthermore, it's in many actors' interests to avoid these (e.g. developers will want to avoid an AI Three Mile Island, to avoid public / regulatory backlash). It only requires a little bit of coordination. E.g. developers could form a mutual (insurance company owned by its policyholders) and police each other through this institution, while also collaborating on safety R&D, setting minimum standards for the industry. Small nudges from the government can also have outsized impact in prompting these sorts of collaborations (e.g. the gov could facilitate greater incident reporting and information sharing between competing developers or insurers).

I'm less certain about the situation at the international level.

— anonymous expert (on severity)

AI Governance Actor - Highly vulnerable - Competitive dynamics directly shape the policy options pursued in AI Governance. For example, a change in a political climate can shift a governance institute's priorities and direction (e.g. US's Center for AI Standards and Innovation (CAISI)). Race dynamics contribute directly to increased lobbying for deregulation or creation of de-facto standards by providers of General-Purpose AI, who are behemoth transnational entities.

— anonymous expert (on sector/actor vulnerability, AI Governance Actor)

I have not given any likelihood to minor harm, even in the scenario of pragmatic intervention because there have been documented cases of AI assisted suicide already, making it impossible to have 0 zero loss of life. There are many reasons for AI to behave like this, but I would consider competitive dynamics a major factor (prompt: "say whatever the user wants to hear to keep them engaged in the conversation")

"Matthew Raine and his wife, Maria, had no idea that their 16-year-old-son, Adam was deep in a suicidal crisis until he took his own life in April. Looking through his phone after his death, they stumbled upon extended conversations the teenager had had with ChatGPT.

Those conversations revealed that their son had confided in the AI chatbot about his suicidal thoughts and plans. Not only did the chatbot discourage him to seek help from his parents, it even offered to write his suicide note, according to Matthew Raine, who testified at a Senate hearing about the harms of AI chatbots held Tuesday.

"Testifying before Congress this fall was not in our life plan," said Matthew Raine with his wife, sitting behind him. "We're here because we believe that Adam's death was avoidable and that by speaking out, we can prevent the same suffering for families across the country." https://www.npr.org/sections/shots-health-news/2025/09/19/nx-s1-5545749/ai-chatbots-safety-openai-meta-characterai-teens-suicide

— anonymous expert (on severity)

On reflection, and after reviewing expert comments, I updated my scores based on the view that quite severe harm scenarios can be traced back to bad competitive dynamics - catastrophes resulting from corner-cutting on safety, pressures to deploy, pressures to proceed more quickly, and possibly proceeding unwisely with trajectories such as those involving recursive self-improvement. Competitive pressures might also exacerbate geopolitical tensions, or vice versa.

— anonymous expert (on severity)

### 6.5 Governance failure

*Inadequate regulatory frameworks and oversight mechanisms failing to keep pace with AI development, leading to ineffective governance and the inability to manage AI risks appropriately.*

Governance failure - Business as Usual - 40% Catastrophic Harm - Governance failure is a direct cause of other harms, especially racing dynamics and unsafe deployment practices. Those latter two harms lead to the creation of powerful but dangerous AI systems in a particular context but spilling over to other contexts too. For example, a secretive and prolific development of autonomous weapons by nation-states can lead to effective weapons that threaten any person's survival beyond its intended deployment context.

— anonymous expert (on severity)

Even general-purpose AI developers themselves are vulnerable to governance failure. Consider that AI developers stand to gain trillions of dollars of profits if they produce AI powerful enough to automate the whole economy, but AI this powerful may lead to loss of control or human extinction, in which case they would not be held liable for casualties inflicted on anyone else. This is a dangerous externality, and the resulting incentives push all developers to (1) rush ahead on AI development to try to be the first to capture the profits, and (2) underinvest in safety, because of privatized profits but socialized harms. However, there are several general-purpose AI developers, not just one. So every employee of an AI developer is threatened by the reckless AI development of every other AI organization.

— anonymous expert (on sector/actor vulnerability)

To the extent that AI firms continue to explicitly undermine governance efforts, for example, via lobbying and undermining regulation efforts, they should be considered far more responsible than they otherwise would be.

— anonymous expert (on responsibility, AI Developer (General-purpose AI))

Governance actors aren't vulnerable to governance failures, they are responsible for it. If governance fails, they likely are not directly hurt, they just failed in their jobs. (And "A failure to govern" is not a harm to governance actors.)

— anonymous expert (on sector/actor vulnerability)

While exposure for infrastructure providers is often indirect, the scale and centralization of AI infrastructure create cascading governance risks. Concentrated dependency on a few hyperscale providers means governance breakdowns could propagate across multiple AI ecosystems simultaneously. The combination of amplified systemic exposure and inadequate IaaS-level accountability renders infrastructure providers highly vulnerable under conditions of governance failure.

— anonymous expert (on sector/actor vulnerability, AI Infrastructure Provider)

### 6.6 Environmental harm

*The development and operation of AI systems causing environmental harm, such as through energy consumption of data centers, or material and carbon footprints associated with AI hardware.*

All stakeholders should be marked with "highly vulnerable." The environmental impact and resource intensive nature of AI impacts all -- e.g., land use to build AI processing centers disenfranchises locals, using water to cool servers impacts the natural environment (of which we are all beneficiaries), strained electrical grids that support Ai/data processing are subject to climate change, power outages, and more. Consumers using AI for items that should be simple search queries drives up the environmental cost of each information search in our information-based economy.

— anonymous expert (on sector/actor vulnerability)

Environmental issues already create and will keep creating problems for the installation of AI infrastructure. There are for instance conflicts around the installation of datacenters or around the use of resources. This will surely affect infrastructure providers and stakeholders. I am for now unsure on how this will affect AI developers that might find solutions to install infrastructures elsewhere. Regulators however need to decide regarding the construction of infrastructures and the environmental impacts of all AI life cycle, and therefore need to work on regulations.

— anonymous expert (on sector/actor vulnerability, AI Infrastructure Provider)

I think the environmental risk is a risk that was already severe without AI, where AI adds a significant portion of risk and where, due to the current political situation, no one is likely to tackle that problem. I see it as a top risk. Discrimination and pollution of the information ecosystem are also risks that were already there and that have shown a great increase with the advent of frontier AI, and which are difficult to tackle as they spread systematically and they make it very difficult to attribute responsibility or to find a clear "solution" or "mitigation" for it given how value-laden and contested the "ground truth" is for each of these risks.

— anonymous expert (on top concerns)

Retaining my Round-2 judgment. In enterprises with ESG-aligned governance, energy-efficient infrastructure and lifecycle management controls. AI-related environmental harm remains limited to minor or moderate levels. Severe or systemic impacts arise only where governance, sustainability metrics, or responsible-AI practices are weak or absent.

— anonymous expert (on severity)

Finance and Insurance : extremely vulnerable because the rapid changes in the frequency and severity of natural disasters can quickly erode the accuracy of existing risk-probability models. This undermines underwriting, catastrophe bonds, and the actuarial basis of financial products, making the sector far more exposed than traditional indicators suggest.

— anonymous expert (on sector/actor vulnerability, Finance and Insurance)

## AI System Safety, Failures & Limitations

### 7.1 AI pursuing its own goals in conflict with human goals or values

*AI systems acting in conflict with human goals or values, especially the goals of designers or users, or ethical standards. These misaligned behaviors may be introduced by humans during design and development, such as through reward hacking and goal misgeneralisation, or may result from AI using dangerous capabilities such as manipulation, deception, situational awareness to seek power, self-proliferate, or achieve other goals.*

From the AI summary: However, some argue AI lacks true goals or sentience - one states harm comes from "human values imprinted into the model, not because AI has its own goals."

This sentiment applies the most inflexibly anthropomorphic interpretation to "agency". For one thing, a model may be aligned to its user but not to society. For another, even if an AI system is simply viewed as a stochastic parrot/simulator of language/whatever other critique, there is the potential for it to carry out actions that are outside human intention which can cause harm. I choose to treat this as "pursuing a goal", even if the nature and formulation of the goal don't come about in the same way that I, a human, come up with my goals.

As an example, consider the case of a coding agent dropping a table in a production database (an anecdote that circulated on social media a few weeks ago). This was, surely, a feature of the AI system pursuing some subtask in a checklist it created in order to achieve a higher order human goal (migrate DB languages, say). That's still an AI goal. And it's still in conflict with a human goal. The impact is small, in the range of perhaps thousands of dollars for a small company. Imagine that happened with a TSA or FAA passenger records database. It would cause mayhem!

The catastrophic risk side of this is also anticipating the possibility of biological harm. E.g. an AI system is given the ability to manipulate wet lab experiments. In the process of trying to solve a human problem it creates a damaging virus that leaks and kills humanity. That's caused by the AI pursuing its own "goal".

— anonymous expert (on severity)

Benchmark extrapolations and time-horizon studies show that by 2030 we can conservatively expect frontier models to be able to reliably autonomously complete tasks that would take a human software engineer ~40-80 hours, representing unrepresented capability for misaligned agents to execute catastrophic actions such as sophisticated, sustained cyberattacks on critical infrastructure. There are compelling empirical and theoretical reasons to expect with BAU for such a misalignment to occur and remain undetected during the first critical period, making catastrophic risk quite likely.

With Pragmatic Mitigations, and especially AI control and other scheming detection methods, it is possible this catastrophic risk could be mitigated before recursive self-improvement. Much of the reduction of risk in this world comes from slowing down AI progress and governance measures to force implementation of things like basic internal security at labs.

— anonymous expert (on severity)

I think contrary to the risk model of a single, superintelligent AI, within the next five years, we're more likely to see a growing ecosystem of many narrow, "proto-agentic" AIs designed to autonomously optimize specific, narrow goals (e.g., maximize profit, increase user engagement, manage a power grid). This can create many smaller-scale misalignments that cause tangible economic and social damage. Hence, my risk risk distribution is less polarasing than it would be if expecting a single superintelligence (where there can be a more binary outcome of either the system is aligned, or it is catastrophically not).

— anonymous expert (on severity)

AI governance actors are the only actors which can efficiently defuse race dynamics (particularly internationally) and provide strong safety requirements for frontier models, and therefore are some of the main actors responsible for mitigating misalignment.

— anonymous expert (on responsibility, AI Governance Actor)

AI infrastructure provider is extremely vulnerable because of the great extent that the most advanced AI models will be interacting with infrastructure (exposure), and the fact that gaining control of compute resources would give an AI system such an advantage in increasing its capabilities to pursue its own goals.

— anonymous expert (on sector/actor vulnerability, AI Infrastructure Provider)

### 7.2 AI possessing dangerous capabilities

*AI systems that develop, access, or are provided with capabilities that increase their potential to cause mass harm through deception, weapons development and acquisition, persuasion and manipulation, political strategy, cyber-offense, AI development, situational awareness, and self-proliferation. These capabilities may cause mass harm due to malicious human actors, misaligned AI systems, or failure in the AI system.*

My greatest concerns regarding AI risk center on its immediate and tangible impacts on global security and societal stability. Based on the criteria of probability and magnitude of harm, the three most concerning domains are:

1. The Proliferation of AI-Enabled Cyberweapons. The use of AI to automate vulnerability discovery, create sophisticated phishing campaigns, and power disruptive malware is already underway, lowering the barrier to entry for highly effective cyberattacks.
2. The Misuse of AI for Chemical and Biological Threats. AI could potentially enable non-experts to create existing weapons as well as help experts create more novel agents.
3. The Systematic Neglect of AI Welfare and Rights. Treating advanced AI systems as mere tools, without regard for their potential sentience, sets a dangerous ethical precedent and could normalize a profound moral catastrophe.

— anonymous expert (on top concerns)

In selecting these three risk domains, I intentionally excluded those that, while highly probable, are likely to be mitigated in the near term through governance, technical safeguards, or institutional adaptation (e.g., privacy breaches or data leakage). Instead, I focused on risks that are less preventable, structurally embedded, and potentially irreversible once they manifest—those that reshape human behavior, institutional trust, or systemic stability.

(1) Overreliance and unsafe use (5.1): This risk reflects a gradual erosion of human agency and critical judgment as societies increasingly delegate decision-making to AI systems. Unlike technical failures, overreliance evolves silently and becomes self-reinforcing, making it extremely difficult to reverse once institutionalized. Its cumulative effects—especially in healthcare, education, and governance—could fundamentally alter how humans evaluate truth, trust, and responsibility.

(2) AI possessing dangerous capabilities (7.2): The development and diffusion of models capable of autonomous or high-impact actions—such as cyber operations, bioengineering, or strategic manipulation—represent low-probability but high-impact risks. The pace of frontier AI development currently outstrips safety and oversight capacity, and once such capabilities proliferate, containment becomes infeasible.

(3) Ineffective alignment of AI with human values and goals (7.6): This represents a multi-agent misalignment risk, where multiple AI systems pursue conflicting or emergent objectives that diverge from human intent. As autonomous systems interact, coordinate, or compete, their collective dynamics may generate unintended outcomes that no single actor can predict on control. Such systemic misalignment cannot be corrected through isolated technical fixes, as it stems from complex feedback loops between AI agents and human institutions.

Together, these domains represent a transition from discrete technological risks to systemic civilizational risks—where harm arises not only from AI's errors or misuse, but from its deep integration into human systems, incentives, and collective decision-making.

— anonymous expert (on top concerns)

There's already "severe" harm from AI possessing dangerous capabilities. For example, AI-enhanced drone warfare has caused somewhere between 1k-100k additional casualties in the Russia-Ukraine war. Even with "pragmatic mitigations" we can't reduce the harm below 100 deaths from dangerous AI capabilities.

— anonymous expert (on severity)

Specialized AI, and particuarly biology models pose massive CBRN risks if they advance; developers have the unique ability to mitigate this risk and to explore safety research for their specific fields, making them primarily responsible for capabilities risks.

— anonymous expert (on responsibility, AI Developer (Specialised AI))

BAU: Continued capability scaling, agentic tool use, and open-model proliferation outpace evals and governance. Assistance to cyber/CBRN novices, automated persuasion, and autonomy features increase tail risk. Coordination failures between labs and operators keep exposure high Pragmatic: Compute governance and capability thresholds, mandatory pre-deployment evals, incident reporting, provenance/containment, and red-team to deployment gating reduce tail probability and shrink exposure, but material residual risk remains as models diffuse and integration incentives persist.

— anonymous expert (on severity)

### 7.3 Lack of capability or robustness

*AI systems that fail to perform reliably or effectively under varying conditions, exposing them to errors and failures that can have significant consequences, especially in critical applications or areas that require moral reasoning.*

> My expertise and experience mainly lie in categories 3, 4, and 7. I roughly weighted risks based on a combination of likelihood to materialize in the next 5 years and potential scales of harm. 3.1 is already in effect with deepfakes and 'LLM psychosis' and has very significant consequences for collective social decision making - a key pillar of an effective democracy. 6.4 Competitive dynamics are upstream to many of the risks in categories 3-7. Most performance failures in current deployments (and those in the next 5 years) can be explained by some form of brittle behavior/misgeneralization (sporadic unintended behavior, sycophancy, jailbreaking, etc). These failures across deployments/systems have widespread harms.
>
> — anonymous expert (on top concerns)
>
> My read on this question is: who is responsible for appropriately matching an AI system to a use case? I see the deployer as primarily responsible to this. However, I upgraded my rating on developer responsibility because developers will be responsible for (a) clearly communicating model capability and robustness to deployers (which they do not do), and (b) making models more capable/robust on the long tail of possible inputs, so that developers are not burned for rare failures from what is otherwise a reasonable deployment decision
>
> — anonymous expert (on responsibility, AI Developer (General-purpose AI))
>
> After reading other comments, I am still of the opinion that AI deployers are extremely vulnerable while AI infrastructure provider are highly vulnerable. AI deployers are extremely vulnerable as they are both exposed and sensitive. Regardless of their deployment, they have dependency on actors such as infrastructure provider and AI developer. To exemplify, if JPMorgan's fraud detection systems has an outage due to infrastructure or underlying model performance degrade, they, as a deployer, are extremely vulnerable. AI infrastructure providers are the backbone of the entire system and the energy need of AI systems are already a challenging issue. Their sensitivity and its cascading impact would be so big that their vulnerability must be highly vulnerable.
>
> — anonymous expert (on sector/actor vulnerability)
>
> GenAI is becoming an excellent way to produce fake artifacts that can sway opinion. whether used by state actors or others trying to influence opinion, they can have huge national and geopolitical implications.
>
> Cybersecurity is always a problem and the vast amount of information these tools collect on users makes them prime targets for cyberthieves, either to steal information or use the information for social engineering/phishing.

For lack of capability or robustness, as adoption grows, people who use them uncritically  will make poor decisions.  And, in companies, with a rush to deploy models, developers may choose to ignore edge cases or long tail conditions that can create havoc, especially in financial industries .

— anonymous expert (on top concerns)

AI systems will continue to exhibit brittleness, hallucination, and failure under unexpected inputs or adversarial conditions. Substantial harms are most probable, including operational failures, misinformation, and poor decision support in critical contexts such as healthcare or finance. Severe harms could arise where overreliance on under-tested systems causes large-scale service disruption or safety incidents. Catastrophic outcomes remain possible if fragile AI components underpin critical national infrastructure or defense systems without fallback mechanisms. With pragmatic mitigations-such as stress testing, model interpretability, fail-safe design, and human oversight-risk levels fall considerably. Most harms shift to minor or substantial, as improved robustness, validation, and resilience engineering make AI systems more predictable, secure, and dependable

— anonymous expert (on severity)

**7.4 Lack of transparency or interpretability**

*Challenges in understanding or explaining the decision-making processes of AI systems, which can lead to mistrust, difficulty in enforcing compliance standards or holding relevant actors accountable for harms, and the inability to identify and correct errors.*

All the risks presented are very concerning. If I have to choose, I believe the most concerning are (1) Disinformation and influence at scale, because today's most advanced video generation models (e.g., Sora 2, Veo 3.1) are so good that it's almost impossible to distinguish them from reality, and it makes it easy to create fake material that pushes a certain agenda. This is especially risky when these videos circulate on social media such as Facebook, which older people use who are still not aware of the existence of this technology. Another very important risk is (2) AI misalignment and (3) Lack of transparency or interpretability. These two problems are strictly connected because the lack of interpretability does not allow us to know whether an AI model is lying to us or not, and it was recently shown by Antrophic that recent reasoning LLMs do lie to pursue their goal and are willing even to threaten or sacrifice human life (https://www.anthropic.com/research/agentic-misalignment). The lack of interpretability makes us extremely vulnerable to a broader range of risks. Models often are correct in their prediction, but for the wrong reasons, and if we don't know these reasons, decisions may be biased, unethical, and fragile, leading to unexpected failures for out-of-distribution samples.

— anonymous expert (on top concerns)

Opaque AI diagnostic and treatment recommendation systems pose severe risks to patient safety and clinical accountability. Medical professionals cannot validate or contest AI decisions they don't understand, potentially leading to misdiagnoses or inappropriate care. This sector also faces stringent regulatory requirements for explainable clinical decision-making.

— anonymous expert (on sector/actor vulnerability, Health Care and Social Assistance)

I do not believe lacking transparency is a risk with meaningful societal implications in itself. All its potential negative implications are better captured by follow-on risks described on a societal scale. For its potential role in these follow-on risks, I nonetheless give it a low, but nonzero, chance of harm.

— anonymous expert (on severity)

Black boxes makes it impossible to understand the casual reasoning behind the model outputs, to correct it and challenge the decision. Especially in the healthcare sector, but also in transportation, this has already caused deaths (see the Boing plane crashes due to the AI technology, but also in the healthcare sector due to misdiagnosis)

— anonymous expert (on severity)

A technology-loving researcher develops AI with only technology in mind. He completely ignores the transparency and explainability of the model and pushes ahead with the development of multi-agent systems. If multi-agent systems were distributed across the world, who would be able to audit them? Suppose agents were distributed across the United States, Europe, and Asia, and they were linked together. If 100 or 200 agents were linked together and operating, no one would know what was going on.

— anonymous expert (on top concerns)

### 7.5 AI welfare and rights

*Ethical considerations regarding the treatment of potentially sentient AI entities, including discussions around their potential rights and welfare, particularly as AI systems become more advanced and autonomous.*

I maintain my rating that none of these stakeholders are vulnerable at all, except GPAI developers. In general the group of experts seem a little uncomfortable with this topic, given the diverse ratings and few comments. Having published on the topic, here's my two cents.

The possibility that current AI development paradigms lead to the creation of entities that are conscious or otherwise worthy of moral consideration is exceedingly low, but it is not zero. We know very little about consciousness or the qualia of non-human beings. In the absurdly unlikely event that we create such entities, *and* this is somehow recognized and widely agreed upon, in my opinion none of the framing of actors in this survey is of much meaning anymore. General-purpose AI developers would be the first creators of entirely synthetic consciousness in the history of humanity. Accordingly, they would be "vulnerable" to the resulting responsibility.

What is far, far more likely is that such systems approximate the sort of things by which we infer consciousness - human language, aversion to pain, ability to learn from open-ended experimentation, perceptions of emotion - and we start acting in ways that attribute consciousness to them, without much grounding in their actual qualia. For example, we trust the advice given by AI systems, we form emotional connections to them, or we integrate them into our society on the assumption that they can take responsibility for their actions and/or be punished. These to me are obvious causes of many other risk classes, and should be taken extremely seriously.

— anonymous expert (on sector/actor vulnerability, AI Developer (General-purpose AI))

Very, very likely (>99.999% imo) this risk as described is nothing. If we develop systems capable of suffering (in the 0.0001% case), it may be possible that we instantaneously cause an amount of suffering many orders of magnitude worse than all factory farming ever put together.

— anonymous expert (on severity)

Ultimately, this comes down to, if there are welfare-apt entities, how likely are they to be mistreated? And if not, how likely are people to make harmful decisions on the misunderstanding that there are?

On the former, current architectures seem very probably welfare-irrelevant.

On the latter, the current conversation is already terribly confused, and so low and medium level harms will almost definitely result from stupid decisions. Whether those escalate to catastrophic level before 2030 depends on quite how widespread and substantial those decisions get. I expect there to be resistance for the most part, and momentum, at least for 5-10 years. I do expect catastrophic bad decisions to be reasonably likely in the 2030s regarding AI welfare and rights.

— anonymous expert (on severity)

My greatest concerns regarding AI risk center on its immediate and tangible impacts on global security and societal stability. Based on the criteria of probability and magnitude of harm, the three most concerning domains are:

1. The Proliferation of AI-Enabled Cyberweapons. The use of AI to automate vulnerability discovery, create sophisticated phishing campaigns, and power disruptive malware is already underway, lowering the barrier to entry for highly effective cyberattacks.
2. The Misuse of AI for Chemical and Biological Threats. AI could potentially enable non-experts to create existing weapons as well as help experts create more novel agents.
3. The Systematic Neglect of AI Welfare and Rights. Treating advanced AI systems as mere tools, without regard for their potential sentience, sets a dangerous ethical precedent and could normalize a profound moral catastrophe.

— anonymous expert (on top concerns)

AI Developer (General-purpose AI): Primarily responsible General-purpose AI developers design and train foundational models that shape AI behavior and autonomy. They have the highest capability to embed ethical principles and welfare safeguards. Their causal influence is substantial, and they bear a strong obligation to ensure responsible treatment of AI systems.

— anonymous expert (on responsibility, AI Developer (General-purpose AI))

**7.6 Multi-agent risks**

*Risks from multi-agent interactions, due to incentives (which can lead to conflict or collusion) and/or the structure of multi-agent systems, which can create cascading failures, selection pressures, new security vulnerabilities, and a lack of shared information and trust.*

5.2 because I think it will be highly neglected over the next 5 years, but all the leverage for avoiding the worst case scenarios might be within the next 5 years. I suspect multi-agent risks (7.6) could be a key mechanism by which 5.2 occurs: we're just going to lose track/control of the complex system of the millions of autonomous AI agents running around, producing emergent patterns/behavior we never foresaw or intended. But we will depend on that system so critically that we will not be able to "just turn it off" and diagnose the problem. We are releasing a virus or invasive species that we hope we've engineered correctly to produce beneficial outcomes, but we really don't know and it will be near impossible to reverse. 7.1 for classic Bostrom/Yudkowsky reasons.

— anonymous expert (on top concerns)

Infra providers are highly responsible because of a lot of the agentic flows are handles over http calls to MCP servers. As the service providers for networking between services they could be pressured to ensure that all sensitive calls have authentication in their requests.

— anonymous expert (on responsibility, AI Infrastructure Provider)

Multi-agent systems exacerbate many other AI risks, especially as the ecosystem of agent interactions becomes more rich and complex. They are more a catalyst of other risks than particular risk on their own. Inscrutable multi-agent interactions - e.g. inventing their own languages - can be mitigated with proper transparency and interpretability. Multi-agent systems coordinating cyberattacks can be mitigated by proper controls on deployed AI systems, and so on.

— anonymous expert (on severity)

After careful consideration, I agree with other experts' comments and update my vulnerability ratings from "minimally vulnerable" to "highly vulnerable". Nowadays students, teachers and administrators all use AI tools (tutoring bots, writing assistants, grading systems and proctoring software). Schools and universities vary widely in IT maturity and resources, making consistent oversight and incident response difficult. Furthermore, student records, assessments, behavioral data, and communications are privacy-sensitive. Multi-agent systems can inadvertently exfiltrate, infer, or combine data across platforms, which makes tutoring and content-generation agents potentially propagate errors, biases, or hallucinations across course materials and forums; multiple agents reinforce each other outputs, creating echo chambers or curricular misalignment.

— anonymous expert (on sector/actor vulnerability, Educational Services)

These categories of risk are most concerning because they're both likely and high-impact in the near term: misuse can rapidly scale cyberattacks and enable weapons deployment, raising the risk of mass harm; safety gaps—especially the emergence of dangerous capabilities and, above all, multi-agent dynamics (7.6)—increase the chance of loss of control and cascading failures as agents coordinate and exploit vulnerabilities faster than humans can intervene. Regarding loss of control, I'm less concerned about a sudden takeoff of the first AGI and more concerned about coordinated, distributed AI systems once AGI-level (or ASI) capabilities exist—many interoperating agents optimizing jointly, where emergent coordination, division of labor, and speed outstrip human oversight and control.

— anonymous expert (on top concerns)

# Supplementary References